\documentclass{emulateapj} 
\usepackage{longtable}
\slugcomment{Accepted to ApJ}

\newcommand{\sm}{M$_{\odot}$}
\newcommand{\sr}{R$_{\odot}$}

\newcommand{\cm}{$~{\rm cm^{\scriptscriptstyle -2}}$}

\newcommand{\msol}{M$_{\odot}$}
\newcommand{\ergs}{$~{\rm erg~s^{\scriptscriptstyle -1}}$}
\newcommand{\ergcm}{$~{\rm erg~cm^{\scriptscriptstyle -2}}$}
\newcommand{\ergcms}{$~{\rm erg~cm^{\scriptscriptstyle -2}~s^{\scriptscriptstyle -1}}$}
\newcommand{\kms}{\ensuremath{{\rm km~s}^{-1}}}

\newcommand{\kmsMpc}{${\rm km~s^{\scriptscriptstyle -1}~Mpc^{\scriptscriptstyle -1}}$}

\newcommand{\xtransient}{XRT~080109}
\newcommand{\snd}{SN~2008D}
\newcommand{\snxray}{XRT~080109/SN~2008D}
\newcommand{\snaj}{XRF~060218/SN~2006aj}
\newcommand{\snbw}{GRB~980425/SN~1998bw}

\newcommand{\ubvri}{\protect\hbox{$U\!BV\!RI$} }
\newcommand{\bvri}{\protect\hbox{$BV\!RI$} }

\newcommand{\degree}{\ensuremath{^\circ}}

\newcommand{\ebvhost}{$E(B-V)_{\rm{Host}} = 0.6~\pm~0.1~{\rm mag}$}

\newcommand{\absmagVsn}{$-17.0 \pm 0.3$}%
\newcommand{\absmagBsn}{$-16.3 \pm 0.4$}%
\newcommand{\distd}{$D = 31\pm2~{\rm Mpc}$} 
\newcommand{\He}{\ion{He}{1}}

\newcommand{\HeFive}{\ion{He}{1}~$\lambda$5876}
\newcommand{\HeSix}{\ion{He}{1}~$\lambda$6678}
\newcommand{\HeSeven}{\ion{He}{1}~$\lambda$7065}

\newcommand{\Oxy}{[\ion{O}{1}]}
\newcommand{\Ca}{[\ion{Ca}{2}]}
\newcommand{\OxySeven}{\ion{O}{1}~$\lambda$7774}

\newcommand{\OxyOne}{[\ion{O}{1}]~$\lambda\lambda$6300, 6363}
\newcommand{\OxyTwo}{[\ion{O}{2}] $\lambda\lambda$7319, 7330}

\newcommand{\CaTwo}{[\ion{Ca}{2}]~$\lambda\lambda$7291, 7324}

\newcommand{\nad}{\ion{Na}{1}} 
\newcommand{\caii}{\ion{Ca}{2}~H\&K}

\newcommand{\otwo}{[\ion{O}{2}]~$\lambda3727$}

\newcommand{\synCo}{$^{56}$\rm{Co}}
\newcommand{\synNi}{$^{56}$\rm{Ni}}

\newcommand{\firstphot}{0.1} 
\newcommand{\lastphot}{35} 

\begin{document}

\title{From Shock Breakout to Peak and Beyond: Extensive Panchromatic Observations of the Type Ib Supernova 2008D associated with {\it Swift} X-ray Transient 080109}

\author{M.~Modjaz\altaffilmark{1,2}, %
W.~Li\altaffilmark{1}, %
N.~Butler\altaffilmark{1,3}, %
R.~Chornock\altaffilmark{1}, %
D.~Perley\altaffilmark{1}, 
S.~Blondin\altaffilmark{4,5}, %
J.~S.~Bloom\altaffilmark{1,6,7}, 
A.~V.~Filippenko\altaffilmark{1}, 
R.~P.~Kirshner\altaffilmark{5}, 
D.~Kocevski\altaffilmark{1}, 
D.~Poznanski\altaffilmark{1}, 
M.~Hicken\altaffilmark{5}, 
R.~J.~Foley\altaffilmark{1,5,8}, 
G.~S.~Stringfellow\altaffilmark{9}, 
P.~Berlind\altaffilmark{5}, 
D.~Barrado~y~Navascues\altaffilmark{10}, 
C.~H.~Blake\altaffilmark{5}, 
H.~Bouy\altaffilmark{1,11}, 
W.~R.~Brown\altaffilmark{5}, 
P.~Challis\altaffilmark{5}, 
H.~Chen\altaffilmark{12}, 
W.~H.~de~Vries\altaffilmark{13,14}, 
P.~Dufour\altaffilmark{15}, 
E.~Falco\altaffilmark{5}, 
A.~Friedman\altaffilmark{5}, 
M.~Ganeshalingam\altaffilmark{1}, 
P.~Garnavich\altaffilmark{16}, 
B.~Holden\altaffilmark{17}; 
G.~Illingworth\altaffilmark{17}; 
N.~Lee\altaffilmark{1}, 
J.~Liebert\altaffilmark{15}, 
G.~H.~Marion\altaffilmark{18}, 
S.~S.~Olivier\altaffilmark{19}, 
J.~X.~Prochaska\altaffilmark{17}, 
J.~M.~Silverman\altaffilmark{1}, 
N.~Smith\altaffilmark{1}, 
D.~Starr\altaffilmark{1,7}, 
T.~N.~Steele\altaffilmark{1}, 
A.~Stockton\altaffilmark{20}, 
G.~G.~Williams\altaffilmark{21}, 
and W.~M.~Wood-Vasey\altaffilmark{22} 
}
\altaffiltext{1}{Department of Astronomy, University of California, Berkeley, CA 94720-3411; mmodjaz@astro.berkeley.edu .}
\altaffiltext{2}{Miller Fellow.}
\altaffiltext{3}{GLAST/Einstein Fellow.}
\altaffiltext{4}{European Southern Observatory (ESO), Karl-Schwarzschild-Str. 2, D-85748 Garching, Germany.}
\altaffiltext{5}{Harvard-Smithsonian Center for Astrophysics, 60 Garden Street, Cambridge, MA, 02138.}
\altaffiltext{6}{Sloan Research Fellow.}
\altaffiltext{7}{Las Cumbres Global Telescope Network, 6740 Cortona Dr., Santa Barbara, CA 93117.}
\altaffiltext{8}{Clay Fellow.}
\altaffiltext{9}{Center for Astrophysics \& Space Astronomy, University of Colorado, Boulder, CO 80309-0389.} 
\altaffiltext{10}{LAEX, Centro de Astrobiolog\'{\i}a, (INTA-CSIC), Apdo. 78, E-28691 Villanueva de la Ca\~nada (Madrid), Spain.}
\altaffiltext{11}{Instituto de Astrofisica de Canarias, C/ Via Lactea S/N, E-38205 La Laguna, Tenerife, Spain.}
\altaffiltext{12}{Department of Astronomy, University of Chicago, 5640 South Ellis Avenue, Chicago, IL 60637.}
\altaffiltext{13}{University of California, Department of Physics, 1 Shields Ave, Davis, CA 95616.}
\altaffiltext{14}{Institute for Geophysics and Planetary Physics, LLNL, L-413, 7000 East Avenue, Livermore, CA 94550.}
\altaffiltext{15}{Department of Astronomy, University of Arizona, 933 N. Cherry Ave. Tucson, AZ 85721.}
\altaffiltext{16}{Department of Physics, 225 Nieuwland Science Hall, University of Notre Dame, Notre Dame, IN 46556.}
\altaffiltext{17}{University of California Observatories - Lick Observatory, University of California, Santa Cruz, CA 95064.}
\altaffiltext{18}{Astronomy Department, University of Texas at Austin, Austin, TX 78712.}
\altaffiltext{19}{Lawrence Livermore National Laboratory, 7000 East Avenue, Livermore, CA 94550.}
\altaffiltext{20}{Institute for Astronomy, University of Hawaii, 2680 Woodlawn Drive, Honolulu, HI 96822.}
\altaffiltext{21}{Steward Observatory, University of Arizona, 933 North Cherry Avenue, Tucson,  AZ 85721.}
\altaffiltext{22}{Department of Physics \& Astronomy, 3941 O'Hara St, University of Pittsburgh, Pittsburgh, PA 15260.}

\begin{abstract}
  We present extensive early photometric (ultraviolet through
  near-infrared) and spectroscopic (optical and near-infrared) data on
  supernova (SN) 2008D as well as X-ray data analysis on the
  associated {\it Swift} X-ray transient (XRT) 080109. Our data span a
  time range of 5 hours before the detection of the X-ray transient to
  150 days after its detection, and detailed analysis allowed us to
  derive constraints on the nature of the SN and its progenitor;
  throughout we draw comparisons with results presented in the
  literature and find several key aspects that differ. We show that
  the X-ray spectrum of XRT 080109 can be fit equally well by an
  absorbed power law or a superposition of about equal parts of both
  power law and blackbody. Our data first established that SN~2008D is
  a spectroscopically normal SN Ib (i.e., showing conspicuous He
  lines), and show that SN~2008D had a relatively long rise time of 18
  days and a modest optical peak luminosity. The early-time light
  curves of the SN are dominated by a cooling stellar envelope (for
  $\Delta t \approx 0.1- 4$ day, most pronounced in the blue bands)
  followed by $^{56}$Ni decay. We construct a reliable measurement of
  the bolometric output for this stripped-envelope SN, and, combined
  with estimates of $E_{\rm{K}}$ and $M_{\rm{ej}}$ from the
  literature, estimate the stellar radius $R_{\star}$ of its probable
  Wolf-Rayet progenitor. According to the model of Waxman et al. and
  of Chevalier \& Fransson, we derive $R_{\star}^{\rm{W07}}=1.2\pm0.7$
  \sr\ and $R_{\star}^{\rm{CF08}}=12\pm7$ \sr, respectively; the
  latter being more in line with typical WN stars. Spectra obtained at
  3 and 4 months after maximum light show double-peaked oxygen lines
  that we associate with departures from spherical symmetry, as has
  been suggested for the inner ejecta of a number of SN Ib cores.

\end{abstract}


\section{INTRODUCTION}\label{intro_sec}

The connection between long-duration gamma-ray bursts (GRBs) and
broad-lined Type Ic supernovae (SNe Ic-bl) has recently piqued interest in
certain SNe as GRB progenitors (see \citealt{woosley06_rev}
for a review). A few SNe connected with GRBs have been fairly
well observed, but there exists little early-time and high-energy data
on the comparison set of core-collapse SNe. 

Massive stars (initial mass $M \ga$ 8 \sm) die violently, and the
ensuing core-collapse supernovae (SNe) are classified into different
spectroscopic SN subtypes. The classification depends on the presence
or absence of hydrogen and helium in the SN spectra, and its sequence
is set by the amount and kind of outer envelopes the progenitor
retained before explosion: Type II SNe show prominent H, while Type
IIb SNe show He and weaker H, Type Ib SNe lack obvious H and
conspicuously show He (but see \citealt{branch06}), and Type Ic SNe
lack both H and obvious He (see \citealt{filippenko97_review} for a
review).  SNe IIb, Ib, and Ic are collectively called
``stripped-envelope SNe'' \citep{clocchiatti96}. While the number of
well-observed, normal, stripped-envelope SNe has grown in the past
decade (see \citealt{matheson01,richardson06,modjaz07_thesis} and
references therein), there are few observations at very early times
after the collapse, especially when the shock breaks out of the
stellar envelope \citep{campana06,gezari08,schawinski08}.

Recently, the {\it Swift} satellite \citep{gehrels04} with its rapid
slewing capabilities, space-based nature, and X-ray and
optical/ultraviolet telescopes has opened a new window for SN
observations soon after their massive progenitors explode. A marvelous
opportunity to closely study the shock breakout and evolution of a
nearby SN was offered by X-ray transient (XRT) 080109/SN~2008D. During
the {\it Swift} follow-up observations of SN~2007uy
\citep{nakano08_07uy,blondin08_07uyid} in NGC~2770, XRT 080109 was
discovered serendipitously
\citep{berger08_discgcn,kong08_atel,soderberg08} in the same galaxy in
data obtained with the {\it Swift} X-ray Telescope ({\it Swift}/XRT),
starting on 2008 Jan. 9.56 (UT dates are used throughout this paper).

After the announcement of the XRT detection, numerous groups obtained
data on the optical counterpart (see \citealt{page08_gcn},
\citealt{soderberg08}, and \citealt{malesani09} for a detailed
account), which revealed it to be a SN. From spectra,
\citet{soderberg08_gcn} reported a featureless continuum with a
suggestion of a broad bump near 5500 \AA\ possibly indicating a SN
feature, and \citet{malesani08_idgcn} originally classified the
optical transient as a SN Ib/c via its lack of H;
\citet{blondin08_idiauc} and \citet{valenti08_idiauc} then classified
the transient as a SN Ic possibly with broad spectral lines. Then we
\citep{modjaz08_08dhe} were the first to suggest a classification as a
spectroscopically ordinary SN Ib, a classification supported by
subsequent reports
\citep{valenti08_08dhe,soderberg08,mazzali08,malesani09}. We promptly
established an extensive monitoring program of \snd\ with facilities
at Mt. Hopkins and at Lick Observatory, extending the observing
campaign that was already underway for SN~2007uy in the same galaxy.
Spectra and photometry were obtained on a nightly
basis (weather permitting) starting 2008 Jan. 10, sometimes with
multiple observations per night. In addition, we reduced and analyzed
public {\it Swift}/XRT and {\it Swift}/UVOT data, as well as {\it Chandra}
X-ray observations.  Furthermore, we obtained optical spectra at the
Apache Point Observatory through the awarding of Director's
Discretionary Time.

The mechanism for removing progressively increasing amounts of
the H and He layers in the progenitors of stripped-envelope SNe is not
fully understood. The origin could either lie in strong winds of the
very massive progenitor ($\ga$ 30 \sm; \citealt{choisi86,woosley93}), sudden
eruptions \citep{smith06}, interactions with a binary companion star
(e.g., \citealt{nomoto95,podsiadlowski04}), or the interplay of all
the above. A larger set of SN data is needed to uncover trends between
the underlying properties of progenitor radii, ejecta mass, and amount
of \synNi\ produced \citep{richardson09}.

In \S\ref{xrayobs_sec} we present X-ray data obtained with {\it
  Swift}/XRT and {\it Chandra} of XRT 080109 associated with SN~2008D, and
in \S\ref{photobs_sec} and \S\ref{specobs_sec} we present our
extensive optical and near-infrared (NIR) photometry and spectroscopy
on SN~2008D. Moreover, we present pre-explosion optical and NIR images
from follow-up observations of SN~2007uy that were
obtained 2.8 and 4.5 hours before the onset of \xtransient , yielding
stringent upper limits. We discuss the photometric properties of \snd\
related to the two phases of evolution in \S\ref{phot_sec} and
construct bolometric light curves in \S\ref{bol_fullsec}.
\S\ref{comp_sec} compares it with the rest of the known SNe with
observed shock breakout and emission from the stellar envelope, and
\S\ref{late_sec} presents nebular spectra that reveal double-peaked
oxygen lines, which we attribute to global core ejecta asphericities. In
\S\ref{xrayinter_sec} we discuss in detail the X-ray properties of
\xtransient\ compared with other kinds of X-ray events. We conclude
with \S\ref{conclusion_sec}. Throughout we draw comparisons with
results presented in the literature
\citep{soderberg08,mazzali08,chevalier08,malesani09} and find several
key aspects that differ. Unique aspects of our work include the very
early-time ($<$ 1 day after outburst) NIR data on SN~2008D, NIR
spectra, and late-time optical spectra.

\begin{figure*}[!ht] 
\plotone{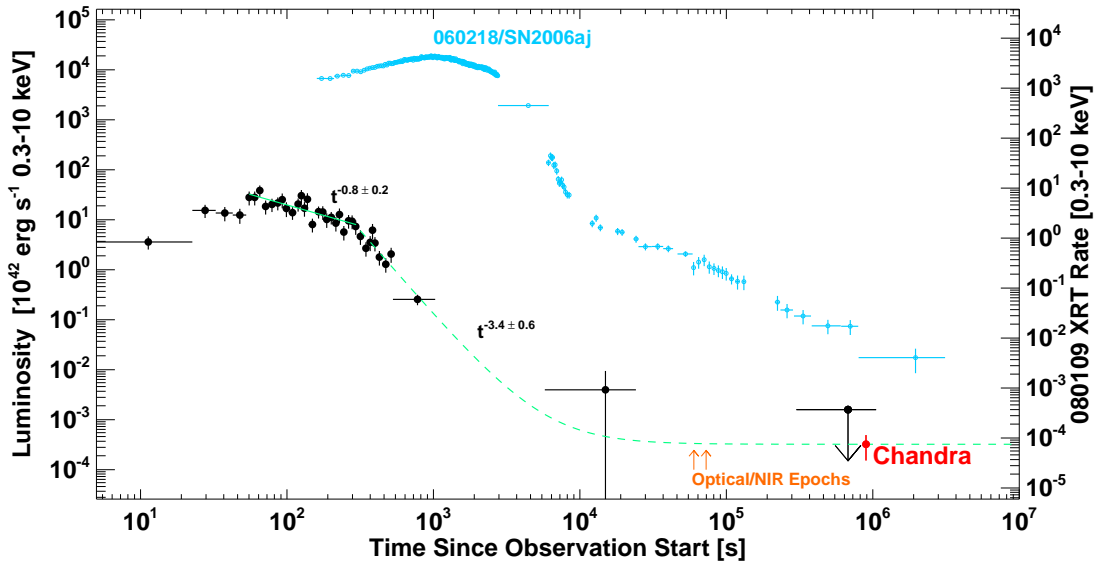}
\caption{Observed temporal evolution of the X-ray luminosity of
  \snxray\ (black points) in the XRT bands (0.3$-$10 keV), since
  $t_0$= 2008-01-09 13:32:49, including the {\it Chandra} observation
  from 2008 Jan. 19. Depending on when one defines the time of onset,
  a range of power-law slopes (green line) can be fit with $\alpha =
  -3.4 \pm 0.6$ (where $L_{\rm{X}} \propto t^{\alpha}$) for data after
  $\sim$ 300 s. We adopt the count rate to luminosity conversion from
  the power-law fit to the spectrum, which is $4.3 \times 10^{42}$
  \ergs\
  per count~s$^{-1}$ .  The epochs on which optical and NIR data were
  obtained are indicated. See text for details. For comparison, we
  also plot the {\it Swift} X-ray light curve of \snaj\ (blue points,
  \citealt{campana06,butler06}), which {\it Swift}/XRT started
  observing $\sim$ 100 s after the gamma-ray trigger.}
\label{xraylc_fig}
\end{figure*}        

\section{X-Ray Photometry and Spectroscopy}\label{xrayobs_sec}

Here, we discuss our independent reduction and analysis of public {\it
  Swift} data and of {\it Chandra} data (PI Pooley;
\citealt{pooley08_gcn}) of \xtransient . While separate reductions and
analysis of the same raw data exist in the literature
\citep{soderberg08,li08,mazzali08,xu08}, our analysis shows that the
data are consistent with a range of combinations for the contributions
of the blackbody and power-law contributions. Constraining the
spectral shape of the X-ray emission and its components has
ramifications for interpreting the X-ray emission mechanism and
directly impacts the question of whether \xtransient\ was produced by
a jet, as for one group \citep{mazzali08} the amount of blackbody
contribution and thus the small derived value for blackbody X-ray
emitting area is a cornerstone for their jet interpretation of the
origin of the X-ray emission.

We downloaded the {\it Swift}/BAT and {\it Swift}/XRT
\citep{burrows05} data from the {\it
  Swift}~Archive\footnote{ftp://legacy.gsfc.nasa.gov/swift/data .}.
The XRT data were processed with version 0.11.4 of the {\tt
  xrtpipeline} reduction script from the
HEAsoft~6.3.1\footnote{http://heasarc.gsfc.nasa.gov/docs/software/lheasoft/
  .} software release.  We employed 4 December 2007 XRT calibration
files. Our reduction of XRT data from cleaned event lists output by
{\tt xrtpipeline} to science-ready light curves and spectra is
described in detail by \citet{butler07}.  In particular, our custom
IDL tools account on a frame-by-frame basis for pileup and source flux
lost due to hot or bad pixels and also perform rejection of pixels
contaminated by nearby field sources.

In order to best account for contamination from nearby sources, we
extract source flux in circular regions of two sizes: 16 and 4.4
pixels radius, corresponding to 90\% and 50\% source containment,
respectively. The spectra were fit using
ISIS\footnote{http://space.mit.edu/CXC/ISIS/ .}. Quoted uncertainties 
below correspond to 90\% confidence intervals.

\subsection{X-ray Photometry: A 520 s Transient}

The X-ray transient \citep{berger08_discgcn} rises in a time $50\pm30$
s, and has a duration $T_{90}=470\pm30$ s or $\Delta t_{\rm
  FWHM}=80\pm30$ s. Consistent with \citet{soderberg08}, we adopt $t_0
$= 2008-01-09 13:32:49 ($\pm$ 5 sec) as the time of onset of the XRT
--- i.e., the start of the {\it Swift}/XRT observations.

In Figure~\ref{xraylc_fig} we plot the X-ray light curve of \snxray ,
after adequately subtracting the other three X-ray field sources as
resolved by {\it Chandra}  observations (\citealt{pooley08_gcn}; see below).
Prior to 1~ks but after the peak time, the X-ray luminosity ($L_{\rm{X}}$)
decay can be fit by a broken power law in time with late-time index
$\alpha= -3.4 \pm 0.6$ (where $L_{\rm{X}} \propto t^{\alpha}$), while data
taken $\Delta t \la$ 300~s are best described by a decaying power law
with $\alpha= -0.8 \pm 0.2$. For comparison, we plot in
Figure~\ref{xraylc_fig} the X-ray light curve of \snaj , the most
recent GRB/XRF-SN
\citep{campana06,mirabal06,modjaz06,pian06,sollerman06}, which was
discovered by {\it Swift} \citep{campana06,butler06}. While the
expression ``X-ray flash'' (XRF, see \citealt{heise01}) is used in the
literature when referring to the high-energy event 060218, the
transient 060218 was qualitatively different from classical X-ray
flashes, in that it likely exhibited a shock breakout component
\citep{campana06,waxman07}. In \S~\ref{xrayinter_sec} we discuss in
detail the properties of the peculiar XRF~060218 and of \xtransient\
compared to other, classical XRFs.

The peak X-ray luminosity of \xtransient\ is $\sim$ 700 times lower
than that of XRF~060218 associated with SN~2006aj. 
 The fluence of the X-ray transient is about $3.0 \times 10^{-8}$
\ergcm\ in the 2$-$10 keV band ($3.7 \times 10^{-8}$ \ergcm\ in the
2$-$18 keV band), where we have adopted the count-rate conversion for
the pure power-law spectral fit (see \S \ref{xrayspec_sec}) of $4.0
\times 10^{-11}$ \ergcms\ , which corresponds to $4.3 \times 10^{42}$
\ergs\ for the adopted distance to \xtransient\ of $D$=31 Mpc.

{\it Chandra}  observations \citep{pooley08_gcn,soderberg08} have shown that
three bright field sources lie near the X-ray transient and likely
contribute substantial counts at late times even for small {\it
  Swift}/XRT extraction regions.  These sources (X1, X2, and X3; see
\citealt{pooley08_gcn}) are 3${\farcs}$8, 1${\farcs}$4, and
2${\farcs}$6 (9.0, 3.2, and 6.1 XRT pixels), respectively, from the
X-ray transient. Using the {\it Chandra}  spectral fitting results and the
{\it Swift}/XRT point-spread function (PSF) from the calibration
database, we determine a field source contamination count rate of
$2\times 10^{-3}$ ($6.9\times 10^{-4}$) counts~s$^{-1}$ for the 16 (4.4) pixel
extraction region.

Correcting to the true source flux level using the PSF model, this
shows that the flux from \xtransient\ (Fig.~\ref{xraylc_fig}) is
dominated by the field source flux for any observations that start 5
ksec after $t_0$ (i.e., the data point at 16 ksec) and is likely
completely overwhelmed by the field source flux after 1 day. We 
therefore use the 16 pixel extraction region for the temporal/spectral
analysis prior to 5 ksec, while we use the 4.4 pixel extraction region
to measure or limit the flux at later times. 

We determine the maximum-likelihood X-ray centroid position and frame
  offset relative to SDSS ($\Delta \alpha$ =0.8\arcsec\ and $\Delta
  \delta$=2.4\arcsec, using 5 X-ray/optical source matches) as
  described by \citet{butler07_astrometry}. Our best-fit XRT position
  is $\alpha$ = 09$^{\rm{h}} 09^{\rm{m}} 30.78^{\rm{s}}$, $\delta =
  +33\degree 08\arcmin 20.1\arcsec$ ($\pm$ 1.6\arcsec, J2000, 90\%
  conf.), which is more uncertain than the optical position of
  SN~2008D with position end figures 30.65$^{\rm{s}}$ and
  20.3$\arcsec$. These positions are consistent with those in
  \citet{soderberg08}.

\subsection{X-ray Spectral Fitting: Power Law vs. Blackbody}\label{xrayspec_sec}

\begin{figure*}[!ht] 
\centerline{\includegraphics[width=4.5in,angle=0]{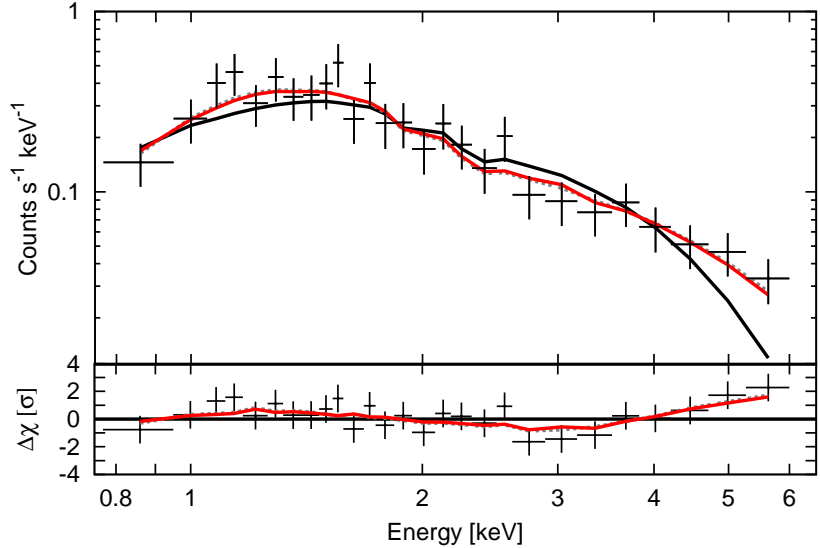}}
\caption{Observed X-ray spectrum of \xtransient\ (crosses) in the {\it
    Swift}/XRT bands (0.3$-$10 keV) averaged over the 520~s duration
  of the event.  \emph{Top}: Best-fit blackbody (BB; black solid
  line), power-law (PL; dashed black line) and combined BB-PL
    (with equal contributions -- red solid line) models. The pure PL
    and the combined PL-BB models are both excellent fits and are
    nearly indistinguishable. The combined PL-BB fit has different
    parameters than the individual pure PL and BB fits.
  \emph{Bottom}: Data residuals (crosses) from the BB model fit (solid
  line), from the PL model fit (dashed line), and from the
  combined BB-PL fit (red solid line). The value for the deviation
  of the respective model fit from the data is the distance between
  the crosses and the respective lines.}
\label{xspecfit_fig}
\end{figure*}        

We restrict our spectral fitting to the time period 0$-$520 s after the
onset of \xtransient , where the contamination from X-ray field
sources is negligible \citep{pooley08_gcn}. 

Different groups, performing analysis of the same {\it Swift}/XRT
data, come to different conclusions: \citet{soderberg08} exclude an
absorbed single blackbody (BB) fit on statistical grounds (with
$\chi^2/\nu=26.0/17$) and thus favor a simple power-law (PL) fit (with
$\chi^2/\nu=7.5/17$); \citet{li08} argues that the data can be fit by
an absorbed two-component BB fit (with $\chi^2/\nu=10.4/16$) as well
as by a pure PL fit, but excludes the two-component BB fit based on
physical grounds arguing that the inferred BB radius ($5\times10^{9}$
cm) is much smaller than the possible progenitor radius.
\citet{mazzali08} claim that for any combined BB and PL fit, the BB
contribution has to be small, namely at most 14\% (whose fit gives
$\chi^2/\nu=20.6/21$), and that the implied small BB radius ($\sim
10^{10}$ cm) is evidence for a GRB jet (see also \S
\ref{xrayinter_sec}). However, we find that the low S/N data of
\xtransient\ alone do not well constrain the contributions from a soft,
thermal excess and a harder, PL continuum. In contrast to
\citet{mazzali08}, we find excellent fits ($\chi^2/\nu=12.57/23$, see
Figure~\ref{xspecfit_fig}.) for a {\it combined} BB and PL fit
assuming a factor unity ratio of PL to bolometric BB flux (i.e., 50\%
bolometric contribution from a BB, see below) without increasing $N_H$
relative to that in the pure PL fit ($\chi^2/\nu=12.73/24$,
$N_H=5.2^{+2.1}_{-1.8} \times 10^{21}$ \cm ). This H-equivalent column
density $N_H$, which is greater than that expected from the Galaxy
($1.7 \times 10^{20}$ \cm ; \citealt{dickey90}), is consistent with
the large reddening value inferred from the optical-NIR data
(\citealt{soderberg08}, \S \ref{bol_fullsec}), given standard
conversions between $N_H$ and reddening \citep{predehl95}.

For the combined BB-PL fit, the time-integrated PL photon index is
$\Gamma = 2.1^{+0.3}_{-0.4}$ (where $N(E) \propto E^{- \Gamma }$), the
BB temperature is $kT=0.10\pm0.01$ keV and $R_{\rm BB}^{\rm X}=
10^{11}$ cm, i.e. a factor of 10 larger than computed by
\citet{mazzali08}. For comparison, the pure BB fit gives
$\chi^2/\nu=28.56/25$ with $R_{\rm BB}^{\rm X}=(1.0\pm 0.2) \times
10^9$ cm and $kT=0.75 \pm 0.07 $keV (Table \ref{xrayvals_table}).
Hence, the apparent low $R_{\rm BB}^{\rm X}$ from the pure BB fit
alone may be an artifact of incorrectly fitting a low signal-to-noise
ratio (S/N) composite BB plus PL spectrum with just a pure BB
model. For completeness, we list in Table \ref{xrayvals_table} our
{\it pure} BB and PL fits, whose parameters are broadly consistent
with the analysis performed by the groups above; specifically,
  our pure PL fit gives a time-integrated photon index of $\Gamma =
  2.1^{+0.3}_{-0.4}$.

Although the exact fraction of soft BB emission in \xtransient\ is
poorly constrained by the data, a case can be made for the possible
presence of BB emission through comparison to XRF~060218.  The
high-S/N X-ray spectrum of XRF~060218 during the proposed shock
breakout phase at $t<6$~ks was predominantly nonthermal but also
required at high confidence a soft ($kT = 0.1$--0.2 keV) BB
\citep{campana06,butler06} with a ratio of about unity between
PL and BB flux \citep{butler06}. If we downsample the spectrum
of XRF~060218 for $t<6$~ks to contain 380 counts as in XRT~080109, we
find that the XRF~060218 spectrum is fit in a strikingly similar
fashion to the XRT~080109 spectrum: a BB fit has
$kT=0.72^{+0.9}_{-0.5}$ keV and local $N_H<3.6 \times 10^{21}$
cm$^{-2}$ ($\chi^2/\nu=35.22/23$), while a PL fit has
$\Gamma=1.7\pm 0.3$ and local $N_H=2.2^{+1.8}_{-1.5} \times 10^{21}$
cm$^{-2}$ ($\chi^2/\nu=9.25/23$). From the low $\chi^2/\nu$, we see
that the PL model with $N_H$ in excess of Galactic also
effectively {\it overfits} the soft spectral excess in XRF~060218,
which requires a PL plus BB given all the counts. We also note
that the light curve of \xtransient\ is quite similar to that observed
for XRF~060218/SN~2006aj, if we allow a time stretch by a factor of
$\sim 0.1$ (Fig.~\ref{xraylc_fig}). We discuss the implications and
interpretation of the X-ray data in \S~\ref{xrayinter_sec}.

\section{Optical Photometry}\label{photobs_sec}

Optical photometry was obtained with the Katzman Automatic Imaging
Telescope \citep[KAIT;][]{filippenko01} and
the 1-m Nickel telescope, both at Lick Observatory, and the 1.2-m
telescope at the Fred Lawrence Whipple Observatory (FLWO).  We
  note that the host galaxy of \snd , NGC~2770, was monitored by KAIT 
  \citep{li08_08diauc}
  during the regular course of the optical Lick Observatory Supernova
  Search \citep{filippenko01,filippenko05} as part of the nearby
  target galaxy sample, and that the regular SN detection software
  successfully detected this SN in data from 2008 January 11.42, which
  is 1.85 day after the X-ray outburst \citep{li08_08diauc}.

\subsection{Photometric Calibrations}

For photometric calibrations, the field of \snd\ was observed in
\ubvri\ during photometric nights on 2008 Jan. 18, 19, and 20 with
KAIT, in \bvri on 2008 Jan. 12 with the Nickel 1-m telescope, and in
$BVr'i'$ on 2008 Jan. 18 with the FLWO 1.2-m telescope. About a dozen
\citet{landolt92} standard-star fields were observed at different
airmasses throughout each photometric night. Photometric solutions to
the Landolt standard stars yield a scatter of $\sim$ 0.02 mag for all
the filters for the Nickel telescope, and about 0.03 mag for KAIT. The
SN 2008D field was also observed for several sets of \ubvri\ images
with different depth in the photometric nights. The photometric
solutions are used to calibrate a set of local standard stars in the
SN 2008D field as listed in Table~\ref{standard_table}, and a finder
chart is shown in Figure~\ref{fchart_fig}.  The calibration based on
the FLWO 1.2-m telescope observations (``FLWO calibration'') is
consistent within the uncertainties with the Lick calibration (from
the KAIT and Nickel telescopes) in the filters that they have in
common ($BV$).

\subsection{Lick Observatory Data Reduction}

\begin{figure}[!ht] 
\plotone{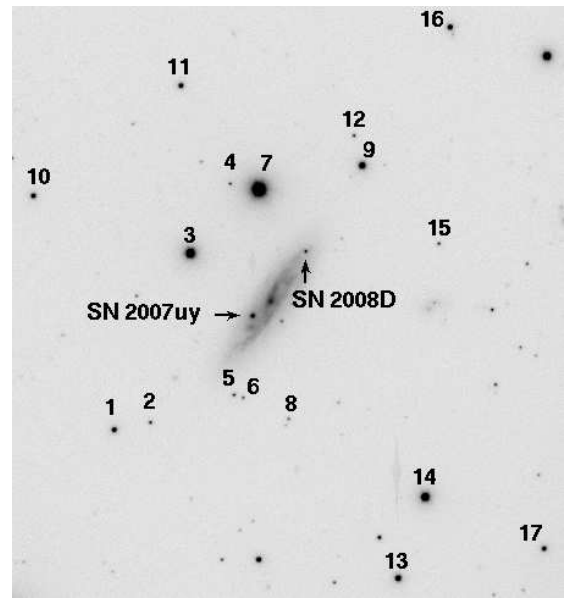} 
\caption[relation] {A finder chart for the local standard stars in the
  field of \snd\ showing both \snd\ and SN~2007uy in NGC~2770. The field of view
  is 10$\arcmin$$\times$11$\arcmin$.  North is up and east is to the
  left. Displayed is the $r'$-band image taken with the FLWO
  1.2-m telescope on 2008 Jan. 29.}
\label{fchart_fig}
\end{figure}

KAIT followed SN 2008D in \bvri and unfiltered mode nightly (weather
permitting) after its discovery.  As SN 2008D occurred on a spiral arm
of NGC~2770, we use image subtraction to remove the contamination of
the host-galaxy emission. For template subtraction, we used
pre-explosion images obtained with the FLWO 1.2-m telescope during the
course of follow-up photometry for SN~2007uy, which also occurred in
NGC~2770 (as shown in Figure~\ref{fchart_fig}).

For KAIT, the image subtraction and subsequent photometry reduction are
performed with the KAIT photometry pipeline (M. Ganeshalingam et al.,
in preparation). Two independent image-subtraction routines were
employed in the pipeline.  One routine is based on the ISIS package
\citep{alard98} as modified by Brian Schmidt for the High-z Supernova
Search Team \citep{riess98}, and the other is based on the IRAF\footnote{IRAF is distributed
  by the National Optical Astronomy Observatory, which is operated by
  the Association of Universities for Research in Astronomy, Inc.,
  under cooperative agreement with the National Science Foundation.}
task PSFMATCH \citep{phillips95}. PSF fitting
photometry is performed on the subtracted images, and the results from
the two routines are averaged whenever appropriate.  Artificial stars
are injected into the original KAIT images and extracted from the
subtracted images to estimate the scatter of the measurements. To
convert the KAIT instrumental magnitudes into the standard Johnson
$BV$ and Cousins $RI$ system, the color terms for the KAIT filters as
determined from many photometric calibrations are used in the
conversion (as detailed by \citealt{modjaz01} and
\citealt{li01_02cx}). The final error bars for the magnitudes are the
scatter in the artificial-star simulation and the calibration error
added in quadrature.

We note that the FLWO 1.2-m telescope template images were taken with
the Johnson $BV$ and Sloan $r\arcmin i\arcmin$ filters, while the KAIT
observations of \snd\ were taken with the Johnson $BV$ and Cousins
$RI$ filters. The difference between the $r\arcmin  i\arcmin$ and $RI$ filter
transmission curves raises the possibility of a systematic uncertainty
during the image-subtraction process. We investigated this issue with the
artificial-star simulations and concluded that the systematic uncertainty
is smaller than $\sim 0.03$ mag (which is not reported in the final uncertainties). 

\snd\ was remotely observed with the Lick Observatory 1-m Nickel
telescope in \bvri on 2008 Jan. 11 and 12. These data were processed
with proper bias and flat-field images, and the image subtraction and
photometry were performed with the KAIT photometry pipeline (but
modified to deal with the Nickel telescope images). The FLWO 1.2-m
telescope template images were used to generate the subtracted images,
and the Lick calibration and the proper color terms for the Nickel
telescope filters are used to convert the instrumental magnitudes into
the standard system.  The final photometry of \snd\ from the KAIT and
Nickel observations is listed in Table~\ref{opt_table}.

\subsection{FLWO 1.2-m Telescope Data Reduction}

The field of SN~2007uy in NGC~2770 was monitored by the FLWO 1.2-m
telescope in the context of the CfA SN monitoring\footnote{See
  http://www.cfa.harvard.edu/supernova/index.html .} efforts when
SN~2008D was discovered, providing us with valuable pre-explosion
images. We followed SN~2008D on a nightly basis (weather permitting)
in $BVr'i'$. For performing photometry of FLWO 1.2-m telescope images,
we adopted the image-analysis pipeline of the SuperMACHO and ESSENCE
collaborations (see \citealt{rest05}, \citealt{garg07}, and
\citealt{miknaitis07} for details). The operations of this optical
photometry pipeline are discussed in depth by \citet{hicken09}. 
In brief, we employed differential photometry by
measuring the brightness of the SN with respect to a set of stars (see
Figure~\ref{fchart_fig}) in the SN field and we employed the DoPHOT
photometry package \citep{schechter93} to measure the flux of the SN
and its comparison stars. We performed image subtraction with the
robust algorithm of Alard \& Lupton (1998; see also \citealt{alard00}, using as
templates the 1.2-m images taken on 2008 Jan. 08 of the field.
The ``FLWO calibration'' and the color terms for Keplercam
\citep{modjaz07_thesis} were used to convert the instrumental
magnitudes into the standard system.

We present the 1.2-m $BVr'i'$ photometry in Table~\ref{opt_table}.
For multiple observations on the same night we report the weighted
mean of those observations. Intranight observations did not reveal
statistically significant variations for \snd . We furthermore derive
upper limits for \snd\ based on 2008 Jan. 9.44 images of the field of
SN~2007uy, three hours before the onset of \xtransient\
(\S~\ref{xrayobs_sec}), and present them in Table~\ref{opt_table}.
The upper limits were estimated by placing fake stars with a range of
magnitudes at the position of \snd\ on the template images and then
determining when they are detected at varying thresholds above
background.

\subsection{UVOT Data Reduction}

The {\it {\it Swift}}/UVOT has followed SN 2008D intensively 
\citep{kong08_gcn,immler08_gcn1,li08_gcn,immler08_gcn2,soderberg08}.
We retrieved the level-2 UVOT \citep{roming05} data for \snd\ from the {\it Swift} data archive in all available filters
  (i.e., in the $UVW2$, $UVM2$, $UVW1$, $U$, $B$, and $V$ filters). 

We used the UVOT observations of SN~2007uy on 2008 Jan. 6 as the
template images to reduce the data of \snd\ (a total exposure time of
808~s in each of the filters). For the $U$, $B$, and $V$ filters, we
employed the photometric reduction and calibration as described by
\citet{li06} to perform photometry of \snd\ on the subtracted
images, while for the Far-UV filters ($UVW2$, $UVM2$, $UVW1$),
  we used a 3$\arcsec$ aperture and employed the calibration of
  \citet{poole08}. Many tests have shown that the two calibrations by
\citet{li06} and \citet{poole08} are consistent with each other.

The UVOT photometry of \snd\ is reported in Table~\ref{uvot_table},
and is presented in Figure~\ref{lcoptir_fig}, along with the overall
optical photometry, with respect to the start of \xtransient\
(\S\ref{xrayobs_sec}). The UVOT $BV$ values reported here are fully
consistent with those in \citet{soderberg08}, while our $U$ photometry
is systematically fainter by 0.2 mag (consistent within
  systematic errors, see below), and our $UVW1, UVW2, $ and $UVM2$ 3$\sigma$ upper
limits are systematically fainter by 0.2--1 mag. Our analysis used larger
binning periods than that of Soderberg et al., producing deeper images
and enabling additional detections in $UVW1$.  The S/N of the UV
detections is poor, and thus precision photometry difficult.
Specifically, the choice of aperture size affects the final SN
magnitude; for example, using a 2\arcsec\ aperture for the $UVW1$ band
gives magnitudes that are $\sim$ 0.23 mag brighter than for the
3$\arcsec$ aperture that we adopt (even after appropriate aperture
corrections). Thus, we estimate the systematic uncertainty of the
final UV photometry to be on the order of $\sim$ 0.25 mag.

\subsection{Near-Infrared Photometry}\label{nirobs_sec}

We obtained NIR ($JHK_s$) photometric measurements
with the refurbished and fully automated 1.3-m Peters Automated
Infrared Telescope (PAIRITEL)\footnote{See http://www.pairitel.org/ .}
located at FLWO.  PAIRITEL is the world's largest NIR imaging telescope
dedicated to time-domain astronomy \citep{bloom06}. The telescope and
camera were part of the Two Micron All Sky Survey (2MASS;
\citealt{skrutskie06}) project.  $J$-, $H$-, and $K_s$-band (1.2, 1.6,
and 2.2~$\mu$m; \citealt{cohen03}) images were acquired simultaneously
with the three NICMOS3 arrays with individual exposure times of 7.8 s. 

The PAIRITEL reduction pipeline software \citep{bloom06} is used to
estimate the sky background from a star-masked median stack of the SN
raw images. After sky subtraction, the pipeline is used to cross-correlate,
stack, and subsample the processed images in order to produce the
final image with an effective scale of 1\arcsec~pixel$^{-1}$ and an
effective field of view (FOV) of 10\arcmin$\times$10\arcmin. The
effective exposure times for the final ``mosaic'' images ranged from
15 to 20 min.  Multi-epoch observations over the course of the night
(totaling between 40 min to 4 hours) were obtained to search for
intranight variations, but none were found at a statistically
significant level.

For the NIR photometry, we used the image-analysis pipeline of the
ESSENCE and SuperMACHO projects \citep{rest05,garg07,miknaitis07}.  We
performed difference photometry (following \citealt{wood-vasey08})
using the image-subtraction algorithm of \citet{alard98}, similar to
our optical photometry. As a template image for the field of \snd\
we used our PAIRITEL images of SN~2007uy taken on 2008 Jan. 9. Thus,
we also derive NIR upper limits for \snd\ based on images of the field
of SN~2007uy, taken on 2008 Jan. 9 08:26:19, 4.5 hours before $t_0$,
the onset of \xtransient\ (see \S\ref{xrayobs_sec}), and present them
in Table~\ref{nir_table}.  The upper limits were estimated in the same
fashion as for the optical data. We performed the final calibration
onto the 2MASS system by using as reference stars the field stars from
the 2MASS catalog, of which there were 20$-$30 in the FOV. No
color-term corrections were required since our natural system is
already in the 2MASS system.  We present our PAIRITEL photometry in
Table~\ref{nir_table}.

\section{Early-Time Broad-Band Photometry}\label{phot_sec}

\begin{figure*}[!ht]   
\epsscale{0.7} 
\plotone{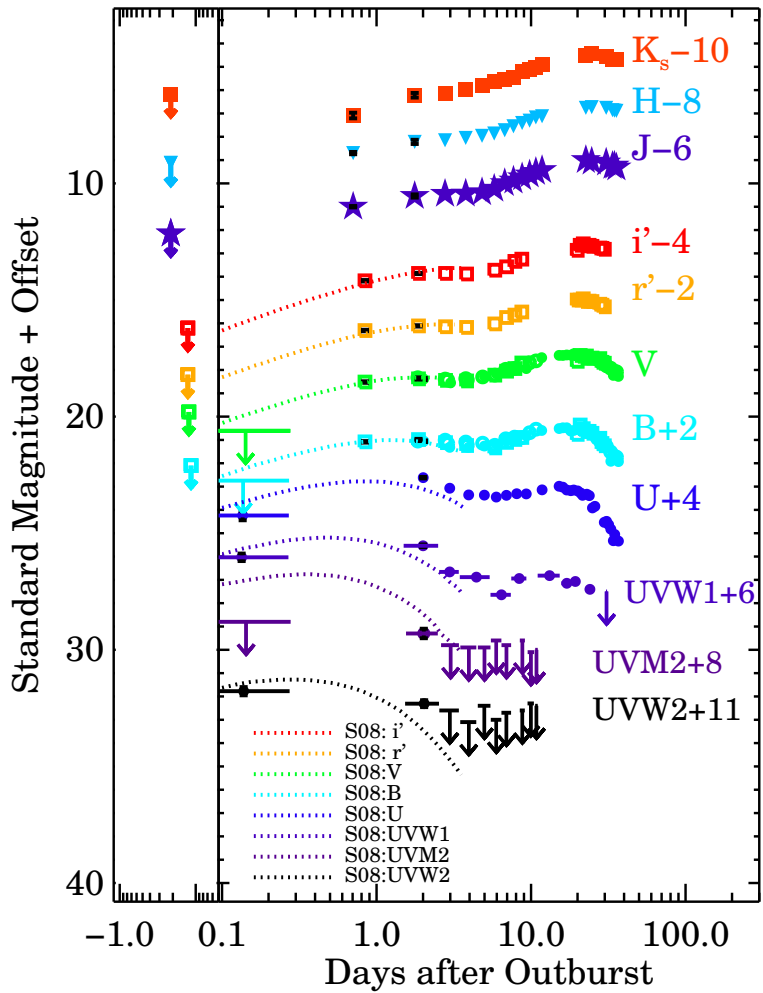}
\caption{Observed optical and NIR light curves of SN 2008D after the
  onset (right panel) of \xtransient\ at $t_0=$ 2008-01-09 13:32:49
  ($\pm$ 5 s), which we adopt as the time of shock breakout. The filled
  circles show $UVW1,UVM2,UVW1,U,B,V$ data from {\it Swift}/UVOT, the
  empty circles are $BV$ data from KAIT (its $RI$ data are not shown
  for sake of clarity), while the empty squares are $BVr'i'$ data from
  the FLWO 1.2-m telescope.  $JHK_s$ data (filled stars, triangles,
  and squares) are from PAIRITEL. Note the very early optical data
  points ($\Delta t=$0.84 day after shock breakout) from the FLWO
  1.2-m telescope, as well as NIR data (at $\Delta t = $0.71 day)
  from PAIRITEL. {\it Swift}/UVOT upper limits are indicated by the
  arrows. We also plot the pre-explosion upper limits derived from the
  1.2-m CfA (at $\Delta t= -$2.8 hr) and the PAIRITEL (at $\Delta t\
  =-$4.5 hr) data (left panel). The data have not been corrected for
  extinction. We note that our earliest ground-based $BVr'i'$ data
  points are consistent with the light-curve fits by
    Soderberg et al.  (2008, S08; dotted lines), who use the envelope
    BB emission model from \citet{waxman07}.  }
\label{lcoptir_fig}
\end{figure*}

Figure~\ref{lcoptir_fig} shows the UV ($UVW2,UVM2,UVW1,U$), optical
($BVr'i'$), and NIR ($JHK_s$) photometric evolution of SN~2008D,
starting as early as \firstphot\ day after burst, and going out to
\lastphot\ day. The data reveal a two-component light curve
\citep{soderberg08}: the first due to the cooling stellar envelope
that is most expressed in the blue bands ($\Delta t$ = 0.1$-$4 day)
and the second due to the standard radioactive decay of \synNi\ and
\synCo\ ($\Delta t \ga 5$ day) with half-life times of 6.1 day and
77.3 day, respectively, which lead to energy deposition behind the
SN photosphere. A number of theoretical studies have computed shock
breakout in the context of SN~II 1987A and corresponding optical light
curves due to the cooling stellar envelope (e.g.,
\citealt{woosley87,ensman92,chevalier92,blinnikov00}), SN~IIb 1993J
(e.g., \citealt{blinnikov98}), and broad-lined SN Ic 2006aj
\citep{waxman07}. In the following, we will concentrate on the models
by \citet{waxman07} as used by \citet{soderberg08} to fit to the {\it
  Swift}/UVOT data and their ground-based data, as well as the most
recent model by \citet{chevalier08}. In \S \ref{radius_sec}, we will
briefly discuss the implications of the differences between those two
models for deriving the progenitor radius.

  The most remarkable aspects are the early-time {\it Swift}/UVOT data and
  our very early optical and NIR data points (at $\Delta t =$ 0.84 and
  0.71 day, respectively), of which the NIR data are the earliest
  broad-band observations reported of a SN Ib. They were obtained as
  part of our observing campaign of SN Ib~2007uy in the same galaxy.
  Here all four ($BVr'i'$) filters indicate a first peak at 0.8--1.8
  day. We note that these earliest $BVr'i'$ data points are consistent
  with the cooling stellar envelope fits (dashed lines in
  Fig.~\ref{lcoptir_fig}) by \citet{soderberg08} to their ground-based
  data, while the fit in the $UVM2$ passband does not agree with the
  photometry as performed by us; the fit predicts brighter magnitudes
  than observed, e.g., by $\sim$ 0.2 mag for the first  $UVM2$ upper limit.

Our $V$-band light curve of \snd\ peaks on 2008 Jan. 27.9 $\pm$ 0.5
(i.e., at $\Delta t$ = 18.3 $\pm$ 0.5 day) at apparent magnitude $m_V$
= 17.33 $\pm$ 0.05 mag. The $B$-band rise time (between shock breakout
and maximum light) is slightly shorter than in the $V$ band, namely
16.8 $\pm$ 0.5 day, peaking on 2008 Jan. 26.4 at $m_B$ = 18.5 $\pm$
0.05 mag. Our apparent peak magnitudes are consistent with other
reports within the uncertainties
\citep{soderberg08,mazzali08,malesani09}.  The rise times of SN~2008D
are at the long end of the range of those observed for
stripped-envelope SNe, and similar to SN~1999ex \citep{stritzinger02}.
After correcting for a Galactic extinction and host-galaxy reddening
of \ebvhost\ (see \S~\ref{bol_fullsec}), these values correspond to a
peak absolute magnitude of $M_V$ = \absmagVsn\ mag and $M_B$ =
\absmagBsn\ mag for \snd , using a distance modulus of $\mu$ = 32.46
$\pm$ 0.15 mag (which corresponds to a distance of \distd).  This
distance modulus is based on the measured heliocentric velocity of
NGC~2770, corrected for the effect of the Virgo cluster and the Great
Attractor by using the local velocity field model given by
\citet{mould00}. These are modest peak luminosities compared to those
of other known SNe Ib and SNe Ic, in particular about 1 mag less
luminous than the mean of the SN Ib sample in \citet{richardson06}
($<M_V>$= $-$17.9 $\pm$ 1.0 mag, recomputed using $H_0=73$ \kmsMpc\
for those SN with known luminosity distances), but within the range of
observed values.  SN~2008D is fainter than SN~1999ex, a well-observed
SN Ib (by $\sim$0.5 mag and 0.8 mag in $V$ and $B$, respectively) and
SN~1993J (by $\sim$ 0.4 mag).

\section{Spectroscopic Observations}\label{specobs_sec}

Optical spectra were obtained with a variety of instruments:
Gemini-South via queue-scheduled observations (GS-2007B-Q-2, PI Chen),
the 6.5-m MMT, the 10-m Keck I and Keck II telescopes, the 6.5 m Clay
Telescope of the Magellan Observatory located at Las Campanas
Observatory (LCO), the 3-m Shane telescope at Lick Observatory, the
Astrophysical Research Consortium 3.5-m telescope at Apache Point
Observatory (APO), and the FLWO 1.5-m telescope. The spectrographs
utilized were GMOS-South \citep{hook03} at Gemini, Blue
Channel \citep{schmidt89} at the MMT, LRIS \citep{oke95} and HIRES
\citep{vogt94} at Keck I, DEIMOS \citep{faber03} at Keck II,
LDSS-3 (Mulchaey \& Gladders 2005) at LCO, the Kast Double
Spectrograph \citep{miller93} at Lick, the DIS spectrograph at the
APO, and the FAST spectrograph \citep{fabricant98} on the FLWO 1.5-m
telescope.  A detailed journal of observations is shown in
Table~\ref{spec_table}.

Almost all optical spectra were reduced and calibrated employing
standard techniques (see, e.g., \citealt{modjaz01}) in IRAF and our own
IDL routines for flux calibration. In addition, the HIRES
spectrum was reduced using HIRedux\footnote{
  http://www.ucolick.org/$\sim$xavier/HIRedux/index.html .}, an IDL package
developed by one of us (J.X.P.) for the HIRES mosaic, and the GMOS-S
and DEIMOS spectra were reduced using slightly different techniques
(see \citealt{foley06,foley07}).

Extractions of the low-resolution spectra were done using the optimal
weighting algorithm \citep{horne86}, and wavelength calibration was
accomplished with HeNeAr lamps taken at the position of the targets.
Small-scale adjustments derived from night-sky lines in the observed
frames were also applied. The spectra were either taken at the
parallactic angle \citep{filippenko82}, at low airmass, or with an
atmospheric dispersion corrector, in order to minimize differential
light loss produced by atmospheric dispersion, except for the
GMOS-South spectrum. Telluric lines were removed following a procedure
similar to that of \citet{wade88} and \citet{matheson00_93j}. The
final flux calibrations were derived from observations of
spectrophotometric standard stars of different colors to correct for
second-order light contamination in the MMT and FAST spectrographs
(see \citealt{modjaz08_Z} for more details) .

Four NIR spectra were obtained on 2008 Jan. 14.45,
21.51, 28.51 and 2008 Mar. 14.48 using the 3.0-m telescope at the NASA Infrared
Telescope Facility (IRTF) with the SpeX medium-resolution spectrograph
in prism (LRS) mode \citep{rayner03}.  The SpeX instrument provides
single-exposure coverage of the wavelength region 0.8--2.5 $\mu$m
in either cross-dispersed or prism mode.  The data were reduced and
calibrated using a package of IDL routines specifically designed for
the reduction of SpeX data (Spextool v3.2; \citealt{cushing04}).
Corrections for telluric absorption were performed using the extracted
spectrum of an A0V star and a specially designed IDL package developed
by \citet{Vacca03}. Additional details of the data acquisition and
reduction methods are given by \citet{marion08}. 
\begin{figure*}[!ht] 
\centerline{\includegraphics[width=4.25in]{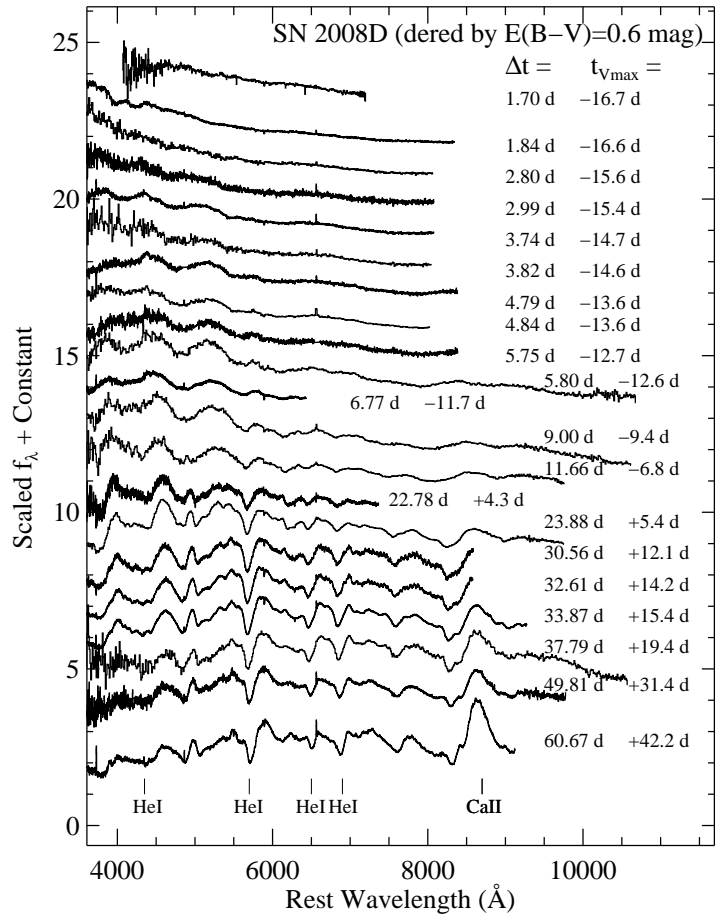}}
\caption{Spectral evolution of \snd , dereddened by
  $E(B-V)_{\rm{Host}}=0.6$ mag and labeled with respect to date of
  shock breakout, $t_0$ = 2008--01-09 13:32:49 (indicated by $\Delta
  t$), and to date of $V$-band maximum, 2008 Jan. 27.9 (indicated by
  $t_{V\rm{max}}$). Note the fleeting double-absorption feature around 4000
  \AA\ in our early spectrum at $\Delta t = 1.84$ day, which is
  discussed in \S~\ref{wfeature_sec}. Some spectra have been binned to
  more clearly visualize their effective S/N. Unfortunately, we have a
  gap in our spectroscopic coverage around maximum light. The
  characteristic optical \ion{He}{1} lines (due to blueshifted 
  He~I $\lambda\lambda$4471, 5876, 6678, 7061)
  become visible starting $\Delta
  t \approx$ 12 day or $t_{V\rm{max}} \approx -$6 day. Note that not
  all spectra are presented in this figure. For late-time
  spectra, see \S~\ref{late_sec}. }
\label{montspecdered_fig}
\end{figure*} 
\subsection{High-Resolution Spectroscopy: Probing the Interstellar Medium}\label{highres_sec}

We obtained one high-resolution spectrum with Keck I on 2008 Jan.
12.5 (i.e., at $\Delta t = 2.95$ day), over the observed wavelength
range 3440$-$6300 \AA. We detect underlying \ion{H}{2} region emission
lines (Balmer lines and \otwo), as well as \nad\ and \caii\ absorption
lines in the host galaxy (see also \citealt{soderberg08,malesani09}).
The \nad\ and \caii\ absorption lines have identical redshifts as the
centroid of the [\ion{O}{2}] and Balmer emission lines, $z=0.00700 \pm 0.00005$.
Since both emission and absorption lines show the same velocity, we
consider the \nad\ and \caii\ absorption lines to have interstellar
origin (as opposed to circumstellar) and attribute the offset of
$\sim$160 \kms\ from the measured redshift of the nucleus of the host
galaxy NGC~2770 ($z_{\rm{nuc}}=0.006451 \pm 0.000087$;
\citealt{falco99}) to the galaxy's rotation curve. Given that NGC 2770
is an inclined, edge-on, typical spiral (Sb/Sc) galaxy
\citep{soderberg08,thoene08}, rotation on the order of 160 \kms\ at
the distance of the SN from the nucleus ($\sim$ 9 kpc) is plausible;
the full rotation curve of NGC~2770 extends to $\sim$400 \kms\
\citep{haynes97}.

Since the absorption lines are saturated, we
can only place lower limits to the column densities of \nad\ and \caii
: $N_{\rm{Na~I}} > 1.0 \times 10^{13}$ \cm\ and $N_{\rm{Ca~II}}
> 1.4 \times 10^{13}$ \cm . We measure an equivalent width (EW) for
\ion{Na}{1} D$_2$ of 0.67 $\pm$ 0.04 \AA\ (at rest), and for
\ion{Ca}{2} K of 0.55 $\pm$ 0.04 \AA\ (at rest). 




\subsection{Low-Resolution Spectroscopy}\label{lowres_sec}

In Figure~\ref{montspecdered_fig}, we present the sequence of early-time
optical spectra after subtracting the SN recession velocity, dereddened
by \ebvhost\ (\S~\ref{bol_fullsec}). Furthermore, the spectra are
labeled according to age ($\Delta t$) with respect to date of shock
breakout ($t_0=$ 2008-01-09 13:32:49) and epoch ($t_{V\rm{max}}$),
defined as days relative to $V$-band maximum (2008 Jan. 27.9 = JD
2454493.4, see \S~\ref{phot_sec}). Knowing both reference dates is
crucial for interpreting the spectra and their temporal evolution.

These are amongst the earliest spectra presented of a normal SN Ib
(starting 1.70 day after shock breakout, which means 17.0 day before
$V$-band maximum) and provide insights into the different layers of
the SN ejecta. As the photosphere recedes with time, it reveals
lower-velocity material since the SN ejecta are in homologous
expansion (where $v \propto R$). Moreover, we obtained late-time
spectra at $t_{V\rm{max}}=+$91 and $+132$ day that show the advent
of the nebular epoch, in which the spectrum is marked by strong
forbidden emission lines of intermediate-mass elements \Oxy\ and \Ca\
(not shown in Fig.~\ref{montspecdered_fig}). We discuss the late-time
spectra in \S~\ref{late_sec}.

At early times our densely time-sampled series shows the dereddened
spectra to consist of a blue, nearly featureless continuum (except around
4000~\AA), with superimposed
characteristic broad SN lines not attributed to hydrogen
\citep{malesani08_idgcn,blondin08_idiauc,valenti08_idiauc,soderberg08}. In
\S~\ref{wfeature_sec} below, we
discuss the fleeting, double-absorption feature around
4000~\AA\ which is only seen in our spectrum at
$\Delta t =$ 1.84 day.

Figure~\ref{montspecdered_fig} shows that the optical helium lines
(blueshifted He~I $\lambda\lambda$4471, 5876, 6678, 7061) become
apparent starting at $\Delta t \approx$ 10--12 day (8--6 day
before $V$-band maximum) and grow stronger over time, allowing this SN
to be classified as a SN Ib 
(\citealt{modjaz08_08dhe,valenti08_08dhe}; see also
\citealt{soderberg08,mazzali08,malesani09}). This is in stark contrast
to all SNe associated with GRBs that have been exclusively broad-lined
SN Ic (e.g.,
\citealt{galama98,stanek03,hjorth03,malesani04,modjaz06,mirabal06,campana06,pian06}).

\subsection{The Double-Absorption Feature around 4000~\AA}\label{wfeature_sec}

\begin{figure}[!ht]
\plotone{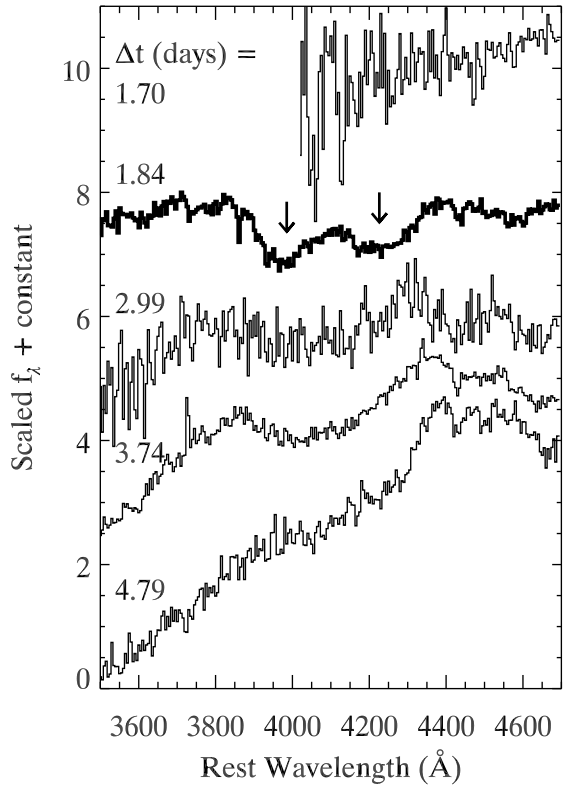}
\caption{
\label{fig:sn2008D_w}
Early-time spectra of SN~2008D in the range 3500--4700\,\AA, from
$\Delta t = 1.70$ to $\Delta t = 4.79$ day past the X-ray outburst,
not corrected for reddening. The transient double-absorption feature
is marked with arrows in the spectrum at $\Delta t = 1.84$ day.
}
\end{figure}

Perhaps the most intriguing aspect of the very early-time spectra of
SN~2008D is the transient, prominent, double-absorption feature around
4000\,\AA\ (Fig.~\ref{fig:sn2008D_w}). It is only present in our second
spectrum, the MMT+Blue Channel spectrum taken on Jan 11.41 ($\Delta t
= 1.84$ day), and was first noted by
\citet{blondin08_idiauc}.

Furthermore, \citet{malesani09} obtained spectra of \snd\ at nearly
the same epoch, Jan. 11.32 (i.e., within two hours of our Gemini and
MMT spectra, see Table~\ref{spec_table}) with wavelength range
extending down to 3700\,\AA , and note that their spectra show the
same double-absorption feature. The double-absorption feature is
apparently absent from our very first spectrum, the Gemini+GMOS
spectrum taken on Jan 11.26, but the spectrum is extremely noisy below
$\sim4500$\,\AA\ due to the particular setup used. The spectrum of
\citet{soderberg08} on the same and subsequent nights apparently does
not extend sufficiently far to the blue to include this feature at
$\sim$4000~\AA\ (their Fig. 3). The double-absorption line is no
longer visible in spectra taken $\ga$ 1 day later (i.e., $\Delta t \ga
2.99$ day) and the two components appear to blend into a single,
broad (FWHM $\approx 23,000$\,\kms) absorption feature (maximum
absorption at rest frame $\sim4100$\,\AA), as clearly seen in the Jan.
13.30 MMT+Blue Channel spectrum ($\Delta t = 3.74$ day) in Figure
\ref{fig:sn2008D_w}. Also, spectra taken by \citet{malesani09} do not
show the feature in spectra taken after Jan. 13.07, i.e., $\Delta t
\ga 3.51$ day.

The double-absorption feature appears to consist of two overlapping
P-Cygni profiles, with rest-frame full width at half-maximum (FWHM) of
$\sim11,500$\,\kms\ and $\sim11,000$\,\kms\ for the bluer (maximum
absorption at rest-frame wavelength $\sim3980$\,\AA) and redder (maximum
absorption at rest-frame wavelength $\sim4240$\,\AA) absorptions, respectively.
Both absorption components are thus significantly narrower than other
absorption features in the same spectrum, whose FWHM ranges between
$\sim15,000$\,\kms\ and $\sim30,000$\,\kms.

\begin{figure}[!ht]
\includegraphics[width=3.25in]{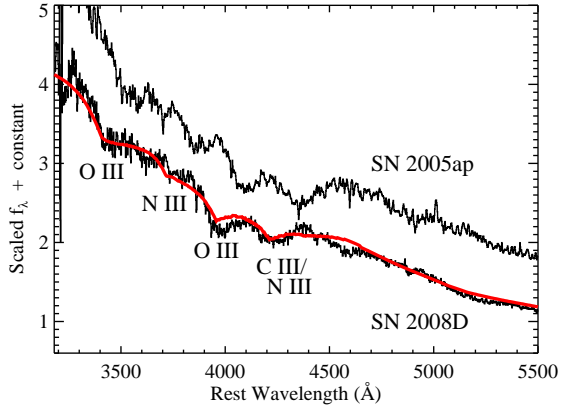} 
\caption{
\label{synow_fig}
The reddening-corrected early-time spectrum of SN~2008D (black) at
$\Delta t = 1.84$ day (2008 Jan 11.41) and for comparison, the
spectrum of SN~2005ap \citep{quimby07}. The fit via the parameterized
spectral synthesis code SYNOW is shown in red: the ``W,''
along with some other weaker features, can be reproduced with a
combination of O III and CIII/NIII.}
\end{figure}

In the following section we attempt to address the line identification
for this feature. Amongst the published supernova spectra, the closest
match to the ``W'' feature that we can find is in the earliest
spectrum of SN 2005ap \citep{quimby07} from 2005 March 7 ($\sim$ 1
week before maximum light), which shows a similar feature at slightly
redder wavelengths (by $\sim$200 \AA).  Quimby et al. used the
parameterized supernova spectrum synthesis code SYNOW
\citep{fisher99,branch02} to fit that spectrum with four ions. They
ascribed the ``W'' feature to a combination of C~III, N~III, and O~III
with a photospheric velocity of 21,000 \kms.  We downloaded the
spectrum of SN~2005ap from the SUSPECT
database\footnote{http://bruford.nhn.ou.edu/~suspect/index1.html .},
generated SYNOW models using only those three ions, and were able to
satisfactorily reproduce the ``W'' feature in SN 2005ap using similar
fitting parameters to those of Quimby et al. We then increased what
SYNOW calls the ``photospheric velocity'' ($v_{\rm{phot}}$) of the fit
to match the observed wavelengths of the ``W'' feature in \snd\ and
found that $v_{\rm{phot}}\approx$ 30,000 \kms\ produced a good match
to the ``W'' as well as to some other weaker features, as shown in
Figure~\ref{synow_fig} (using a continuum temperature of $\sim$13,000
K, consistent with the BB temperature as derived from our photometry
data, Table~\ref{BBfit_table}).  SYNOW did not produce any other
strong features that are associated with these ions, matching the full
extent of the observed MMT spectrum.

While our SYNOW fit produces a good match to the data, we note
that SYNOW fits do not produce unique line identifications and the
suggested identification here needs to be confirmed by detailed
spectral modeling \citep{mazzali98,baron99,dessart07}. Furthermore,
while the ``W'' feature is no longer visible in spectra of \snd\ taken
$\sim$ 1 day later (see also \citealt{malesani09}), it seemed to
linger for $\sim$ 1 week in the spectrum of SN~2005ap (at slightly
lower velocities).

At face value, an identification of the ``W'' feature with C III, N
III, and O III at high velocities seems plausible. We expect the early
supernova spectrum to have high velocities based on an extrapolation
of the observed velocities at late times back to the earliest
observations. In addition, what SYNOW designates as the ``excitation
temperature'' is high ($T_{\rm{ex}} \approx$ 17,000 K) at these early
times\footnote{The apparent incongruency between the excitation and
  continuum temperatures probably arises from the LTE (local
  thermodynamic equilibrium) assumption in the SYNOW code
  \citep{branch02}, or because we underestimated the reddening, but
  this is less likely.} and many of the elements that will contribute
to the later spectra are more highly ionized. The transient nature of
the feature might then be explainable because the elements involved
will recombine to lower ionization stages as the ejecta cool (at
$\sim$15,000 K where most of doubly ionized elements recombine to
singly ionized species).

Alternatively, the reason for the short-lived nature of this ``W''
feature could be asphericity in the SN ejecta. We have evidence from
late-time spectra for asphericity in at least the inner SN core (see
\S~\ref{late_sec}), which may be connected to some inhomogeneity in
the SN ejecta density or chemical abundance profile at the outermost
radii that the early-time spectra are probing. If indeed asphericity
is involved, then existing spectral synthesis codes (such as SYNOW
which assumes spherical symmetry) are challenged to accommodate that
aspect.\footnote{The MMT spectra (including the one displaying the
double-absorption feature) will be made publicly available through the
CfA Supernova Archive at {\tt
    http://www.cfa.harvard.edu/supernova/SNarchive.html.}}

\section{Bolometric Light Curves}\label{bol_fullsec}

\begin{figure*}[!ht] 
\vspace*{-6mm}
\epsscale{1.1}  
\plottwo{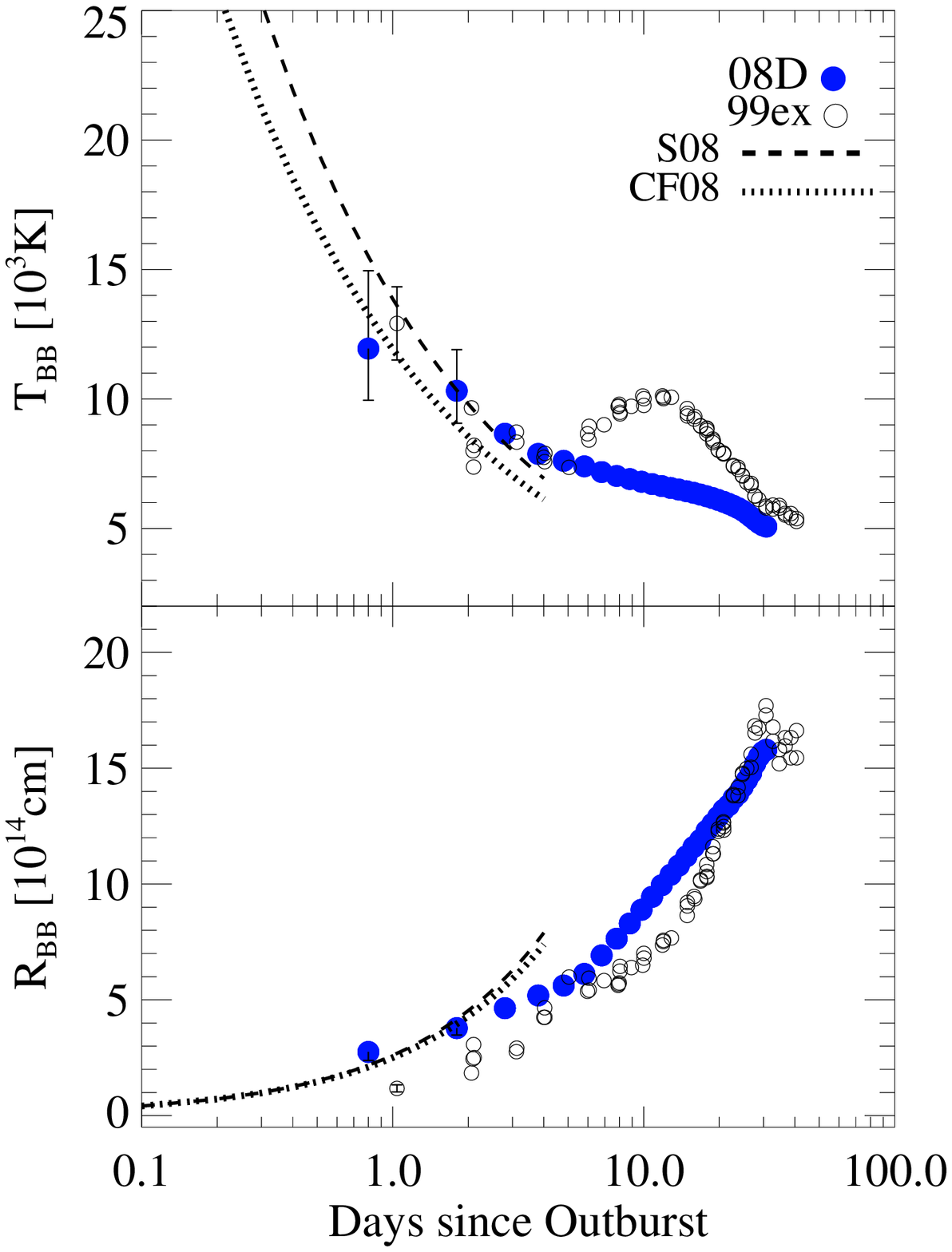}{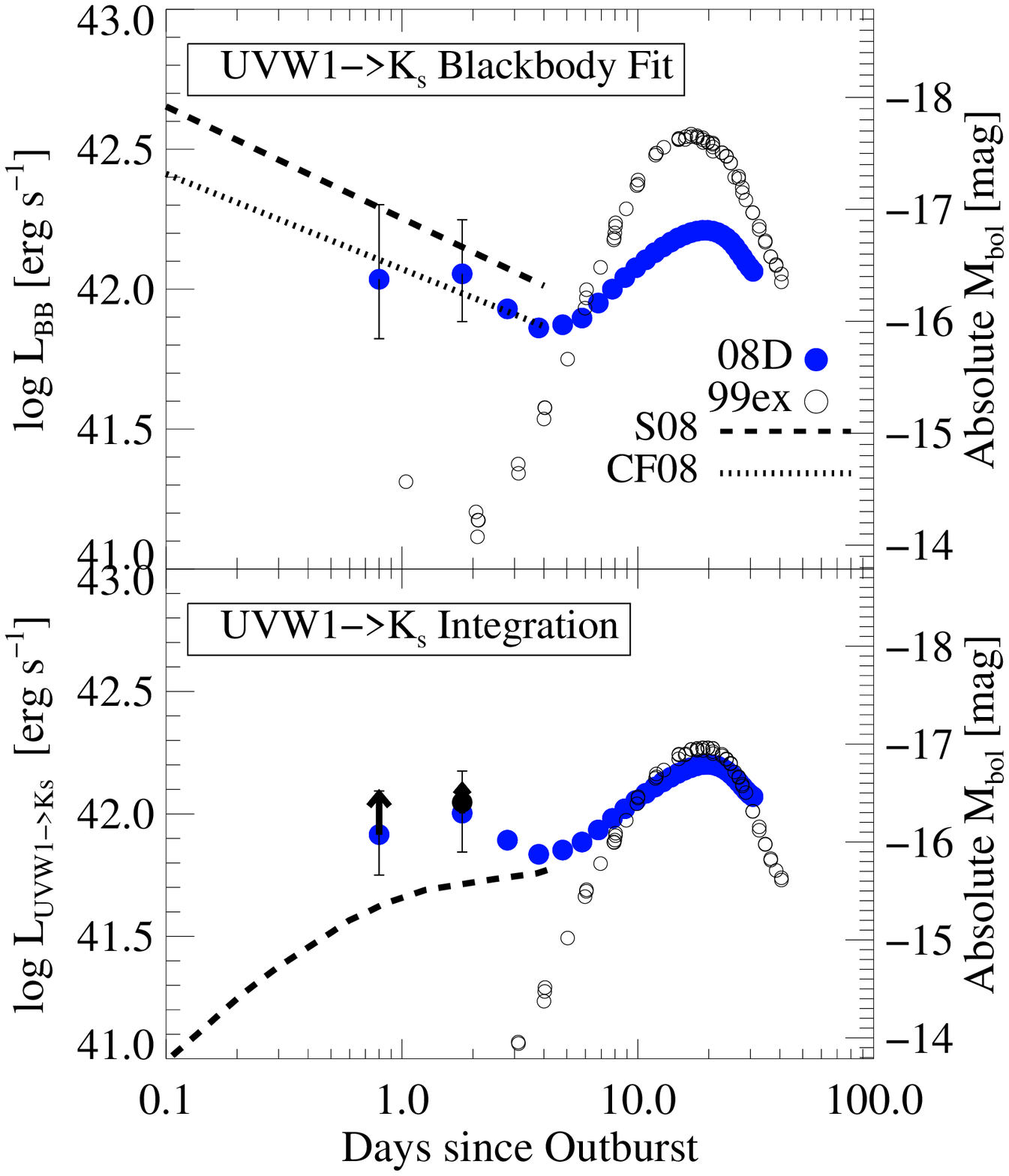}
\vspace*{-2mm}
\caption{ {\it Left:} Temperature (\emph{top}) and radius
  (\emph{bottom}) derived from a BB fit to the optical and NIR SED of
  \snxray\ (filled circles) as a function of time, assuming \distd\
  and corrected for reddening (\ebvhost). The error bars include the
  uncertainties as reported from the formal fits, as well as the
  effect of uncertain extinction.  SN~1999ex (open circles,
  \citealt{stritzinger02}) is shown for comparison.
  We compare our \snd data to
  the values of the stellar cooling envelope model (dashed
  lines) in Soderberg et al. (2008, S08), using the same parameters (in particular, $R_{\star}\approx$ \sr )
  as well as to the model of Chevalier \& Fransson (2008,
    CF08, with their suggested $R_{\star}\approx$9 \sr).
{\it Right:}
  Bolometric light curves of \snxray\ (filled circles) computed via
  two different techniques: fitting a BB to the reddening-corrected
  $UVW1 \rightarrow K_s$ fluxes ($L_{\rm{BB}}$, \emph{top}) and via direct
  integration of the reddening-corrected broad-band photometry ($L_{UVW1
    \rightarrow K_s}$, \emph{bottom}). The black data point at 1.8 day 
   includes the detections in the Far-UV filters and thus represents 
   $L_{UVW2\rightarrow K_s}$. 
   We note that the value for the bolometric magnitude $M_{\rm{bol}}$
   of \snd\ will vary depending on the passband used over which
   different authors (S08, \citealt{malesani09}) integrate its flux.}
\label{bbfit-tr_fig}
\end{figure*} 

Many lines of evidence indicate significant host-galaxy extinction
toward \snd: the high-resolution spectrum (\S~\ref{highres_sec}) shows
strong interstellar \ion{Na}{1}~D and \ion{Ca}{2} lines at the
recession velocity of the host galaxy (with EW[\ion{Na}{1}~D] = 0.7
\AA), while low-resolution spectra show in addition diffuse
interstellar bands \citep{herbig75}. From \citet{soderberg08}, we
adopt $E(B-V)_{\rm{Host}}$=0.6 mag for SN~2008D, and a Milky Way
extinction law \citep{fitzpatrick99} with $R_V = 3.1$.  This value is
used to deredden all of our spectra and is consistent with the value
adopted by \citet{mazzali08}, but 0.2 mag smaller than that used by
\citet{malesani09}. The choice of extinction law does not
strongly affect our results.

This value for the reddening is consistent with fitting an absorbed BB
spectrum to the spectral energy distribution (SED) of \snd\
constructed from our simultaneous broad-band
($UVW2 \rightarrow K_s$) photometry (see \S~\ref{bol_sec}).

In addition, comparison of spectra and $B-V$ colors with those of
other SNe Ib at comparable epochs, such as SN~1999ex
\citep{stritzinger02,hamuy02}, yields $E(B-V) = 0.6$--0.7 mag.
In particular, we compared SN~2008D with SN~2005hg, a normal
  SN Ib that suffered very little host-galaxy extinction
  \citep{modjaz07_thesis}. Comparison of the spectra
  (Fig.~\ref{specomp_fig}) and the broad-band SEDs constructed from
  our $UBVr'i'JKH_s$ photometry for both SNe indicates a host-galaxy
  extinction for SN~2008D of $E(B-V) \approx 0.65$ mag, if both SNe
  have similar color evolution.

\subsection{Constructing Bolometric Light Curves}\label{bol_sec}

The broad-band dataset (spanning 2200--24,100~\AA) presented
here lends itself uniquely to computing the bolometric output of \snd.
We determine it in two different ways: (a) by fitting a BB to the
broad-band photometry and deriving the equivalent BB luminosity
($L_{BB}$) by analytically computing the corresponding Planck
function, and (b) by performing a direct integration of the
$UVW1BVRr'Ii'JHK_S$ broad-band magnitudes ($L_{UVW1 \rightarrow
  K_s}$), including the far-UV filters $UVW2$ and $UVM2$ in
  case of detections. In both cases, the {\it Swift} UV and our NIR data
are crucial, as they provide a factor of at least two in wavelength
coverage and leverage for the SEDs. While theoretical considerations
might justify using a simple BB approximation for the SN emission
shortly after shock breakout \citep{waxman07,chevalier08}, we do note
that without detailed models of SN Ib/c atmospheres, it proves
difficult to say how well BB fits approximate the true SN luminosity
over the full light curve of \snd. We address this question more in
the remainder of the section, when comparing the bolometric luminosities
computed via BB fits and those via direct integration.

We start by interpolating the light curves by {\it Swift}/UVOT (in
$UVW1,UBV$), FLW0 1.2 m (in $r'i'$), and PAIRITEL (in $JHK_s$) onto a
grid spaced regularly in time, as the observing times were different
for different telescopes. We then sampled those light curves at
intervals closest to the actual observing time of {\it Swift}/UVOT and
the FLWO telescopes. After removing extinction (\ebvhost\ and assuming
a the Milky Way extinction law parameterization with $R_V = 3.1$;
\citealt{fitzpatrick99}), we converted our photometry into
monochromatic fluxes at the effective wavelengths of the broad-band
filters, using zeropoints from \citet{fukugita95} (for $UBVr'i'$,
since the UVOT $UBV$ photometric system is close to the Johnson $UBV$
with Vega defining the zeropoint; \citealt{poole08}) and
\citet{cohen03} (for $JHK_s$). We included 0.05 mag of systematic
error in the UVOT filters \citep{li06} and 4\% conversion error when
using \citet{fukugita95}.

For our first method, we perform a least-squares fit of a BB
spectrum to the SED of each epoch and constrain the derived BB
temperature ($T_{BB}$), BB radius ($R_{BB}$), and BB
luminosity ($L_{BB}$) as a function of age (see
Table~\ref{BBfit_table} and Fig.~\ref{bbfit-tr_fig}). The error bars
include the uncertainties from the formal fits as well as
uncertainties introduced by the uncertain reddening ($\pm$0.1 mag),
which affect the derived BB fit parameters differently, depending on
the underlying SED. 

We do not include the systematic uncertainty in distance (of order
$\sim$20\%). At later times ($\Delta t \ga$ 8--10 day) the
assumption of a BB continuum is expected to break down, based on the
optical spectra where distinct absorption lines develop (see
Fig.~\ref{montspecdered_fig}). While the $\chi^2$ values of the fits
are strictly not acceptable except at early times (1 $ < \Delta t <$ 5
day), the overall fits are reasonable even beyond this time with no
obvious residual trends, and we consider the BB measured temperatures,
radii, and luminosities to still be generally quite reliable.
Thus, the UVOPTIR data are well fit by a BB, especially at
  early times (and $\Delta t <$ 5 day), and we do not detect any NIR
  excess (e.g., due to potential dust formation). For clarification,
we note that the optical/NIR data to which we fit a BB were obtained
much later ($\Delta t \ga 10^5$ s) than when the main X-ray emission
period of \xtransient\ occurred ($\Delta t \la 10^3$ s,
\S~\ref{xrayobs_sec}). For $\Delta t$= 1.8 day, we detect the
  SN in the {\it Swift} $UVW2$ and $UVM2$ band and include those data in the
  BB fit, though their inclusion does not have a significant effect on
  the outcome (the parameters change by less than 2\%).

We present the resultant BB luminosity evolution in
Figure~\ref{bbfit-tr_fig} (right, top panel) and include that of
SN~1999ex for comparison (from Table 8 in \citet{stritzinger02}, for
their adopted values of $E(B-V) = 0.28$ mag, and explosion date of JD
2,451,480.5), but with a luminosity distance recomputed with $H_0 = 73$
\kmsMpc . Both SNe exhibit the initial dip from the cooling stellar
envelope emission, with SN~2008D showing a much less pronounced
contrast between dip and peak due to heating of \synNi\ than in SN~1999ex.

The second method for computing the bolometric light output is by
direct integration of the observed broad-band photometry. Here we are
in the favorable position of having NIR data, as the NIR contribution
to the bolometric output is usually unknown for SNe Ib/c and can be
quite large (30--50\%) for certain SNe at late time ($>$20 day)
\citep{tominaga05,tomita06,modjaz07_thesis}, and even larger for
dust-producing SNe (SN~2006jc, e.g., \citealt{smith08_dust,dicarlo08,modjaz07_thesis}).


We integrate the total optical and NIR monochromatic fluxes in the
region between the effective wavelengths of the $UVW1$ and $K_s$
filters via the trapezoid approximation. For this purpose, we
extrapolate the SED to zero at the blue $UVW1$ (2220 \AA) and red
$K_s$ (24,210 \AA) edges of the total set of filters. We plot the
resultant bolometric light curve ($L_{UVW1 \rightarrow K_s}$) in
Figure~\ref{bbfit-tr_fig} (right, bottom panel) and, for comparison,
also that of SN~1999ex (based on $U \rightarrow z$ integration, \citealt{stritzinger02}).

For SN~2008D, we find that the bolometric luminosity based on direct
$UVW1 \rightarrow K_s$ integration is very similar to that based on
the BB fits (within $\la$ 0.04 dex or $\la~8\%$), except for the first
three data points at $\Delta t$ = 0.8, 1.8, and 2.8 day (different by
a factor 1.32, 1.13, and 1.09, respectively). SN~2008D reaches
  bolometric peak at $\Delta t =19.2\pm 0.3$ days with a bolometric
  peak luminosity of $\rm{log}L_{\rm{bol}} = 42.2 \pm 0.1$ \ergs , i.e.,
  $M_{\rm{bol}} =-16.8 \pm 0.1$ mag.

At $\Delta t$ = 0.84 day, the peak of the SED is at wavelengths bluer
than the observed wavelength coverage and explains the missing flux in
the direct integration, which is the absolute lower limit to the total
light output. We note that for SN~2008D, the agreement between
bolometric values derived from BB fits and those based on direct
integration is much better than for SN~1999ex; if the emission from
SN~1999ex can be as well described by a BB as that of SN~2008D, then
the true bolometric peak luminosity of SN~1999ex is not $-$17.0 mag
(Fig.~\ref{bbfit-tr_fig}), but rather $-$17.7 mag; i.e., the SN is more
luminous than previously thoughtm because it was not observed
  in the UV nor in the NIR whose data could have been used to
  construct a full bolometric output.

\begin{figure}[!ht] 
\centerline{\includegraphics[width=3.in,angle=90]{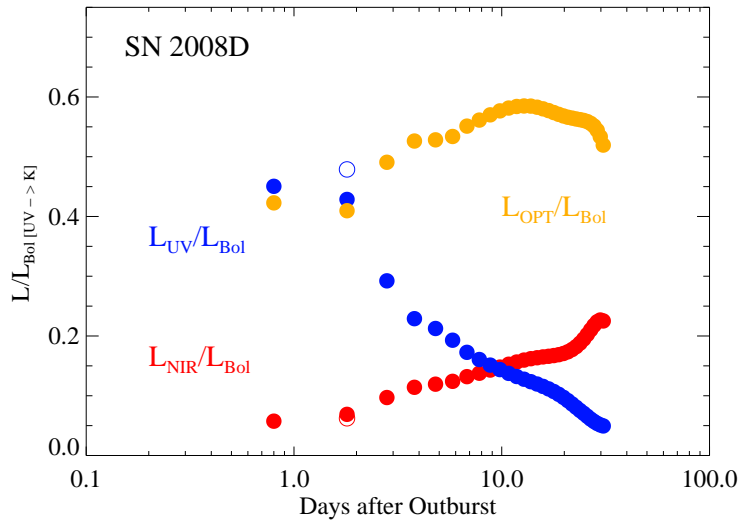}}
\caption{Contribution of the UV ($UVW1,U$: blue circles), optical
  ($BVr'i'$: orange circles), and NIR ($JHK_s$: red circles)
  passbands to the bolometric luminosity $L_{\rm{int}}$ as function of
  age for SN~2008D. The empty circles at 1.8 day denote the
  contributions when including the Far-UV detections (in the $UVM2, UVW2$
  filters). }
\label{lcontribution_fig}
\end{figure}

In Figure~\ref{lcontribution_fig}, we present the contribution
  of different passbands to the full bolometric luminosity output
  $L_{UVW1 \rightarrow K_s}$ as a function of SN age. At early times,
  the UV is 40\% of the bolometric luminosity.  At later times (30
  day after outburst), the NIR is 20\%.  Most supernova observations
  lack these measurements, so their derived bolometric flux has
  significant uncertainties.

\subsection{Constraining the Progenitor Size of
  \snxray}\label{radius_sec}

Now we compare our data to the fits of \citet{soderberg08} via
the cooling BB envelope model of \citet{waxman07}, and to the
  predictions of \citet{chevalier08} as the source for the early-time
emission, where the stellar envelope expands and cools after the
passage and breakout of the shock wave.

In both models, the time evolution of the radius and the temperature
of the cooling envelope are weak functions of the ejecta mass
($M_{\rm{ej}}$) and ejecta kinetic energy ($E_{\rm{K}}$), and a stronger
function of the radius of the progenitor before explosion
($R_{\star}$) to the power of 1/4.  Both models assume that the
photophere is in the outer shock-accelerated part of the SN ejecta
density profile, but their methods differ.  In the following, we use
our early-time light-curve data to provide constraints on $R_{\star}$
if the ratio $E_{\rm{K}}/M_{\rm{ej}}$ can be measured independently from the
SN light curves and spectra \citep{soderberg08,mazzali08,tanaka08}.

Analytically, we can rewrite the equations for the radius and the
temperature of the cooling envelope as given in W07 and in CF08
(equations [18] \& [19], and equations [2] \& [5], respectively), and
generalize such that

\begin{equation}
  \frac{R_{\star}}{ 10^{11}~\rm{cm}} = C ~T_{\rm{BB,ev}}^4~R_{\rm{BB,14}}^{0.4} ~\Delta t_{\rm{d}}^{\alpha}~~\left(\frac{E_{\rm{K,51}}} {M_{\rm{ej,\sun}} }\right)^{\beta},
\label{rstar_eq}
\end{equation}

with 
\begin{eqnarray}
C & = 0.286  ~\rm{for~ W07~or~} & 3.028   ~\rm{for~ CF08}, \nonumber \\
\alpha & = 1.69 ~\rm{for~ W07~or~} & 1.61  ~\rm{for~ CF08}, \nonumber \\
\beta & = -0.24 ~\rm{for~ W07~or~}  & -0.27   ~\rm{for~ CF08},   \nonumber 
\end{eqnarray}

\noindent where $T_{\rm{BB,ev}}$ is the BB temperature of the
photosphere in eV, $R_{\rm{BB}}= 10^{14} R_{\rm{BB,14}}$~cm, $\Delta
t_{\rm{d}}$ time in days after shock breakout, $E_{\rm{K}} = 10^{51}
E_{{\rm K},51}$~erg, and $M_{\rm{ej}} = M_{\rm{ej,\sun}}$~\sm.
Thus, the derived BB fits to the early-time data yield
  independent measurements of $R_{\star}$ for each epoch, for a given
  $E_{\rm{K}}/M_{\rm{ej}}$, whose uncertainty does not influence the
  value of $R_{\star}$ significantly, given the very weak dependence
  of $R_{\star}$ on $E_{\rm{K}}/M_{\rm{ej}}$. While the power-law
  scalings of the parameters are fairly similar for the two models,
  their largest discrepancy lies in a factor of $\sim$10 difference in
  the constant $C$, which is primarily caused by the factor of
  $\sim$2 difference in the temperature equations, as noted by CF08,
  and by the high sensitivity of $R_{\star}$ on $T_{\rm{BB}}$.  

  Here we explicitly calculate $R_{\star}$ for the two models with our
  own data combined with published values of $E_{\rm{K}}/M_{\rm{ej}}$
  and include all relevant sources of uncertainties; using the
  slightly different values for $E_{\rm{K}}/M_{\rm{ej}}$ for SN~2008D
  from the literature \citep{soderberg08,mazzali08,tanaka08}, and our
  estimates for $T_{\rm{BB}}$ and $R_{\rm{BB}}$
  (Table~\ref{BBfit_table}), we compute $R_{\star}$ for each of the
  first four epochs (0.8 $<\Delta t<$ 3.8 day), and for the range of
  adopted extinction values (\ebvhost). Taking an average over the
  epochs and using the different reddening values as bounds, we obtain
  for the W07 model $R_{\star}^{\rm{W07}} = (8.6 \pm \sigma_{\rm{Ex}}
  + \sigma_{\rm{E/M}})\times 10^{10}$ cm, where $\sigma_{\rm{Ex}}$ and
  $\sigma_{\rm{E/M}}$ refer to the systematic error due to
  uncertain extinction and uncertain $E_{\rm{K}}/M_{\rm{ej}}$,
  respectively, with $\sigma_{\rm{Ex}} = 4.8$ and $\sigma_{\rm{E/M}}
  = 0.4$. In other words, $R_{\star}^{W07} = 1.2 \pm 0.7 \pm 0.06 $
  \sr .  Likewise for the CF08 model, $R_{\star}^{\rm{CF08}} = (86 \pm
  45 \pm 4) \times 10^{10}$ cm $= 12 \pm 7 \pm 0.6 $ \sr .  Thus,
  $R_{\star}$ as derived from the cooling stellar envelope emission is
  10 times larger according to the CF08 model than according to the
  W07 model.  CF08 had noted this fact, and furthermore, that their
  larger progenitor size is consistent with $R_{\star}^{\rm{X-ray}}
  \approx$ 9 \sr\ , based on their simple interpretation of the
  published X-ray spectrum of \xtransient\ as a purely thermal X-ray
  breakout spectrum.

 In addition to the literal application of both models, it is
  important to keep in mind the caveats and simplifications that enter
  them, as both models are consistent with the basic idea of emission
  from a post-breakout cooling stellar envelope. Any simplifications
  that lead to a slight mis-estimation of $T_{\rm{BB,ev}}$ will lead
  to a large effect onto $R_{\star}$ due to its large
  temperature-sensitivity. Caveats here include equating effective
  temperature (as given by the models) with observed color
  temperature, and the fact that for temperatures during the time of
  observations ($10^4$ K), the opacity is probably not set by simple
  Thomson scattering as assumed in the models, but is composition
  dependant (E. Waxman, 2009, private communication). These aspects
  mark areas of improvement that need to be addressed if measurements
  like the ones for \snd\ are to be used to routinely and precisely
  measure SN progenitor sizes in the future.

For illustration, we overplot in Figure~\ref{bbfit-tr_fig} the
predicted values for the Waxman model using the parameters put forth
by \citet{soderberg08} for \snd, namely $R_{\star}=10^{11}$ cm,
$M_{\rm{ej}}=5$ \sm, and $E_{\rm{K}} = 2 \times 10^{51}$ erg (see their
caption to Figure 3). Our values are consistent with their fits,
within the uncertainties. For comparison we also plot the predicted
values for the model of \citet{chevalier08} with $R_{\star}=9$ \sr\ .

We can now directly compare our estimated $R_{\star}$ of \snxray\ to
stellar radii of Wolf-Rayet (WR) stars, which possess dense winds (see
\citealt{abbott87,crowther07} for reviews), and are the most likely
progenitors of stripped-envelope SNe (e.g., \citealt{filippenko91}),
including that of \snxray\ \citep{soderberg08}. Specifically, $bona
fide$ WN stars --- the nitrogen-sequence spectral subtype of WR stars
with their helium layer intact but without H (see
\citealt{smith08_wnh} for the distinction between WN and WNH, i.e.,
H-rich WN stars) --- are the likely progenitors of some SNe Ib. Their
radii are in the range 1--20 \sr\, with typical sizes between 5$-$10
\sr\ \citep{herald01,hamman06,crowther07}. Provided that those models
and their data are accurate, $R_{\star}^{\rm{W07}} \approx 0.5-$1.9
\sr\ as suggested by W07 is smaller by a factor of 5$-$10 than typical
values for such WN radii, while the size of $R_{\star}^{\rm{CF08}} =
5-20$ \sr\ as suggested by CF08 is more in line with data of typical
WN stars.

\begin{figure*}[!ht]  
\includegraphics[width=4.5in,angle=270]{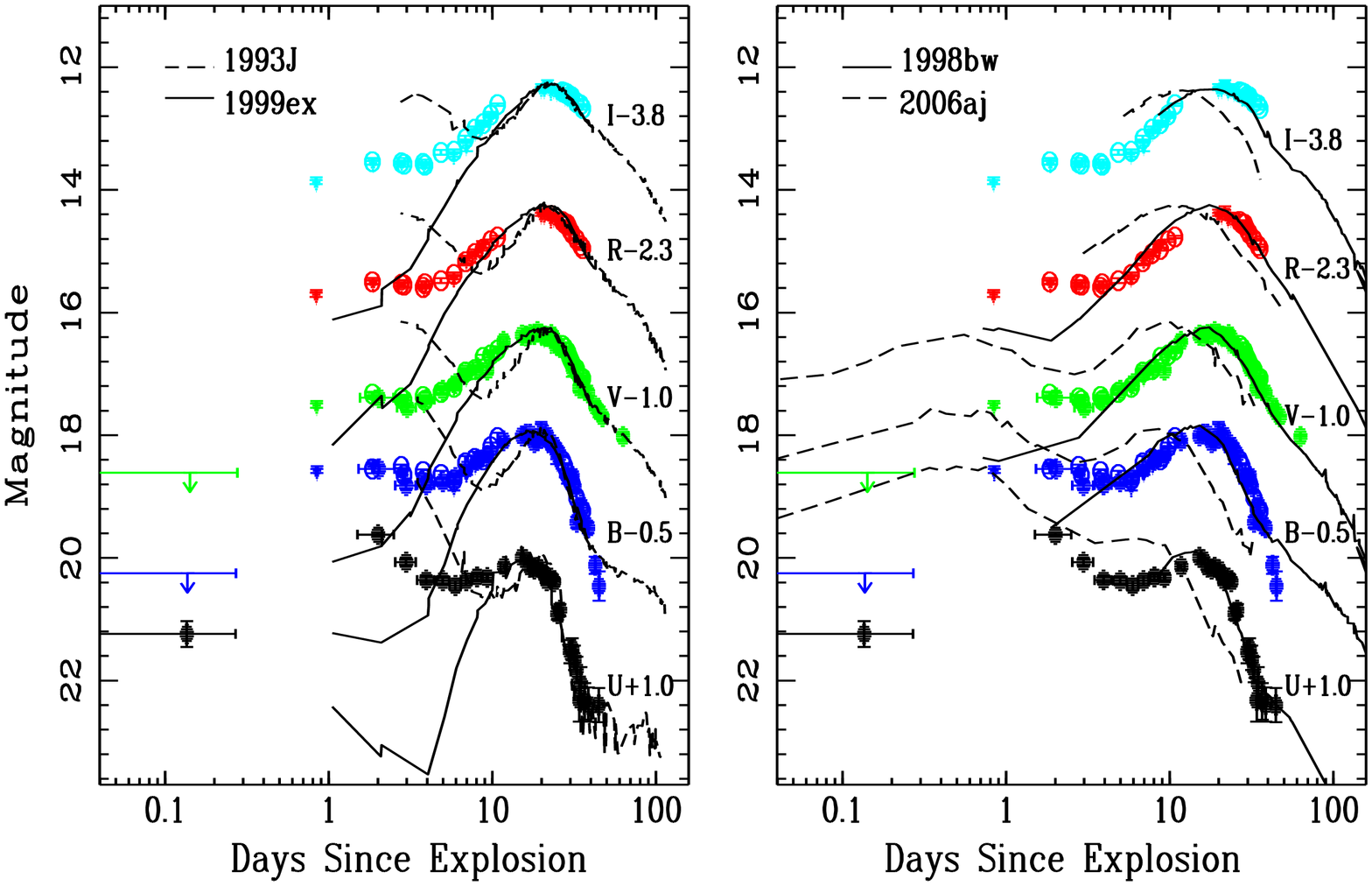}
\caption{Comparison of the optical {\it Swift}/UVOT (filled circles)
  and ground-based (empty circles and triangles) light curves of \snd\
  with those of other normal SNe (\emph{left}) and SNe associated with
  low-luminosity GRBs (\emph{right}, but see \S~\ref{xrayinter_sec}), for which
  shock breakout was observed or the time of explosion is well known.
  Data are from \citet{richmond94} for SN1993J, \citet{stritzinger02}
  for SN~1999ex, \citet{galama98} for SN~1998bw, and \citet{campana06}
  ($UBV$) and \citet{modjaz06} ($r'i'$) for SN~2006aj. The offsets
  refer to the $UBVRI$ light curves of \snd , while its $r'i'$ light
  curves are offset by $r'-2.6$ and $i'-4.3$ to match $RI$.  The rest
  of the SNe are offset to match the peak magnitude of \snd\ in the
  respective bands. For SN~1998bw and SN~2006aj, the SN explosion date
  was taken to be coincident with the onset of GRB~980425 and XRF~060218,
  respectively \citep{galama98,campana06}. For SN~1993J and SN~1999ex,
  the explosion dates were estimated from optical nondetections on
  the night before (for SN~1999ex, JD 2,451,480.5;
  \citealt{stritzinger02}), and on the night during discovery (for SN~1993J, JD
  2,449,074.5 $\pm$ 0.05; \citealt{wheeler93}).  }

\label{lccompexp_fig}
\end{figure*} 

Independent of the considerations above, \citet{soderberg08} infer
from their extensive radio observations that the progenitor had a
steady wind with a mass-loss rate consistent with those of WR stars.

\section{Comparison with other supernovae }\label{comp_sec}

The optical signature of postshock breakout, namely a cooling stellar
envelope (extremely blue color and dip in the first few days
after explosion) has been observed in two stripped-envelope normal
SNe: SN IIb 1993J \citep{schmidt93,richmond94} and SN Ib 1999ex
\citep{stritzinger02}. Other SNe with early-time observations (at
$\Delta t \approx 1$ day or earlier) include the broad-lined SN
Ic~2006aj connected with low-luminosity GRB/XRF~060218
(\citealt{campana06}, but see \S~\ref{xrayinter_sec}), \snbw\ (for the
$V$ and $R$ filters, \citealt{galama98}), the famous SN II~1987A
\citep{hamuy88}, and most recently, two SN II found by the Supernova
Legacy Survey and detected with the UV satellite {\it GALEX}
\citep{schawinski08,gezari08}.

\subsection{Photometry:  Cooling Stellar Envelope and Radioactive Decay as Power Sources}\label{compphot_sec}

\begin{figure}[!ht] 
\centerline{\includegraphics[width=4.5in,angle=270]{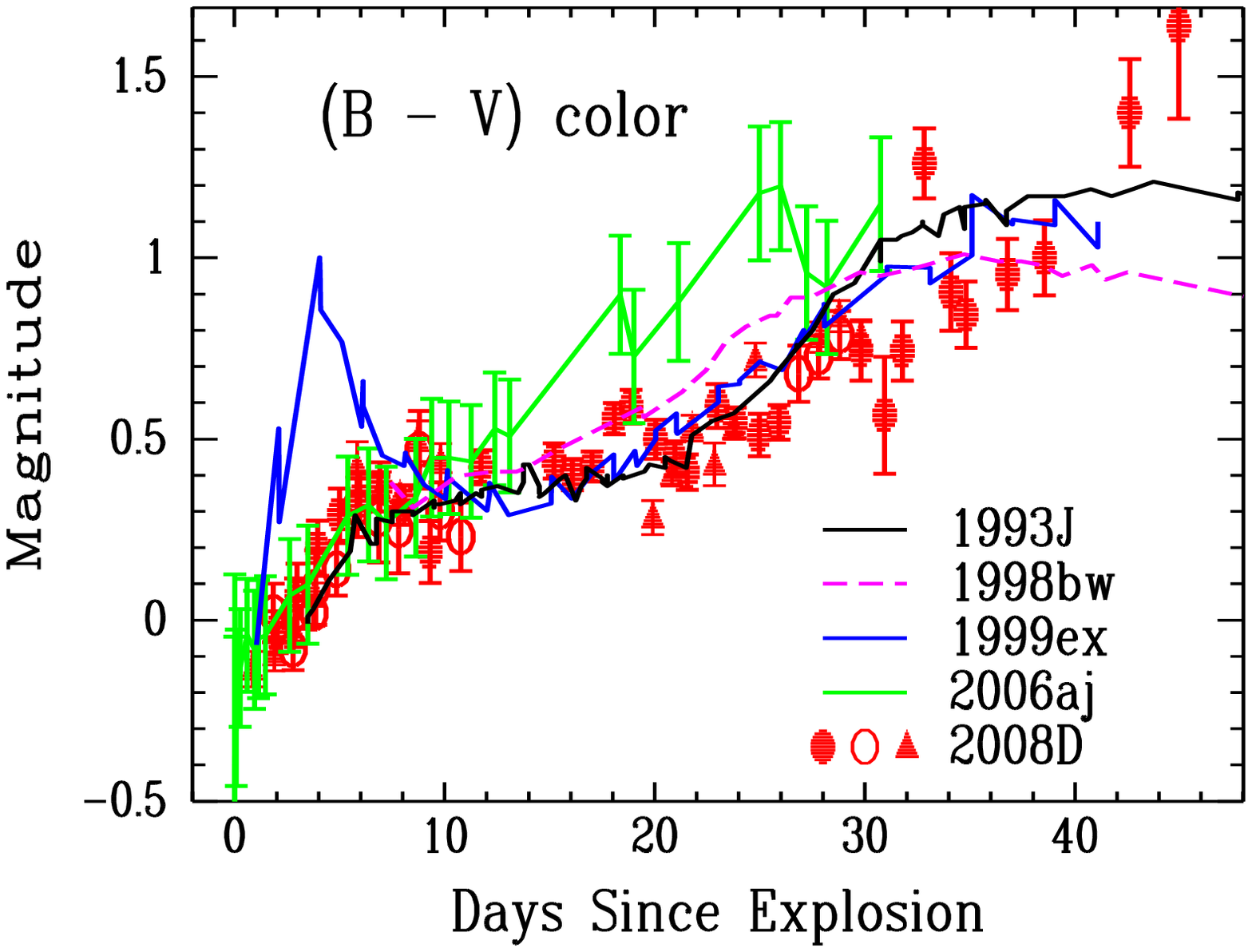}}
\caption{Comparison of the $B-V$ color curve of \snd\, dereddened by
  $E(B-V)$ = 0.6 mag, with those of other SNe. The $B-V$ color of SN
  IIb 1993J \citep{richmond94} is dereddened by $E(B-V) = 0.08$ mag (we
  adopted the lower value derived by Richmond et al. and not their higher
  alternative of $E(B-V) = 0.28$ mag), and
  that of SN Ib 1999ex by $E(B-V) = 0.3$ mag
  \citep{stritzinger02}.
}
\label{colorcomp_fig}
\end{figure}

Figure~\ref{lccompexp_fig} compares the optical light curves of
SN~2008D to those of other SNe with either observed shock breakout or
subsequent phase of cooling stellar envelope emission, where the SNe
are offset to match the peak magnitude of SN~2008D in the respective
bands. Careful comparison with other stripped-envelope SNe caught
shortly ($\sim$ 1 day) after shock breakout (SN~1993J, SN~1999ex)
reveals that while the light curves (and spectra) of SN~2008D and
SN~1999ex are similar during the \synNi -decay phase, the light curves
related to the cooling stellar envelope are different among these
three SNe. Furthermore, we note that the $B-V$ color evolution of
\snd\ is very similar to that of SN~1999ex during the time of \synNi\
decay, but very different during the shock-breakout phase
(Figure~\ref{colorcomp_fig}). The color curve of SN~1999ex
  has a steep rise to and decline from the red during 1$< \Delta t < $ 10 days,
  peaking at $B-V$=1 mag, while that of SN~2008D evolves gradually
  during this time and reaches $B-V$=0.5 mag at most.

We leave it as a future topic to model the diversity of the
very early-time light curves of SNe 1993J, 1999ex, 2006aj, and 2008D in
terms of progenitor radius, $E_{\rm{K}}$, and $M_{\rm{ej}}$, and to compare these
values with those derived from other methods (e.g., light curve and
nebular models). We note, however, that the temporal evolution of the
inferred BB temperature and radius of \snd\ are quite different
from those derived for SN~1999ex (Fig.~\ref{bbfit-tr_fig}, left panel, \citealt{stritzinger02}). 

Comparing the light-curve shapes of the different SNe with respect to
date of $V$-band maximum, we find that SN~2008D declines more slowly
than any of the other SNe and has a wider peak. One route of
parameterizing the SN light-curve shape is using the $\Delta
m_{15}(X)$ parameter, a formalism initially developed for SNe Ia
(\citealt{phillips93}), which is defined as the decline (in
magnitudes) during the first 15 days after maximum brightness in the
passband $X$.  For \snd , we measure $\Delta m_{15}(U)=1.2$ mag,
$\Delta m_{15}(B)=0.8$ mag, and $\Delta m_{15}(V)=0.6$ mag.

\subsection{Optical Spectroscopy: Emergence of \He\ Lines }\label{compoptspec_sec}

\begin{figure}[!ht]
\centerline{\includegraphics[width=3.1in,angle=+90]{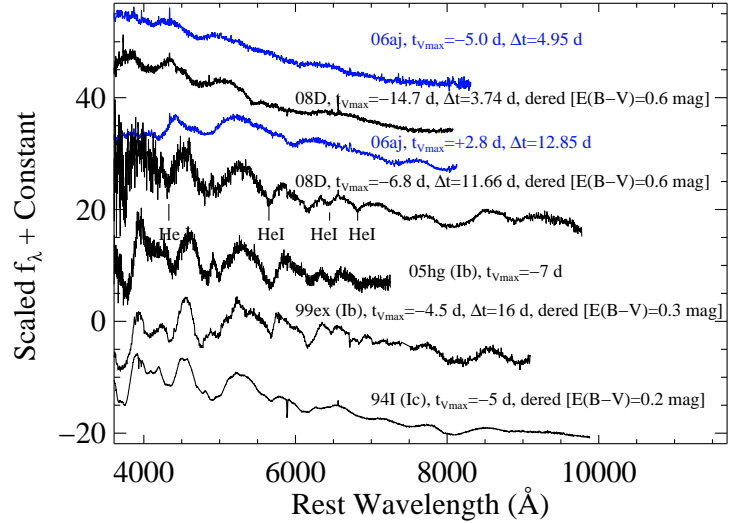}}  
\caption{Comparison of the MMT and APO spectra of \snd\ taken at
  $\Delta t=$ 3.74 day and 11.66 day, respectively, with those of
  SN~2006aj at similar ages ($\Delta t=$ 4.95 and 12.85 day,
  respectively), and with those of other SN Ib and SN Ic. SN~2006aj
  was the SN associated with XRF~060218 \citep{modjaz06}. 
  The early-time spectra of other SNe Ib and Ic include, dereddened by the
  indicated amounts: SN~1999ex \citep{stritzinger02,hamuy02},
  SN~2005hg \citep{modjaz07_thesis}, and SN Ic 1994I
  \citep{filippenko95}. The optical series of \He\ lines, defining
  \snd\ as a SN~Ib, are indicated (namely, He~I $\lambda\lambda$4471, 
  5876, 6678, 7061). We note the similarity between \snd\ and SNe
  1999ex and 2005hg.}
\label{specomp_fig}
\end{figure}

The very early spectra ($\Delta t \approx 3-5$ day) actually resemble
those of SN~2006aj (see top two spectra in Figure~\ref{specomp_fig};
see also \citealt{mazzali08,malesani09}), the SN connected with
GRB/XRF~060218, which initially led to the identification of this SN
as a broad-lined SN \citep{blondin08_08Did,valenti08_idgcn} (but see
\citealt{malesani08_idgcn}). This spectral resemblance suggests that
in both systems, the material constituting the photosphere at early
times is at high velocity \citep{mazzali08}, as expected from
homologous expansion.  Indeed, the velocity we determined via SYNOW,
$v_{\rm phot} \approx$ 30,000 \kms\ at $\Delta t = 1.8$ day in the spectrum 
with the double-absorption feature (see \S\ref{wfeature_sec}), is similar to
those of broad-lined SNe~Ic at the same epoch measured in a slightly
different manner \citep{mazzali08}.

However, while SN~2006aj and other broad-lined SNe~Ic continue to have
a significant amount of mass at high velocity (at least 0.1 \msol\ at
20,000--30,000 \kms\ of the total of 2 \msol\ ejected mass as
derived from models in \citealt{mazzali06_06aj}), at $\Delta t \approx 12$
day, the receding photosphere in SN~2008D reveals the lower-velocity
layers and the development of \He\ lines which SN~2006aj and other
broad-lined SNe~Ic do not show (Fig.~\ref{specomp_fig}).
Spectral synthesis modeling yields an estimate of $\sim$0.03
  \msol\ at $v >0.1c$ for \snd\ \citep{mazzali08,tanaka08}, a factor
  of 10$-$50 less than for GRB-SNe.


As Figure~\ref{specomp_fig} shows, the later spectral evolution of
\snd\ resembles that of a normal SN Ib, where the \He\ lines are as
pronounced as in SN~2005hg and more so than in SN~1999ex. 

We now investigate the temporal behavior of the \He\ lines, which are
the spectral hallmarks of SNe Ib. In \snd\ and in a larger sample of SNe
Ib, we determine the blueshifts in the maximum absorption in the \He\
lines via the robust technique of measuring SN line profiles as
presented by \citet{blondin06}. 

\begin{figure}[!ht]
\includegraphics[width=3.in,angle=+90]{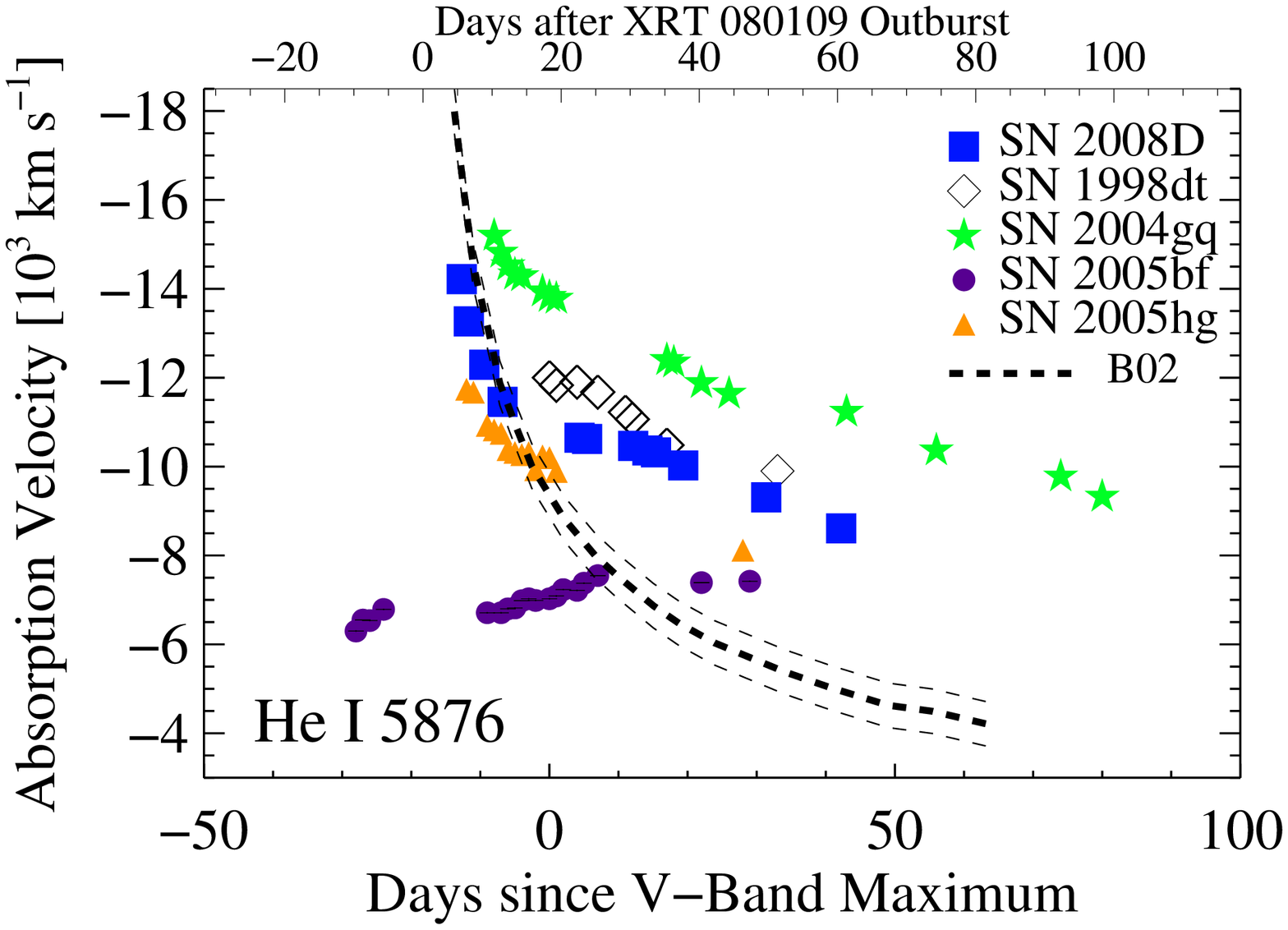} 
\caption{Temporal velocity evolution of \HeFive\ in SNe~Ib: SN~1998dt
  \citep{matheson01}, SN~2005bf
  (\citealt{tominaga05,modjaz07_thesis}), SN 2004gq and SN 2005hg
  \citep{modjaz07_thesis}, and SN~2008D (this work). \HeFive\ is
  usually the strongest \He\ line. The absorption velocities decrease
  over time, except for the maverick SN~Ib 2005bf (see text). The
  velocities for SN~2008D lie within the range of observed values,
  with the earliest measurements before maximum for a normal SN
  Ib. The thick dashed line is the functional form for the
  photospheric velocity as fit by Branch et al. (2002, B02) to SN Ib
  data, The thin dashed lines demarcate the extent in their
  uncertainty.  See text for details. }
\label{hevels_fig}
\end{figure}

 We refer to our velocity measurements as
``\He\ absorption-line velocities'' in the rest of this work. In
Figure~\ref{hevels_fig} (see also Table~\ref{hevels_table}) we present
the measured values for \HeFive , the strongest optical \He\ line in
SN~2008D, along with absorption-line velocities of a sample of other
SNe Ib (see \citealt{modjaz07_thesis}), measured in the same fashion.

In SN~2008D, the \He\ lines become apparent as early as $\Delta
t\approx$ 6 day (i.e., $t_{V\rm{max}}\approx -13$ day) and are fully
identified at $\Delta t \approx$ 11.7 day ($t_{V\rm{max}}\approx
-6.9$ day). Figure~\ref{hevels_fig}
reveals that the velocities span a large range for normal SNe Ib;
e.g., at maximum light, they range from $-$14,000 \kms\ (for
SN~2004gq) to $\sim$ $-$10,000 \kms\ (for SN~2005hg), and seem to
follow a power-law decline. For comparison, \citet{branch02} showed
that the \He\ velocities in the SNe Ib of their limited sample
followed a standard pattern and traced fairly closely (within
2000~\kms\ for the same epoch) the photospheric velocities derived
from synthetic fits to \ion{Fe}{2} lines, which we overplot in
Figure~\ref{hevels_fig} (dashed line, using their power-law fit
$v_{\rm{phot}}\propto t^{-2/(n-1)}$ with $n = 3.6$). However, the SNe Ib
in our sample show a larger departure (up to 6000~\kms) from the
corresponding standard photospheric velocity, larger than those of
SN~1998dt, which \citet{branch02} had declared as the exception in
their sample. In addition, both SNe~2008D and 2005hg show \He\ line
velocities that are \emph{lower} than the corresponding photospheric
velocity for dates before maximum light. This might mean that the
line-emitting material is located at radii smaller than the bulk of
the material that makes up the photosphere, or that there are 
optical-depth effects at play.

It has been long suggested that the \He\ lines are due to nonthermal
excitation and ionization of gamma-ray photons produced during the
radioactive decay of $^{56}$Ni and $^{56}$Co
\citep{harkness87,lucy91,mazzali98}. The fact that the gamma-ray
photons need time to diffuse through the ejecta to reach the He layer
has been invoked to explain the late emergence (after maximum light)
of \He\ in some SNe Ib \citep[e.g.,][and references 
therein]{filippenko97_review}. A
challenge for those who study the complicated nonthermal excitation
effects in SN line formation \citep{baron99,kasen06,dessart07} would
  be to self-consistently explain for \snd\ its early emergence of
  \He\ lines ($\sim$5--6 day after explosion, and $\sim$12--13 day before
  maximum light) by excitation of the relatively small amount of
  synthesized \synNi\ (0.05$-$0.1 \msol,
  \citealt{soderberg08,mazzali08,tanaka08}).

\subsection{NIR Spectroscopy as a Possible Discriminant}\label{compirspec_sec}

\begin{figure}[!ht]
\centerline{\includegraphics[width=3.5in,angle=+0]{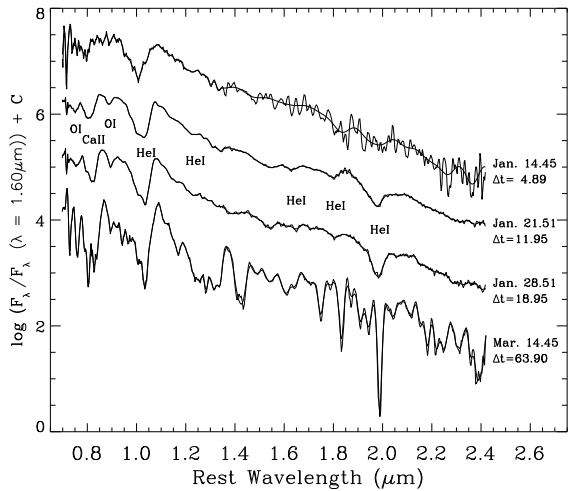}} 
\caption{Three NIR spectra of \snd\ taken with the IRTF. The dates of
  observation are indicated, along with SN ages ($\Delta t$) with
  respect to time of shock breakout (in days). The five detected \He\
  lines are indicated by their rest wavelengths. The spectra are shown
  as both raw and Fourier smoothed, and are normalized and offset by
  arbitrary amounts for clarity. Note the strong presence of
  blueshifted \ion{He}{1} 1.0830 at $\sim$ 1.02$-$1.05 $\mu$m and
  blueshifted \ion{He}{1} 2.0581 $\mu$m at $\sim$ 1.97$-$1.98 $\mu$m.
  The broad absorption at 1.05 $\mu$m probably includes an additional
  blend of other lines (mostly \ion{Fe}{2}, but also \ion{Mg}{2},
  \ion{C}{1}, and \ion{Si}{1}). }
\label{nirspec_fig}
\end{figure}

Figure~\ref{nirspec_fig} shows the four NIR spectra of SN~2008D
(taken with the IRTF) in which we detect five prominent blueshifted
\He\ lines. We tabulate their velocities as inferred from maximum
absorption in Table~\ref{hevels_table}.

\begin{figure}[!ht]
\centerline{\includegraphics[width=2.5in,angle=+0]{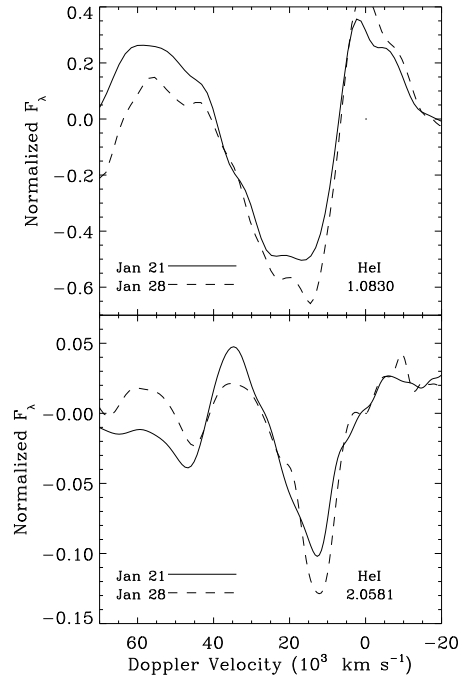}} 
\caption{The two main helium lines detected in NIR spectra of \snd\ in
  velocity space for two epochs: 2008 Jan. 21.51 ($\Delta t$ = 11.95 day; 
  solid lines) and Jan. 28.51 ($\Delta t$ = 18.95 day; dashed lines). 
  The top plot shows the NIR spectra with respect
  to \He\ 1.083 $\mu$m as zero velocity, and the bottom one with
  respect to \He\ 2.058 $\mu$m. The largest difference between the two
  lines is the effect of blending of other elements with the \He\
  1.083 $\mu$m line. The blueshifted \He\ 2.058
  $\mu$m line is narrower and more well-defined than the blend around
  1 $\mu$m, of which we take the red absorption trough to be due to
  \He\ 1.083 $\mu$m. Both \He\ lines have similar velocities based on
  the blueshifted maximum absorption (see Table~\ref{hevels_table}).
  The spectra are normalized to the local continuum.}
\label{nirhe_fig}
\end{figure}

Comparison of the NIR spectra of proper SNe Ib 1999ex, 2001B, and
2008D with those of SNe Ic (see \citealt{taubenberger06} and
references therein) shows that all SNe display a strong feature at
$\sim$1.04 $\mu$m but \rm{\emph{only}} SNe Ib show additional
absorption at 1.95$-$2.0 $\mu$m due to blueshifted \ion{He}{1} 2.058
$\mu$m, while none of the SNe~Ic do. For the few SNe~Ib in general
\citep{gerardy04} and for SN 2008D in particular, the absorption
velocity of the strong blueshifted \He\ 2.058 $\mu$m is very similar
to the strong blueshifted \He\ 1.083 $\mu$m if we identify the redder
absorption trough of the multi-component feature at 1 $\mu$m as \He\
1.083 $\mu$m. As Figure~\ref{nirhe_fig} shows, the blueshifted
  \ion{He}{1} 2.058 $\mu$m line is narrower and more well-defined than
  the broad blend at 1 $\mu$m.


In conclusion, SN~2008D gives supporting evidence to the argument that
the broad absorption at $\sim$1.04 $\mu$m is a poor diagnostic for
helium, since it is most likely a blend of \ion{Fe}{2}, \ion{C}{1},
\ion{Ca}{2}, and \ion{S}{1} at the same wavelengths
\citep{millard99,gerardy04,sauer06} and \He\ in SNe~Ib.
However, claims of significant He in optically classified SN
  Ic rely on identifying the same strong absorption feature at
  $\sim$1.04 $\mu$m exclusively with \ion{He}{1} 1.083 $\mu$m
  \citep{filippenko95,clocchiatti96,patat01}. Our study of SN~2008D
  implies that determining unambiguously if indeed helium is present
  in SNe normally classified as Type Ic requires a $K$-band spectrum
  to test for the presence of the clean and unblended \He\ 2.058
  $\mu$m feature.

\section{Double-Peaked Oxygen Lines And Aspherical  Explosion Geometry}\label{late_sec}

\begin{figure}[!ht]
\centerline{\includegraphics[width=2.6in,angle=+90]{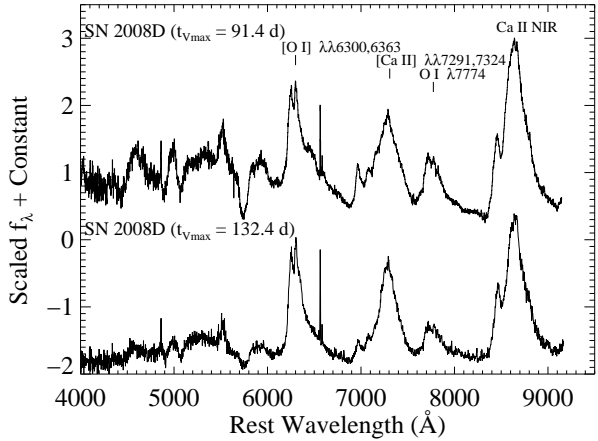}} 
\caption{Spectra of \snd\ at 3 and 4 months after maximum light (April
  28, and June 7, 2008). The main emission lines in the spectrum are
  marked, and their rest wavelengths indicated. We caution that the
  line at 7300~\AA\ could be a blend of \OxyTwo\ and some Fe lines, in
  addition to \CaTwo , as marked. }
\label{latespec_fig}
\end{figure}

In Figure~\ref{latespec_fig} we present the Keck spectra of \snd\
taken at $t_{V\rm{max}} = 91.4$ day (i.e., $\Delta t = 109$ day) on
April 28, 2008, and $t_{V\rm{max}} = 132.4$ day, on June 7,
2008. They show nebular emission lines on top of some residual,
underlying photospheric spectrum, which is more pronounced in
  the earlier spectrum. This is expected, as SNe appear to turn fully
transparent $\sim$200 day after maximum light, and some SNe take
up to 1 year to complete the transition \citep{mazzali04}. Here, we
are witnessing this transition period for \snd . Below we analyze the
nebular emission lines of \snd .

At sufficiently late times, the SN ejecta become fully optically thin
in the continuum; hence, spectra obtained during this time period
afford a deeper view into the core of the explosion than spectra taken
during the early photospheric phase. In the absence of hydrogen,
oxygen is the primary coolant in the ejecta of stripped-envelope SNe
at late epochs when the gas is neutral or at most singly ionized
\citep{uomoto86,fransson87}, and when densities are sufficiently low
for forbidden lines to be the strongest ones in the spectrum. In
\snd\, we detect \OxyOne\ and \OxySeven\, along with \CaTwo\ and the
strong Ca II NIR triplet.

\begin{figure}

\centerline{\includegraphics[width=3in,angle=0]{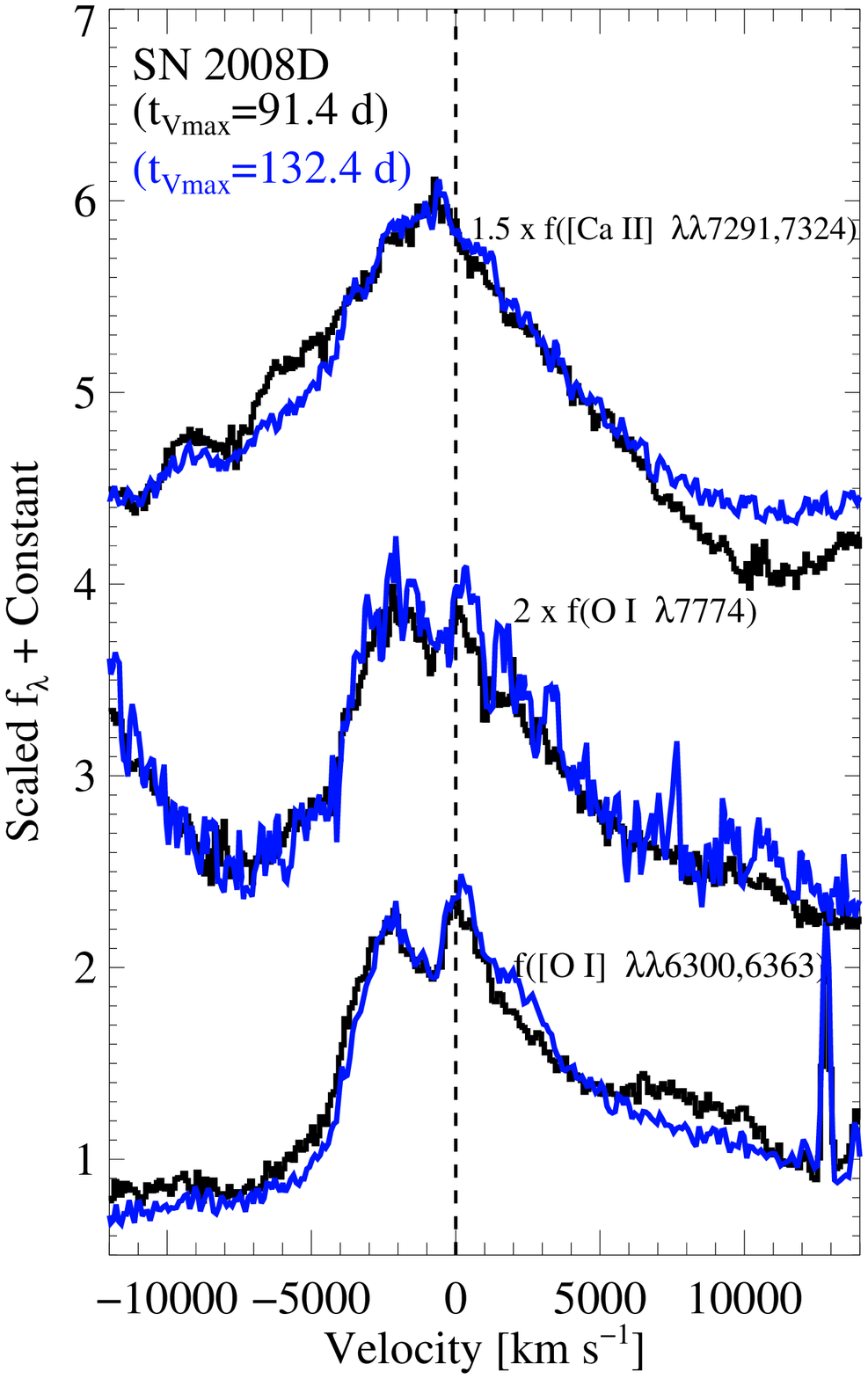}} 
\caption{The main late-time emission lines of \snd\ in velocity space
  in the rest frame of NGC~2770.  The zeropoint for \OxyOne\ here is
  6300 \AA\, while that of \CaTwo\ is taken at 7307.5 \AA\ (straight
  mean of the two calcium lines). We caution that \Ca\ might be a
  blend with [\ion{O}{2}] and Fe.  Zero velocity is indicated by the dashed line.  The two
  oxygen lines show the same conspicuous double-peaked profile, indicating
  that the two horns cannot be due to the doublet nature of
  \OxyOne . In both cases, the redder peak is at zero velocity, and the
  trough between the two peaks is shifted by $\sim -800$ \kms\ from
  zero velocity for both spectra. With the chosen zeropoint for \CaTwo, its peak is
  also blueshifted by $\sim $800 \kms, though blending could be
  affecting the line shape and the exact location of zero velocity.
  See text for more details. }
\label{latespecvel_fig}
\end{figure}

To study the line profiles of \snd\ more closely, we plot the main
relatively unblended emission lines (\OxyOne , \OxySeven, and \CaTwo ), 
in velocity space in Figure~\ref{latespecvel_fig}. The two oxygen
lines show the same unusual double-peaked profile, while \CaTwo\ does
not exhibit it. For both oxygen lines, the redder peak is at
zero velocity, and the trough between the two peaks is shifted by
$\sim -800$ \kms\ from zero velocity. The total separation between the
two peaks is $\sim$ 2000 \kms\ for both the forbidden and the
permitted oxygen lines. With the chosen zeropoint for \CaTwo\ (namely
7307.5 \AA , the straight mean of the two calcium lines), its peak is
also blueshifted by $\sim $800 \kms , but blending with other lines
(\OxyTwo, Fe) could be affecting the line shape and the exact
determination of zero velocity.

While optical-depth effects sometimes generate double-peaked \OxyOne\
because of its doublet nature (most notably in SN~1987A;
\citealt{spyromilio91,leibundgut91,li92}), it is not the culprit in
\snd , since (a) the separation of the doublet (3000 \kms) is larger
than the observed separation between the double peaks (2000 \kms), and
(b) \OxySeven , a multiplet with very small separations (at most
3~\AA, i.e., 130 \kms, smaller than our spectral resolution), shows
the same double-peaked profile as \OxyOne\ in \snd. The
  \CaTwo\ line does not show the same double-peaked profile, but this
  may well be because the emission comes from smoothly distributed,
  preexisting (natal) calcium similar to what was seen in SN IIb
  1993J \citep{matheson00_93jdetail}.

  Such double-peaked oxygen emission lines have recently been observed
  in a number of stripped-envelope SNe
  (\citealt{mazzali05_03jd,maeda07,modjaz08_doubleoxy,maeda08,taubenberger09}),
  and have been interpreted in the framework of global asphericities,
  where the emitting oxygen-rich ejecta are situated in a ring- or
  torus-like structure viewed along the equatorial plane.  Both
  \citet{modjaz08_doubleoxy} and \citet{maeda08} suggest that such
  global asphericities are common during core collapse.
  \citet{maeda08} specifically suggest that their observed fraction of
  double-peaked profiles ($40\pm10$\%) is consistent with the
  hypothesis that all stripped core-collapse events are mildly
  aspherical, given their models in \citet{maeda06}, whereas
  \citet{taubenberger09} find a lower observed fraction (5$-$18\%).
  Recently, \citet{milisavljevic09} have argued that it is difficult
  to explain all line profiles observed in their recent SN sample via
  purely geometric effects. However, their conclusions only concern
  SNe whose double-peaked profile has a velocity separation close to
  that of the doublet \OxyOne\ and does not show it in lines other
  than the doublet; those features are not observed in \snd\
  (see Fig.~\ref{latespec_fig} and above).

While other SNe IIb and Ib have been shown to exhibit some small-scale
structure in oxygen, indicating some clumping in the density
distribution of oxygen probably due to instabilities
\citep{filippenko86,spyromilio94_93j,sollerman98,matheson00_93jdetail,elmhamdi04},
the clear symmetric double peaks in \snd\ are different and
may rather indicate a global scale for the anisotropy. The
mechanism giving rise to such a global anisotropy could lie in the
intrinsic explosion physics of core collapse, since each of the
various theoretical core-collapse models, be it neutrino driven
\citep{scheck06}, acoustic \citep{burrows06}, or magneto driven
\citep{burrows07,dessart08}, finds that symmetry breaking is essential
for a successful explosion. 
  
Alternatively, in case the progenitor of \snd\ is part of a binary,
binary interaction or a merger might be modulating the geometry of the
explosion, though this speculation will be very hard to verify.
Nevertheless, other lines of evidence, such as polarization studies of
SNe II, Ib, and Ic (e.g., \citealt{leonard05}), neutron-star kick
velocities \citep{wang06_nskicks}, and young SN remnant morphologies
(e.g., Cas A; \citealt{fesen06_casabi}) suggest that asphericities are
generic to the core-collapse process.

We note the blueshift from zero velocity (of the order $\sim$ 800
\kms, Figure~\ref{latespecvel_fig}) in the oxygen trough, if the
trough between the two peaks is regarded as the symmetry center. The
same blueshift might be also present in the calcium peak, depending on
the choice of zero velocity, which is more uncertain than for the
oxygen lines due to possible blending with other lines. This
blueshift, if real, could be caused by residual opacity before
reaching the full nebular phase \citep{taubenberger09}, peculiar
distribution of \synCo\ which is powering the emission at such later
epochs, large-scale asymmetry, or offset of the SN rest-frame velocity
from that of NGC~2770 by 800 \kms. However, given that the rotation
curve of the host galaxy, NGC 2770, extends to only $\sim$400 \kms
\citep{haynes97}, the last option is least likely.

Future multi-epoch late-time spectra (especially those taken
at $t_{V\rm{max}} \ga$ 200 day) are needed to differentiate between the
various scenarios as each one predicts a different behavior of the
blueshift over time. If it is due to residual opacity, the most
probable case, we predict that the blueshift will go to zero for
spectra at $t_{V\rm{max}} \ga$ 200 day. We can exclude dust
  formation as the culprit, since the apparent blueshift does {\it
    not} increase over time, contrary to the behavior of blueshifted
  lines in SN where dust had been formed (e.g.,
  \citealt{smith08_dust}). If the blueshift continues to remain
constant over time (as seen in SN~1996N; \citealt{sollerman98}), then
bulk blueshifts or unipolar blob ejections might need to be invoked as
suggested for SN~2005bf \citep{maeda07}.

In summary, we infer from the presence of double peaks in the two
different oxygen lines in \snd\ that global asphericities were
probably involved during the explosive event of \snd, regardless of the
specific mechanism that actually leads to a successful explosion.

\section{X-ray Interpretation}\label{xrayinter_sec}

The origin of the X-ray emission of \xtransient\ is highly debated.
\citet{soderberg08} and \citet{chevalier08} attribute it fully to the
processes related to core-collapse shock-breakout emission, while
others \citep{xu08,li08,mazzali08} claim it as emission caused by a
stifled or weak GRB jet. \citet{soderberg08} adopt the pure power-law
X-ray spectrum fit and use the calculations of \citet{wang07_06aj} to
explain the power-law shape in terms of upscattering (i.e.,
Comptonization) by the circumstellar wind medium of thermal shock
breakout photons with energies $< 0.1$ keV.  On the other hand,
\citet{chevalier08} favor the pure blackbody fit and explain the
artificially small BB radius as the product of a scattering
atmosphere, where the emission is generated in a relatively deep layer
but effectively emitted at a larger radius.  Even for \snaj, which had
a much higher S/N spectrum, the exact origin of the X-ray emission is
still debated
\citep{campana06,ghisellini07,soderberg06_06aj,waxman07}.

While the above authors attempt to test the system of \xtransient\ for
emission from relativistic ejecta, their conclusions differ and are
heavily model dependent. In the next section, we explore a
phenomenological approach where we compare the observed properties of
\xtransient\ with those of so-called X-ray flashes \citep{heise01},
high-energy transient that are likely softer analogues of cosmological
GRBs \citep{sakamoto08}. We note that recently, \citet{bietenholz09}
find no evidence for a relativistic, off-axis jet in SN~2008D based on
their VLBI observations of SN~2008D. Very
recently, \citet{katz09} have argued that non-thermal X-ray spectra may be
common during the breakout of radiation-mediated non-relativistic
shocks, as suggested to be the case for \xtransient .

\subsection{Comparison with X-ray Flashes}\label{xraycomp_sec}

Here we compare three properties of \xtransient\ with those observed
in a sample of classical XRFs: duration (as parameterized by
$T_{90}$), the shape of the X-ray spectrum, and X-ray-to-optical flux
ratio. In addition we discuss its energy output.

While in each case \xtransient\ is not necessarily an obvious outlier
in the observed distribution, together all three considerations paint
a strong picture that \xtransient, along with \snaj, might be
different phenomenologically from the rest of classical XRFs. However,
future studies trying to uncover the inherent distribution will need
to carefully consider detector effects and rest-frame properties.

1) Duration: XRFs observed by HETE-2 and {\it Swift} are seen to have
similar temporal profiles and durations compared to long-duration GRBs, 
exhibiting a FRED (i.e., fast rise exponential decline) profile.  
\xtransient\ can be adequately
fit by such a FRED profile (\S~\ref{xraycomp_sec}), but the
transient's duration is an order of magnitude longer than the median
$T_{90}$ for other XRFs and X-ray-rich GRBs (XRRs), which is roughly
30 s \citep{sakamoto05,sakamoto08}. Both \xtransient\ and \snaj\ have
the longest durations ($T_{90}= 470\pm30$ s and 2100 $\pm 100$ s,
respectively). We note the caveat that satellites are typically
inhibited from triggering on long timescales due to background
modulations, and XRT 080109 could not have been detected with any other
survey X-ray telescope before {\it Swift}.

2) X-ray spectrum: \snaj\ has a significant X-ray component that (a)
is thermal and (b) appears to soften over time, with very high S/N
\citep{campana06,butler07_06aj}. The X-ray spectrum of \xtransient\ is
of much lower S/N, and is well-fit by a combined PL+BB
(\S~\ref{xrayobs_sec}).  While the X-ray emission of XRF~060218 fades
at a rate comparable to that often observed for GRB X-ray afterglows,
its spectrum is markedly softer (e.g., \citealt{soderberg06_06aj}) and
also softens strongly in time \citep{butler07_06aj}. This is in
contrast to GRB X-ray afterglows, which exhibit little spectral
evolution at the same epoch: GRB and XRF X-ray afterglows have photon
indices of $\Gamma \approx$ 2, while that of \snaj\ was $\Gamma = 3.41
\pm 0.13$ \citep{butler07_hardness}.  There is evidence that the
\xtransient\ late-time X-ray spectrum (from the {\it Chandra} data)
was also softer with $\Gamma = 3.6^{+1.7}_{-1.4}$
\citep{pooley08_gcn,soderberg08} than the early-time spectrum.

3) We also compare the radiation output in \xtransient\ for different
wavelengths to that of XRFs and their afterglows. 

At the dates of our optical and NIR photometry (at $t-t_o = \Delta t$
= 0.84 and 0.71 day, i.e., $\sim$ 70 and 60~ks), the X-ray luminosity
is in the range $\sim 2 \times 10^{38-39}$ \ergs . Thus, the X-ray
luminosity (for 0.3$-$10.0 keV) is 3--4 orders of magnitude lower than
the luminosity inferred from the optical observations (see
\S~\ref{bol_sec}), thus, $L_{\rm X}/L_{\rm opt} \approx 10^{-3} -
10^{-4}$ at 70~ks for \snxray . This corresponds to a ratio of fluxes
$f_{\rm{O/X}}= 4 \times (10^{2}$--$10^{3})$, while the observed ratio
is $ f_{\rm{O/X}} < 70$ for classical XRF afterglows
\citep{dalessio06} and GRB afterglows \citep{depasquale03} at
comparable times (using the same definition and units of
$f_{\rm{O/X}}$ as in \citealt{dalessio06}). In other words, the
optical to X-ray ratio for \snxray\ is a factor of $\sim 10^3$ larger
than for XRF afterglows. For \snaj, the ratio cannot be precisely
determined since it was not observed in the desired bandpass at such
early time.

The two main features commonly observed in XRFs are not measured for
\xtransient , namely detection of gamma-ray emission and peak of the
energy distribution ($E_{\rm peak}$).

In summary, \xtransient, along with \snaj, appears qualitatively
different from the observed distribution of classical XRFs, when
considering all three discussed properties together (duration,
spectral evolution, and optical-to-X-ray ratio). Finally, the
prompt high-energy emission released in \xtransient\ ($E_{\rm X,\rm{iso}}
\approx 6 \times 10^{45}$ erg) is 3--4 orders of magnitude lower than in XRFs \citep{soderberg08}.

\section{Conclusions}\label{conclusion_sec}

We have presented extensive early-time photometric
($UVW2,UVM2,UVW1,UBVRr'Ii'JHK_s$) and spectroscopic (optical and NIR)
data on SN 2008D as well as X-ray data analysis on the associated
\xtransient. Our data span a time range of 17 hr to \lastphot\ day
after the detection of the X-ray transient and establish that SN~2008D
is a spectroscopically normal SN Ib at maximum light, with a
relatively long rise time (18 day) and a modest optical peak
luminosity of $M_V =$ \absmagVsn\ mag and $M_B =$ \absmagBsn\ mag,
after correcting for \ebvhost .

We also present the earliest NIR ($JHK_s$) detections (at 0.71 day).
Our densely time-sampled optical spectral sequence, along with our NIR
spectra, uncovers the emergence epoch of the helium lines. The
early-time spectra reveal high expansion velocities (15,0000$-$30,000
\kms) up to 6--7 day after outburst, and one specific spectrum (at
$\Delta t =1.84$ day) shows a transient, prominent double-absorption
feature at blue wavelengths ($\sim$ 4000 \AA ) that is no longer
visible in spectra taking 1 day later. In spectra taken 3 and
  4 months after maximum light, we detect double-peaked profiles in
the oxygen lines that we take to indicate global
asphericities in the SN core ejecta, and argue that they cannot be due
to optical-depth effects.  Furthermore, we recommend for future
searches for helium in SNe Ic that $J$- and $K$-band spectra are
obtained to test for the presence of both \He\ 1.083 $\mu$m and \He\
2.058 $\mu$m.

The comprehensive data presented here allow us to construct a reliable
measurement of the bolometric output for this stripped-envelope SN,
including UV and NIR contributions that are usually lacking for other
SNe. For data $\Delta t >$ 3 days, we find that the bolometric
luminosity based on direct $UVW1 \rightarrow K_s$ integration is very
similar to that based on the blackbody fits to the broad-band
photometry (within $\la$0.04 dex or $\la 8\%$), such that the UV-NIR
emission of SN 2008D may well be described by that of a black-body.
We furthermore compare our early-time light curves to fits by
\citet{soderberg08}, who use the calculations of \citet{waxman07} for
the cooling stellar envelope blackbody model, and test the predictions
of the alternative cooling envelope model of \citet{chevalier08}.
Combining our comprehensive data with estimates of $E_{\rm{K}}$ and
$M_{\rm{ej}}$ from the literature, we estimate the stellar radius
$R_{\star}$ of its probable Wolf-Rayet progenitor. According to the
models of \citet{waxman07} and of \citet{chevalier08}, we derive
$R_{\star}^{\rm{W07}}=1.2\pm0.7$ \sr\ and
$R_{\star}^{\rm{CF08}}=12\pm7$ \sr , respectively. The larger size of
the progenitor of SN~2008D based on the model of Chevalier \& Fransson
is somewhat more consistent with typically observed radii of WN stars
than the size required by the Waxman model.

Furthermore, we find that the {\it Swift} X-ray spectrum can be fit
equally well by an absorbed power law or a superposition of about
equal parts of both power law and blackbody, which has not been
suggested before in the literature, and has implications for
  determining the mechanism of the X-ray emission.  Specifically, our
  derived blackbody radius from the PL+BB fit is larger (by a factor
  of 10) than that of \citet{mazzali08}, whose jet interpretation
  of the origin of the X-ray emission is based largely on the small
  inferred BB area.

\xtransient\ is different from the apparent distribution of classical
X-ray flashes as evidenced by its long duration, softening of the
X-ray spectrum, and very high optical to X-ray luminosity ratio,
though future studies of the inherent distribution will need to fold
in detector effects and consider rest-frame properties.  Future
detections are expected to show unambiguously whether these two events
are part of a continuous distribution of XRFs, or members of a
different physical class of events.

Additional multi-epoch late-time observations (especially for more
than 200 days after maximum light) of \snd\ will be useful to monitor
the behavior of the double-peaked oxygen lines and their associated
possible blueshift, and, as the SN becomes fully transparent, allow for
closely constraining the geometry of the explosion as revealed by the
core ejecta.  Moreover, such observations are important to monitor for
any late-time interactions between SN ejecta and the presumably
prior-shed hydrogen envelope to further study the mode of mass
loss and search for episodes of sudden mass loss.

\acknowledgements

M.M. thanks Roger Chevalier, Phil Chang, Chris Matzner, Peter Nugent,
Alceste Bonanos, Andrew MacFadyen, and Tsvi Piran, and R.P.K. thanks
Eli Waxman for stimulating discussions. M.M. is supported by a
fellowship from the Miller Institute for Basic Research in Science. We
thank S.  S. Piatek, T.~Pryor, and E.~Olszewski for obtaining a few of
the spectra presented here.  Some of the observations reported here
were obtained at the MMT Observatory, a joint facility of the
Smithsonian Institution and the University of Arizona, and at the W.
M. Keck Observatory, which was made possible by the generous financial
support of the W. M. Keck Foundation. We wish to extend special thanks
to those of Hawaiian ancestry on whose sacred mountain we are
privileged to be guests. The Apache Point Observatory 3.5-m telescope
is owned and operated by the Astrophysical Research Consortium. This
work is based in part on observations obtained at the Gemini
Observatory, which is operated by the Association of Universities for
Research in Astronomy, Inc., under a cooperative agreement with the
National Science Foundation (NSF) on behalf of the Gemini partnership:
the NSF (US), the Particle Physics and Astronomy Research Council
(UK), the National Research Council (Canada), CONICYT (Chile), the
Australian Research Council (Australia), CNPq (Brazil), and CONICET
(Argentina).  PAIRITEL is operated by the Smithsonian Astrophysical
Observatory (SAO) and was made possible by a grant from the Harvard
University Milton Fund, a camera loan from the University of Virginia,
and continued support of the SAO and UC Berkeley. The PAIRITEL project
is further supported by NASA/{\it Swift} Guest Investigator grant
NNG06GH50G.  KAIT was made possible by generous donations from Sun
Microsystems, Inc., the Hewlett Packard Company, AutoScope
Corporation, Lick Observatory, the NSF, the University of California,
and the Sylvia and Jim Katzman Foundation; we also thank the TABASGO
Foundation for continued support. H.M. is a visiting astronomer at the
Infrared Telescope Facility, which is operated by the University of
Hawaii under Cooperative Agreement no. NCC 5--538 with the National
Aeronautics and Space Administration (NASA), Science Mission
Directorate, Planetary Astronomy Program. N.R.B. is supported through
the GLAST Fellowship Program (NASA Cooperative Agreement: NNG06DO90A)

Supernova research at UC Berkeley is supported in part by the TABASGO
Foundation and NSF grant AST--0607485 to A.V.F. Supernova research at
Harvard University is supported by NSF grant AST--0606772 to R.P.K.
D.P.  and N.R.B. were partially supported through the DOE SciDAC grant
DE--FC02--06ER41453. J.S.B. was partially supported by the Hellman
Faculty Fund. J.X.P. is partially supported by NASA/{\it Swift} grants
NNG06GJ07G and NNX07AE94G and an NSF CAREER grant (AST--0548180). Work
by W.H.dV. and S.S.O. was performed under the auspices of the U.S.
Department of Energy by Lawrence Livermore National Laboratory under
Contract DE--AC52--07NA27344. G.S.S. wishes to thank M. Drosback, B.
Keeney, M.  Moe, and K. O'Malia for relinquishing some University of
Colorado APO time, S. Hawley for awarding APO Director's Discretionary
Time, R.  McMillan for ``a little more time'' to complete the data
acquisition on one night, and the APO Observing Specialists for their
assistance. G.S.S. is partially supported by NASA grants NNX06AG43G
and NNX06AG44G, and B.P.H. and G.D.I.  are partially supported by NASA
grant NAG5--7697. C.H.B. acknowledges support from the Harvard Origins
of Life Initiative. H.M. and P.G. thank Alan Tokunaga for arranging
the IRTF observations on short notice.

This research has made use of NASA's Astrophysics Data System
Bibliographic Services (ADS) and the NASA/IPAC Extragalactic Database
(NED) which is operated by the Jet Propulsion Laboratory, California
Institute of Technology, under contract with NASA.  We thank the {\it
  Swift} team and the observers who provided their data and analysis
through the GCN. We collectively thank the staffs of the MMT, APO,
Keck, Lick, and Gemini Observatories for their assistance. We thank
the arXiv.org server that made it possible to communicate our results to
the interested reader in a timely fashion.

\bibliographystyle{apj}
\bibliography{refs}

\begin{thebibliography}{180}
\expandafter\ifx\csname natexlab\endcsname\relax\def\natexlab#1{#1}\fi

\bibitem[{{Abbott} \& {Conti}(1987)}]{abbott87}
{Abbott}, D.~C. \& {Conti}, P.~S. 1987, \araa, 25, 113

\bibitem[{{Alard}(2000)}]{alard00}
{Alard}, C. 2000, \aaps, 144, 363

\bibitem[{{Alard} \& {Lupton}(1998)}]{alard98}
{Alard}, C. \& {Lupton}, R.~H. 1998, \apj, 503, 325

\bibitem[{{Baron} {et~al.}(1999){Baron}, {Branch}, {Hauschildt}, {Filippenko},
  \& {Kirshner}}]{baron99}
{Baron}, E., {Branch}, D., {Hauschildt}, P.~H., {Filippenko}, A.~V., \&
  {Kirshner}, R.~P. 1999, \apj, 527, 739

\bibitem[{{Berger} \& {Soderberg}(2008)}]{berger08_discgcn}
{Berger}, E. \& {Soderberg}, A. 2008, GCN Circ. 7159

\bibitem[{{Bietenholz} {et~al.}(2009){Bietenholz}, {Soderberg}, \&
  {Bartel}}]{bietenholz09}
{Bietenholz}, M.~F., {Soderberg}, A.~M., \& {Bartel}, N. 2009, \apjl, 694, L6

\bibitem[{{Blinnikov} {et~al.}(2000){Blinnikov}, {Lundqvist}, {Bartunov},
  {Nomoto}, \& {Iwamoto}}]{blinnikov00}
{Blinnikov}, S., {Lundqvist}, P., {Bartunov}, O., {Nomoto}, K., \& {Iwamoto},
  K. 2000, \apj, 532, 1132

\bibitem[{{Blinnikov} {et~al.}(1998){Blinnikov}, {Eastman}, {Bartunov},
  {Popolitov}, \& {Woosley}}]{blinnikov98}
{Blinnikov}, S.~I., {Eastman}, R., {Bartunov}, O.~S., {Popolitov}, V.~A., \&
  {Woosley}, S.~E. 1998, \apj, 496, 454

\bibitem[{{Blondin} {et~al.}(2008{\natexlab{a}}){Blondin}, {Matheson},
  {Modjaz}, \& {Berlind}}]{blondin08_idiauc}
{Blondin}, S., {Matheson}, T., {Modjaz}, M., \& {Berlind}, P.
  2008{\natexlab{a}}, Central Bureau Electronic Telegrams, 1205, 1

\bibitem[{{Blondin} {et~al.}(2008{\natexlab{b}}){Blondin}, {Matheson},
  {Modjaz}, \& {Berlind}}]{blondin08_08Did}
---. 2008{\natexlab{b}}, Central Bureau Electronic Telegrams, 1205, 1

\bibitem[{{Blondin} {et~al.}(2006)}]{blondin06}
{Blondin}, S. {et~al.} 2006, \aj, 131, 1648

\bibitem[{{Blondin} {et~al.}(2008{\natexlab{c}})}]{blondin08_07uyid}
---. 2008{\natexlab{c}}, \iaucirc, 8908, 2

\bibitem[{{Bloom} {et~al.}(2006){Bloom}, {Starr}, {Blake}, {Skrutskie}, \&
  {Falco}}]{bloom06}
{Bloom}, J.~S., et~al.\ 2006, in ASP Conf. Ser. 351: Astronomical Data Analysis Software and  Systems XV, ed. C.~{Gabriel}, C.~{Arviset}, D.~{Ponz}, \& S.~{Enrique},  751--+

\bibitem[{{Branch} {et~al.}(2002){Branch}, {Benetti}, {Kasen}, {Baron},
  {Jeffery}, {Hatano}, {Stathakis}, {Filippenko}, {Matheson}, {Pastorello},
  {Altavilla}, {Cappellaro}, {Rizzi}, {Turatto}, {Li}, {Leonard}, \&
  {Shields}}]{branch02}
{Branch}, D., et~al.\ 2002, \apj, 566, 1005

\bibitem[{{Branch} {et~al.}(2006){Branch}, {Jeffery}, {Young}, \&
  {Baron}}]{branch06}
{Branch}, D., {Jeffery}, D.~J., {Young}, T.~R., \& {Baron}, E. 2006, \pasp,
  118, 791

\bibitem[{{Burrows} {et~al.}(2007){Burrows}, {Dessart}, {Livne}, {Ott}, \&
  {Murphy}}]{burrows07}
{Burrows}, A., {Dessart}, L., {Livne}, E., {Ott}, C.~D., \& {Murphy}, J. 2007,
  \apj, 664, 416

\bibitem[{{Burrows} {et~al.}(2006){Burrows}, {Livne}, {Dessart}, {Ott}, \&
  {Murphy}}]{burrows06}
{Burrows}, A., {Livne}, E., {Dessart}, L., {Ott}, C.~D., \& {Murphy}, J. 2006,
  \apj, 640, 878

\bibitem[{{Burrows} {et~al.}(2005){Burrows}, {Hill}, {Nousek}, {Kennea},
  {Wells}, {Osborne}, {Abbey}, {Beardmore}, {Mukerjee}, {Short}, {Chincarini},
  {Campana}, {Citterio}, {Moretti}, {Pagani}, {Tagliaferri}, {Giommi},
  {Capalbi}, {Tamburelli}, {Angelini}, {Cusumano}, {Br{\"a}uninger}, {Burkert},
  \& {Hartner}}]{burrows05}
{Burrows}, D.~N., et~al.\ 2005, Space Science Reviews, 120, 165

\bibitem[{{Butler}(2007{\natexlab{a}})}]{butler07_06aj}
{Butler}, N.~R. 2007{\natexlab{a}}, \apj, 656, 1001

\bibitem[{{Butler}(2007{\natexlab{b}})}]{butler07_astrometry}
---. 2007{\natexlab{b}}, \aj, 133, 1027

\bibitem[{{Butler} \& {Kocevski}(2007{\natexlab{a}})}]{butler07}
{Butler}, N.~R. \& {Kocevski}, D. 2007{\natexlab{a}}, \apj, 663, 407

\bibitem[{{Butler} \& {Kocevski}(2007{\natexlab{b}})}]{butler07_hardness}
---. 2007{\natexlab{b}}, \apj, 668, 400

\bibitem[{{Butler} {et~al.}(2006){Butler}, {Li}, {Perley}, {Huang}, {Urata},
  {Prochaska}, {Bloom}, {Filippenko}, {Foley}, {Kocevski}, {Chen}, {Qiu},
  {Kuo}, {Huang}, {Ip}, {Tamagawa}, {Onda}, {Tashiro}, {Makishima},
  {Nishihara}, \& {Sarugaku}}]{butler06}
{Butler}, N.~R., et~al.\ 2006, \apj, 652, 1390

\bibitem[{{Campana} {et~al.}(2006)}]{campana06}
{Campana}, S. {et~al.} 2006, \nat, 442, 1008

\bibitem[{{Chevalier}(1992)}]{chevalier92}
{Chevalier}, R.~A. 1992, \apj, 394, 599

\bibitem[{{Chevalier} \& {Fransson}(2008)}]{chevalier08}
{Chevalier}, R.~A. \& {Fransson}, C. 2008, \apjl, 683, L135

\bibitem[{{Chiosi} \& {Maeder}(1986)}]{choisi86}
{Chiosi}, C. \& {Maeder}, A. 1986, \araa, 24, 329

\bibitem[{{Clocchiatti} {et~al.}(1996){Clocchiatti}, {Wheeler}, {Brotherton},
  {Cochran}, {Wills}, {Barker}, \& {Turatto}}]{clocchiatti96}
{Clocchiatti}, A., {Wheeler}, J.~C., {Brotherton}, M.~S., {Cochran}, A.~L.,
  {Wills}, D., {Barker}, E.~S., \& {Turatto}, M. 1996, \apj, 462, 462

\bibitem[{{Cohen} {et~al.}(2003){Cohen}, {Wheaton}, \& {Megeath}}]{cohen03}
{Cohen}, M., {Wheaton}, W.~A., \& {Megeath}, S.~T. 2003, \aj, 126, 1090

\bibitem[{{Crowther}(2007)}]{crowther07}
{Crowther}, P.~A. 2007, \araa, 45, 177

\bibitem[{{Cushing} {et~al.}(2004){Cushing}, {Vacca}, \& {Rayner}}]{cushing04}
{Cushing}, M.~C., {Vacca}, W.~D., \& {Rayner}, J.~T. 2004, \pasp, 116, 362

\bibitem[{{D'Alessio} {et~al.}(2006){D'Alessio}, {Piro}, \&
  {Rossi}}]{dalessio06}
{D'Alessio}, V., {Piro}, L., \& {Rossi}, E.~M. 2006, \aap, 460, 653

\bibitem[{{De Pasquale} {et~al.}(2003){De Pasquale}, {Piro}, {Perna}, {Costa},
  {Feroci}, {Gandolfi}, {Zand}, {Nicastro}, {Frontera}, {Antonelli}, {Fiore},
  \& {Stratta}}]{depasquale03}
{De Pasquale}, M., et~al.\ 2003, \apj, 592, 1018

\bibitem[{{Dessart} {et~al.}(2008){Dessart}, {Burrows}, {Livne}, \&
  {Ott}}]{dessart08}
{Dessart}, L., {Burrows}, A., {Livne}, E., \& {Ott}, C.~D. 2008, \apjl, 673,
  L43

\bibitem[{{Dessart} \& {Hillier}(2008)}]{dessart07}
{Dessart}, L. \& {Hillier}, D.~J. 2008, \mnras, 383, 57

\bibitem[{{Di Carlo} {et~al.}(2008){Di Carlo}, {Corsi}, {Arkharov}, {Massi},
  {Larionov}, {Efimova}, {Dolci}, {Napoleone}, \& {Di Paola}}]{dicarlo08}
{Di Carlo}, E., et~al.\ 2008, \apj,  684, 471

\bibitem[{{Dickey} \& {Lockman}(1990)}]{dickey90}
{Dickey}, J.~M. \& {Lockman}, F.~J. 1990, \araa, 28, 215

\bibitem[{{Elmhamdi} {et~al.}(2004){Elmhamdi}, {Danziger}, {Cappellaro}, {Della
  Valle}, {Gouiffes}, {Phillips}, \& {Turatto}}]{elmhamdi04}
{Elmhamdi}, A., {Danziger}, I.~J., {Cappellaro}, E., {Della Valle}, M.,
  {Gouiffes}, C., {Phillips}, M.~M., \& {Turatto}, M. 2004, \aap, 426, 963

\bibitem[{{Ensman} \& {Burrows}(1992)}]{ensman92}
{Ensman}, L. \& {Burrows}, A. 1992, \apj, 393, 742

\bibitem[{{Faber} {et~al.}(2003){Faber}, {Phillips}, {Kibrick}, {Alcott},
  {Allen}, {Burrous}, {Cantrall}, {Clarke}, {Coil}, {Cowley}, {Davis}, {Deich},
  {Dietsch}, {Gilmore}, {Harper}, {Hilyard}, {Lewis}, {McVeigh}, {Newman},
  {Osborne}, {Schiavon}, {Stover}, {Tucker}, {Wallace}, {Wei}, {Wirth}, \&
  {Wright}}]{faber03}
{Faber}, S.~M., et~al.\ 2003, in Presented at the  Society of Photo-Optical Instrumentation Engineers (SPIE) Conference, Vol.  4841, Instrument Design and Performance for Optical/Infrared Ground-based  Telescopes. Edited by Iye, Masanori; Moorwood, Alan F. M. Proceedings of the  SPIE, Volume 4841, pp. 1657-1669 (2003)., ed. M.~{Iye} \& A.~F.~M.  {Moorwood}, 1657--1669

\bibitem[{{Fabricant} {et~al.}(1998){Fabricant}, {Cheimets}, {Caldwell}, \&
  {Geary}}]{fabricant98}
{Fabricant}, D., {Cheimets}, P., {Caldwell}, N., \& {Geary}, J. 1998, \pasp,
  110, 79

\bibitem[{{Falco} {et~al.}(1999){Falco}, {Kurtz}, {Geller}, {Huchra}, {Peters},
  {Berlind}, {Mink}, {Tokarz}, \& {Elwell}}]{falco99}
{Falco}, E.~E., et~al.\ 1999, \pasp,  111, 438

\bibitem[{{Fesen} {et~al.}(2006){Fesen}, {Hammell}, {Morse}, {Chevalier},
  {Borkowski}, {Dopita}, {Gerardy}, {Lawrence}, {Raymond}, \& {van den
  Bergh}}]{fesen06_casabi}
{Fesen}, R.~A., et~al.\ 2006, \apj, 645, 283

\bibitem[{{Filippenko}(1982)}]{filippenko82}
{Filippenko}, A.~V. 1982, \pasp, 94, 715

\bibitem[{{Filippenko}(1991)}]{filippenko91}
{Filippenko}, A.~V. 1991, in IAU Symposium, Vol. 143, Wolf-Rayet Stars and
  Interrelations with Other Massive Stars in Galaxies, ed. K.~A. {van der
  Hucht} \& B.~{Hidayat}, 529--+

\bibitem[{{Filippenko}(1997)}]{filippenko97_review}
---. 1997, \araa, 35, 309

\bibitem[{{Filippenko}(2005)}]{filippenko05}
{Filippenko}, A.~V. 2005, in Astronomical Society of the Pacific Conference
  Series, Vol. 332, The Fate of the Most Massive Stars, ed. R.~{Humphreys} \&
  K.~{Stanek}, 33--+

\bibitem[{{Filippenko} {et~al.}(2001){Filippenko}, {Li}, {Treffers}, \&
  {Modjaz}}]{filippenko01}
{Filippenko}, A.~V., {Li}, W.~D., {Treffers}, R.~R., \& {Modjaz}, M. 2001, in
  ASP Conf. Ser. 246: IAU Colloq. 183: Small Telescope Astronomy on Global
  Scales, ed. B.~{Paczynski}, W.-P. {Chen}, \& C.~{Lemme}, 121--+

\bibitem[{{Filippenko} \& {Sargent}(1986)}]{filippenko86}
{Filippenko}, A.~V. \& {Sargent}, W.~L.~W. 1986, \aj, 91, 691

\bibitem[{{Filippenko} {et~al.}(1995)}]{filippenko95}
{Filippenko}, A.~V. {et~al.} 1995, \apjl, 450, L11+

\bibitem[{{Fisher} {et~al.}(1999){Fisher}, {Branch}, {Hatano}, \&
  {Baron}}]{fisher99}
{Fisher}, A., {Branch}, D., {Hatano}, K., \& {Baron}, E. 1999, \mnras, 304, 67

\bibitem[{{Fitzpatrick}(1999)}]{fitzpatrick99}
{Fitzpatrick}, E.~L. 1999, \pasp, 111, 63

\bibitem[{{Foley} {et~al.}(2006){Foley}, {Perley}, {Pooley}, {Prochaska},
  {Bloom}, {Li}, {Cobb}, {Chen}, {Aldering}, {Bailyn}, {Blake}, {Falco},
  {Green}, {Kowalski}, {Perlmutter}, {Roth}, \& {Volk}}]{foley06}
{Foley}, R.~J., et~al.\ 2006, \apj, 645, 450

\bibitem[{{Foley} {et~al.}(2007){Foley}, {Smith}, {Ganeshalingam}, {Li},
  {Chornock}, \& {Filippenko}}]{foley07}
{Foley}, R.~J., {Smith}, N., {Ganeshalingam}, M., {Li}, W., {Chornock}, R., \&
  {Filippenko}, A.~V. 2007, \apjl, 657, L105

\bibitem[{{Fransson} \& {Chevalier}(1987)}]{fransson87}
{Fransson}, C. \& {Chevalier}, R.~A. 1987, \apjl, 322, L15

\bibitem[{{Fukugita} {et~al.}(1995){Fukugita}, {Shimasaku}, \&
  {Ichikawa}}]{fukugita95}
{Fukugita}, M., {Shimasaku}, K., \& {Ichikawa}, T. 1995, \pasp, 107, 945

\bibitem[{{Galama} {et~al.}(1998)}]{galama98}
{Galama}, T.~J. {et~al.} 1998, \nat, 395, 670

\bibitem[{{Garg} {et~al.}(2007){Garg}, {Stubbs}, {Challis}, {Wood-Vasey},
  {Blondin}, {Huber}, {Cook}, {Nikolaev}, {Rest}, {Smith}, {Olsen}, {Suntzeff},
  {Aguilera}, {Prieto}, {Becker}, {Miceli}, {Miknaitis}, {Clocchiatti},
  {Minniti}, {Morelli}, \& {Welch}}]{garg07}
{Garg}, A., et~al.\ 2007, \aj, 133, 403

\bibitem[{{Gehrels} {et~al.}(2004){Gehrels}, {Chincarini}, {Giommi}, {Mason},
  {Nousek}, {Wells}, {White}, {Barthelmy}, {Burrows}, {Cominsky}, {Hurley},
  {Marshall}, {M{\'e}sz{\'a}ros}, {Roming}, {Angelini}, {Barbier}, {Belloni},
  {Campana}, {Caraveo}, {Chester}, {Citterio}, {Cline}, {Cropper}, {Cummings},
  {Dean}, {Feigelson}, {Fenimore}, {Frail}, {Fruchter}, {Garmire}, {Gendreau},
  {Ghisellini}, {Greiner}, {Hill}, {Hunsberger}, {Krimm}, {Kulkarni}, {Kumar},
  {Lebrun}, {Lloyd-Ronning}, {Markwardt}, {Mattson}, {Mushotzky}, {Norris},
  {Osborne}, {Paczynski}, {Palmer}, {Park}, {Parsons}, {Paul}, {Rees},
  {Reynolds}, {Rhoads}, {Sasseen}, {Schaefer}, {Short}, {Smale}, {Smith},
  {Stella}, {Tagliaferri}, {Takahashi}, {Tashiro}, {Townsley}, {Tueller},
  {Turner}, {Vietri}, {Voges}, {Ward}, {Willingale}, {Zerbi}, \&
  {Zhang}}]{gehrels04}
{Gehrels}, N., et~al.\ 2004, \apj, 611, 1005

\bibitem[{{Gerardy} {et~al.}(2004){Gerardy}, {Fesen}, {Marion}, {H{\"o}flich},
  {Wheeler}, {Nomoto}, \& {Motohara}}]{gerardy04}
{Gerardy}, C.~L., et~al.\ 2004, in Cosmic explosions  in three dimensions, ed. P.~{H{\"o}flich}, P.~{Kumar}, \& J.~C. {Wheeler},  57--+

\bibitem[{{Gezari} {et~al.}(2008){Gezari}, {Dessart}, {Basa}, {Martin},
  {Neill}, {Woosley}, {Hillier}, {Bazin}, {Forster}, {Friedman}, {Le Du},
  {Mazure}, {Morrissey}, {Neff}, {Schiminovich}, \& {Wyder}}]{gezari08}
{Gezari}, S., et~al.\ 2008, \apjl, 683, L131

\bibitem[{{Ghisellini} {et~al.}(2007){Ghisellini}, {Ghirlanda}, \&
  {Tavecchio}}]{ghisellini07}
{Ghisellini}, G., {Ghirlanda}, G., \& {Tavecchio}, F. 2007, \mnras, 382, L77

\bibitem[{{Hamann} {et~al.}(2006){Hamann}, {Gr{\"a}fener}, \&
  {Liermann}}]{hamman06}
{Hamann}, W.-R., {Gr{\"a}fener}, G., \& {Liermann}, A. 2006, \aap, 457, 1015

\bibitem[{{Hamuy} {et~al.}(2002){Hamuy}, {Maza}, {Pinto}, {Phillips},
  {Suntzeff}, {Blum}, {Olsen}, {Pinfield}, {Ivanov}, {Augusteijn}, {Brillant},
  {Chadid}, {Cuby}, {Doublier}, {Hainaut}, {Le Floc'h}, {Lidman},
  {Petr-Gotzens}, {Pompei}, \& {Vanzi}}]{hamuy02}
{Hamuy}, M., et~al.\ 2002, \aj, 124, 417

\bibitem[{{Hamuy} {et~al.}(1988){Hamuy}, {Suntzeff}, {Gonzalez}, \&
  {Martin}}]{hamuy88}
{Hamuy}, M., {Suntzeff}, N.~B., {Gonzalez}, R., \& {Martin}, G. 1988, \aj, 95,
  63

\bibitem[{{Harkness} {et~al.}(1987){Harkness}, {Wheeler}, {Margon}, {Downes},
  {Kirshner}, {Uomoto}, {Barker}, {Cochran}, {Dinerstein}, {Garnett}, \&
  {Levreault}}]{harkness87}
{Harkness}, R.~P., et~al.\ 1987, \apj, 317, 355

\bibitem[{{Haynes} {et~al.}(1997){Haynes}, {Giovanelli}, {Herter}, {Vogt},
  {Freudling}, {Maia}, {Salzer}, \& {Wegner}}]{haynes97}
{Haynes}, M.~P., {Giovanelli}, R., {Herter}, T., {Vogt}, N.~P., {Freudling},
  W., {Maia}, M.~A.~G., {Salzer}, J.~J., \& {Wegner}, G. 1997, \aj, 113, 1197

\bibitem[{{Heise} {et~al.}(2001){Heise}, {in't Zand}, {Kippen}, \&
  {Woods}}]{heise01}
{Heise}, J., {in't Zand}, J., {Kippen}, R.~M., \& {Woods}, P.~M. 2001, in
  Gamma-ray Bursts in the Afterglow Era, ed. E.~{Costa}, F.~{Frontera}, \&
  J.~{Hjorth}, 16--+

\bibitem[{{Herald} {et~al.}(2001){Herald}, {Hillier}, \&
  {Schulte-Ladbeck}}]{herald01}
{Herald}, J.~E., {Hillier}, D.~J., \& {Schulte-Ladbeck}, R.~E. 2001, \apj, 548,
  932

\bibitem[{{Herbig}(1975)}]{herbig75}
{Herbig}, G.~H. 1975, \apj, 196, 129

\bibitem[{{Hicken} {et~al.}(2009){Hicken}, {Challis}, {Jha}, {Kirsher},
  {Matheson}, {Modjaz}, {Rest}, \& {Wood-Vasey}}]{hicken09}
{Hicken}, M., {Challis}, P., {Jha}, S., {Kirsher}, R.~P., {Matheson}, T.,
  {Modjaz}, M., {Rest}, A., \& {Wood-Vasey}, W.~M. 2009, In press by AJ
  (2009arXiv0901.4787)

\bibitem[{{Hjorth} {et~al.}(2003)}]{hjorth03}
{Hjorth}, J. {et~al.} 2003, \nat, 423, 847

\bibitem[{{Hook} {et~al.}(2003){Hook}, {Allington-Smith}, {Beard}, {Crampton},
  {Davies}, {Dickson}, {Ebbers}, {Fletcher}, {Jorgensen}, {Jean}, {Juneau},
  {Murowinski}, {Nolan}, {Laidlaw}, {Leckie}, {Marshall}, {Purkins},
  {Richardson}, {Roberts}, {Simons}, {Smith}, {Stilburn}, {Szeto}, {Tierney},
  {Wolff}, \& {Wooff}}]{hook03}
{Hook}, I., et~al.\ 2003, in  Proceedings of the SPIE., ed. M.~{Iye} \& A.~F.~M. {Moorwood}, Vol. 4841,  1645--1656

\bibitem[{{Horne}(1986)}]{horne86}
{Horne}, K. 1986, \pasp, 98, 609

\bibitem[{{Immler} {et~al.}(2008{\natexlab{a}})}]{immler08_gcn1}
{Immler}, E. {et~al.} 2008{\natexlab{a}}, GCN Circ. 7168

\bibitem[{{Immler} {et~al.}(2008{\natexlab{b}})}]{immler08_gcn2}
---. 2008{\natexlab{b}}, GCN Circ. 7185

\bibitem[{{Kasen} {et~al.}(2006){Kasen}, {Thomas}, \& {Nugent}}]{kasen06}
{Kasen}, D., {Thomas}, R.~C., \& {Nugent}, P. 2006, \apj, 651, 366

\bibitem[{{Katz} {et~al.}(2009){Katz}, {Budnik}, \& {Waxman}}]{katz09}
{Katz}, B., {Budnik}, R., \& {Waxman}, E. 2009, submitted (arXiv:0902.470)

\bibitem[{{Kong} \& {Maccarone}(2008)}]{kong08_atel}
{Kong}, E. \& {Maccarone}, T.~J. 2008, ATEL 1355

\bibitem[{{Kong} {et~al.}(2008{\natexlab{a}})}]{valenti08_idgcn}
{Kong}, E. {et~al.} 2008{\natexlab{a}}, GCN Circ. 7171

\bibitem[{{Kong} {et~al.}(2008{\natexlab{b}})}]{kong08_gcn}
---. 2008{\natexlab{b}}, GCN Circ. 7164

\bibitem[{{Landolt}(1992)}]{landolt92}
{Landolt}, A.~U. 1992, \aj, 104, 340

\bibitem[{{Leibundgut} {et~al.}(1991){Leibundgut}, {Kirshner}, {Pinto},
  {Rupen}, {Smith}, {Gunn}, \& {Schneider}}]{leibundgut91}
{Leibundgut}, B., {Kirshner}, R.~P., {Pinto}, P.~A., {Rupen}, M.~P., {Smith},
  R.~C., {Gunn}, J.~E., \& {Schneider}, D.~P. 1991, \apj, 372, 531

\bibitem[{{Leonard} \& {Filippenko}(2005)}]{leonard05}
{Leonard}, D.~C. \& {Filippenko}, A.~V. 2005, in ASP Conf. Ser. 342: 1604-2004:
  Supernovae as Cosmological Lighthouses, ed. M.~{Turatto}, S.~{Benetti},
  L.~{Zampieri}, \& W.~{Shea}, 330--+

\bibitem[{{Li} \& {McCray}(1992)}]{li92}
{Li}, H. \& {McCray}, R. 1992, \apj, 387, 309

\bibitem[{{Li}(2008)}]{li08}
{Li}, L.-X. 2008, \mnras, 388, 603

\bibitem[{{Li} {et~al.}(2008){Li}, {Chornock}, {Foley}, {Filippenko}, {Modjaz},
  {Poznanski}, \& {Bloom}}]{li08_gcn}
{Li}, W., {Chornock}, R., {Foley}, R.~J., {Filippenko}, A.~V., {Modjaz}, M.,
  {Poznanski}, D., \& {Bloom}, J.~S. 2008, GRB Coordinates Network, 7176, 1

\bibitem[{{Li} \& {Filippenko}(2008)}]{li08_08diauc}
{Li}, W. \& {Filippenko}, A.~V. 2008, Central Bureau Electronic Telegrams,
  1202, 3

\bibitem[{{Li} {et~al.}(2001){Li}, {Filippenko}, {Gates}, {Chornock},
  {Gal-Yam}, {Ofek}, {Leonard}, {Modjaz}, {Rich}, {Riess}, \&
  {Treffers}}]{li01_02cx}
{Li}, W., et~al.\ 2001, \pasp, 113, 1178

\bibitem[{{Li} {et~al.}(2006){Li}, {Jha}, {Filippenko}, {Bloom}, {Pooley},
  {Foley}, \& {Perley}}]{li06}
{Li}, W., {Jha}, S., {Filippenko}, A.~V., {Bloom}, J.~S., {Pooley}, D.,
  {Foley}, R.~J., \& {Perley}, D.~A. 2006, \pasp, 118, 37

\bibitem[{{Lucy}(1991)}]{lucy91}
{Lucy}, L.~B. 1991, \apj, 383, 308

\bibitem[{{Maeda} {et~al.}(2008){Maeda}, {Kawabata}, {Mazzali}, {Tanaka},
  {Valenti}, {Nomoto}, {Hattori}, {Deng}, {Pian}, {Taubenberger}, {Iye},
  {Matheson}, {Filippenko}, {Aoki}, {Kosugi}, {Ohyama}, {Sasaki}, \&
  {Takata}}]{maeda08}
{Maeda}, K., et~al.\ 2008, Science, 319, 1220

\bibitem[{{Maeda} {et~al.}(2006){Maeda}, {Nomoto}, {Mazzali}, \&
  {Deng}}]{maeda06}
{Maeda}, K., {Nomoto}, K., {Mazzali}, P.~A., \& {Deng}, J. 2006, \apj, 640, 854

\bibitem[{{Maeda} {et~al.}(2007){Maeda}, {Tanaka}, {Nomoto}, {Tominaga},
  {Kawabata}, {Mazzali}, {Umeda}, {Suzuki}, \& {Hattori}}]{maeda07}
{Maeda}, K., et~al.\ 2007, \apj,  666, 1069

\bibitem[{{Malesani} {et~al.}(2009){Malesani}, {Fynbo}, {Hjorth}, {Leloudas},
  {Sollerman}, {Stritzinger}, {Vreeswijk}, {Watson}, {Gorosabel},
  {Micha{\l}owski}, {Th{\"o}ne}, {Augusteijn}, {Bersier}, {Jakobsson},
  {Jaunsen}, {Ledoux}, {Levan}, {Milvang-Jensen}, {Rol}, {Tanvir}, {Wiersema},
  {Xu}, {Albert}, {Bayliss}, {Gall}, {Grove}, {Koester}, {Leitet}, {Pursimo},
  \& {Skillen}}]{malesani09}
{Malesani}, D., et~al.\ 2009, \apjl, 692, L84

\bibitem[{{Malesani} {et~al.}(2004)}]{malesani04}
{Malesani}, D. {et~al.} 2004, \apjl, 609, L5

\bibitem[{{Malesani} {et~al.}(2008)}]{malesani08_idgcn}
---. 2008, GCN Circ. 7169

\bibitem[{{Marion} {et~al.}(2008){Marion}, {H\"oflich}, {Gerardy}, {Vacca},
  {Wheeler}, \& {Robinson}}]{marion08}
{Marion}, G.~H., {H\"oflich}, P., {Gerardy}, C.~L., {Vacca}, W.~D., {Wheeler},
  J.~C., \& {Robinson}, E.~L. 2008, AJ, submitted

\bibitem[{{Matheson} {et~al.}(2000{\natexlab{a}}){Matheson}, {Filippenko},
  {Ho}, {Barth}, \& {Leonard}}]{matheson00_93jdetail}
{Matheson}, T., {Filippenko}, A.~V., {Ho}, L.~C., {Barth}, A.~J., \& {Leonard},
  D.~C. 2000{\natexlab{a}}, \aj, 120, 1499

\bibitem[{{Matheson} {et~al.}(2001){Matheson}, {Filippenko}, {Li}, {Leonard},
  \& {Shields}}]{matheson01}
{Matheson}, T., {Filippenko}, A.~V., {Li}, W., {Leonard}, D.~C., \& {Shields},
  J.~C. 2001, \aj, 121, 1648

\bibitem[{{Matheson} {et~al.}(2000{\natexlab{b}})}]{matheson00_93j}
{Matheson}, T. {et~al.} 2000{\natexlab{b}}, \aj, 120, 1487

\bibitem[{{Mazzali} {et~al.}(2004){Mazzali}, {Deng}, {Maeda}, {Nomoto},
  {Filippenko}, \& {Matheson}}]{mazzali04}
{Mazzali}, P.~A., {Deng}, J., {Maeda}, K., {Nomoto}, K., {Filippenko}, A.~V.,
  \& {Matheson}, T. 2004, \apj, 614, 858

\bibitem[{{Mazzali} {et~al.}(2006){Mazzali}, {Deng}, {Nomoto}, {Sauer}, {Pian},
  {Tominaga}, {Tanaka}, {Maeda}, \& {Filippenko}}]{mazzali06_06aj}
{Mazzali}, P.~A., et~al.\ 2006, \nat,  442, 1018

\bibitem[{{Mazzali} {et~al.}(2005){Mazzali}, {Kawabata}, {Maeda}, {Nomoto},
  {Filippenko}, {Ramirez-Ruiz}, {Benetti}, {Pian}, {Deng}, {Tominaga},
  {Ohyama}, {Iye}, {Foley}, {Matheson}, {Wang}, \& {Gal-Yam}}]{mazzali05_03jd}
{Mazzali}, P.~A., et~al.\ 2005, Science, 308, 1284

\bibitem[{{Mazzali} \& {Lucy}(1998)}]{mazzali98}
{Mazzali}, P.~A. \& {Lucy}, L.~B. 1998, \mnras, 295, 428

\bibitem[{{Mazzali} {et~al.}(2008){Mazzali}, {Valenti}, {Della Valle},
  {Chincarini}, {Sauer}, {Benetti}, {Pian}, {Piran}, {D'Elia}, {Elias-Rosa},
  {Margutti}, {Pasotti}, {Antonelli}, {Bufano}, {Campana}, {Cappellaro},
  {Covino}, {D'Avanzo}, {Fiore}, {Fugazza}, {Gilmozzi}, {Hunter}, {Maguire},
  {Maiorano}, {Marziani}, {Masetti}, {Mirabel}, {Navasardyan}, {Nomoto},
  {Palazzi}, {Pastorello}, {Panagia}, {Pellizza}, {Sari}, {Smartt},
  {Tagliaferri}, {Tanaka}, {Taubenberger}, {Tominaga}, {Trundle}, \&
  {Turatto}}]{mazzali08}
{Mazzali}, P.~A., et~al.\ 2008, Science, 321, 1185

\bibitem[{{Miknaitis} {et~al.}(2007){Miknaitis}, {Pignata}, {Rest},
  {Wood-Vasey}, {Blondin}, {Challis}, {Smith}, {Stubbs}, {Suntzeff}, {Foley},
  {Matheson}, {Tonry}, {Aguilera}, {Blackman}, {Becker}, {Clocchiatti},
  {Covarrubias}, {Davis}, {Filippenko}, {Garg}, {Garnavich}, {Hicken}, {Jha},
  {Krisciunas}, {Kirshner}, {Leibundgut}, {Li}, {Miceli}, {Narayan}, {Prieto},
  {Riess}, {Salvo}, {Schmidt}, {Sollerman}, {Spyromilio}, \&
  {Zenteno}}]{miknaitis07}
{Miknaitis}, G., et~al.\ 2007,  \apj, 666, 674

\bibitem[{{Milisavljevic} {et~al.}(2009){Milisavljevic}, {Fesen}, {Gerardy},
  {Kirshner}, \& {Challis}}]{milisavljevic09}
{Milisavljevic}, D., {Fesen}, R., {Gerardy}, C., {Kirshner}, R., \& {Challis},
  P. 2009, submitted (arXiv:0904.4256)

\bibitem[{{Millard} {et~al.}(1999){Millard}, {Branch}, {Baron}, {Hatano},
  {Fisher}, {Filippenko}, {Kirshner}, {Challis}, {Fransson}, {Panagia},
  {Phillips}, {Sonneborn}, {Suntzeff}, {Wagoner}, \& {Wheeler}}]{millard99}
{Millard}, J., et~al.\ 1999, \apj, 527, 746

\bibitem[{{Miller} \& {Stone}(1993)}]{miller93}
{Miller}, J.~S. \& {Stone}, R.~P.~S. 1993, Lick Obs. Tech. Rep. 66 (Santa Cruz:
  Lick Obs.)

\bibitem[{{Mirabal} {et~al.}(2006){Mirabal}, {Halpern}, {An}, {Thorstensen}, \&
  {Terndrup}}]{mirabal06}
{Mirabal}, N., {Halpern}, J.~P., {An}, D., {Thorstensen}, J.~R., \& {Terndrup},
  D.~M. 2006, \apjl, 643, L99

\bibitem[{{Modjaz}(2007)}]{modjaz07_thesis}
{Modjaz}, M. 2007, PhD thesis, Harvard University

\bibitem[{{Modjaz} {et~al.}(2008{\natexlab{a}}){Modjaz}, {Chornock}, {Foley},
  {Filippenko}, {Li}, \& {Stringfellow}}]{modjaz08_08dhe}
{Modjaz}, M., {Chornock}, R., {Foley}, R.~J., {Filippenko}, A.~V., {Li}, W., \&
  {Stringfellow}, G. 2008{\natexlab{a}}, Central Bureau Electronic Telegrams,
  1222, 1

\bibitem[{{Modjaz} {et~al.}(2008{\natexlab{b}}){Modjaz}, {Kewley}, {Kirshner},
  {Stanek}, {Challis}, {Garnavich}, {Greene}, {Kelly}, \&
  {Prieto}}]{modjaz08_Z}
{Modjaz}, M., et~al.\ 2008{\natexlab{b}}, \aj, 135, 1136

\bibitem[{{Modjaz} {et~al.}(2008{\natexlab{c}}){Modjaz}, {Kirshner}, {Blondin},
  {Challis}, \& {Matheson}}]{modjaz08_doubleoxy}
{Modjaz}, M., {Kirshner}, R.~P., {Blondin}, S., {Challis}, P., \& {Matheson},
  T. 2008{\natexlab{c}}, \apjl, 687, L9

\bibitem[{{Modjaz} {et~al.}(2001){Modjaz}, {Li}, {Filippenko}, {King},
  {Leonard}, {Matheson}, {Treffers}, \& {Riess}}]{modjaz01}
{Modjaz}, M., {Li}, W., {Filippenko}, A.~V., {King}, J.~Y., {Leonard}, D.~C.,
  {Matheson}, T., {Treffers}, R.~R., \& {Riess}, A.~G. 2001, \pasp, 113, 308

\bibitem[{{Modjaz} {et~al.}(2006){Modjaz}, {Stanek}, {Garnavich}, {Berlind},
  {Blondin}, {Brown}, {Calkins}, {Challis}, {Diamond-Stanic}, {Hao}, {Hicken},
  {Kirshner}, \& {Prieto}}]{modjaz06}
{Modjaz}, M., et~al.\ 2006, \apjl,  645, L21

\bibitem[{{Mould} {et~al.}(2000){Mould}, {Huchra}, {Freedman}, {Kennicutt},
  {Ferrarese}, {Ford}, {Gibson}, {Graham}, {Hughes}, {Illingworth}, {Kelson},
  {Macri}, {Madore}, {Sakai}, {Sebo}, {Silbermann}, \& {Stetson}}]{mould00}
{Mould}, J.~R., et~al.\ 2000, \apj, 529, 786

\bibitem[{{Nakano} {et~al.}(2008){Nakano}, {Kadota}, {Itagaki}, \&
  {Corelli}}]{nakano08_07uy}
{Nakano}, S., {Kadota}, K., {Itagaki}, K., \& {Corelli}, P. 2008, \iaucirc,
  8908, 2

\bibitem[{{Nomoto} {et~al.}(1995){Nomoto}, {Iwamoto}, \& {Suzuki}}]{nomoto95}
{Nomoto}, K., {Iwamoto}, K., \& {Suzuki}, T. 1995, \physrep, 256, 173

\bibitem[{{Oke} {et~al.}(1995){Oke}, {Cohen}, {Carr}, {Cromer}, {Dingizian},
  {Harris}, {Labrecque}, {Lucinio}, {Schaal}, {Epps}, \& {Miller}}]{oke95}
{Oke}, J.~B., et~al.\ 1995, \pasp, 107, 375

\bibitem[{{Page} {et~al.}(2008)}]{page08_gcn}
{Page}, K.~L. {et~al.} 2008, GCN Report 110.1

\bibitem[{{Patat} {et~al.}(2001)}]{patat01}
{Patat}, F. {et~al.} 2001, \apj, 555, 900

\bibitem[{{Phillips} \& {Davis}(1995)}]{phillips95}
{Phillips}, A.~C. \& {Davis}, L.~E. 1995, in Astronomical Society of the
  Pacific Conference Series, Vol.~77, Astronomical Data Analysis Software and
  Systems IV, ed. R.~A. {Shaw}, H.~E. {Payne}, \& J.~J.~E. {Hayes}, 297--+

\bibitem[{{Phillips}(1993)}]{phillips93}
{Phillips}, M.~M. 1993, \apjl, 413, L105

\bibitem[{{Pian} {et~al.}(2006)}]{pian06}
{Pian}, E. {et~al.} 2006, \nat, 442, 1011

\bibitem[{{Podsiadlowski} {et~al.}(2004){Podsiadlowski}, {Langer},
  {Poelarends}, {Rappaport}, {Heger}, \& {Pfahl}}]{podsiadlowski04}
{Podsiadlowski}, P., {Langer}, N., {Poelarends}, A.~J.~T., {Rappaport}, S.,
  {Heger}, A., \& {Pfahl}, E. 2004, \apj, 612, 1044

\bibitem[{{Poole} {et~al.}(2008){Poole}, {Breeveld}, {Page}, {Landsman},
  {Holland}, {Roming}, {Kuin}, {Brown}, {Gronwall}, {Hunsberger}, {Koch},
  {Mason}, {Schady}, {vanden Berk}, {Blustin}, {Boyd}, {Broos}, {Carter},
  {Chester}, {Cucchiara}, {Hancock}, {Huckle}, {Immler}, {Ivanushkina},
  {Kennedy}, {Marshall}, {Morgan}, {Pandey}, {de Pasquale}, {Smith}, \&
  {Still}}]{poole08}
{Poole}, T.~S., et~al.\ 2008, \mnras, 383,  627

\bibitem[{{Pooley} \& {Soderberg}(2008)}]{pooley08_gcn}
{Pooley}, D. \& {Soderberg}, A. 2008, GCN Circ. 7216

\bibitem[{{Predehl} \& {Schmitt}(1995)}]{predehl95}
{Predehl}, P. \& {Schmitt}, J.~H.~M.~M. 1995, \aap, 293, 889

\bibitem[{{Quimby} {et~al.}(2007){Quimby}, {Wheeler}, {H{\"o}flich}, {Akerlof},
  {Brown}, \& {Rykoff}}]{quimby07}
{Quimby}, R.~M., {Wheeler}, J.~C., {H{\"o}flich}, P., {Akerlof}, C.~W.,
  {Brown}, P.~J., \& {Rykoff}, E.~S. 2007, \apj, 666, 1093

\bibitem[{{Rayner} {et~al.}(2003){Rayner}, {Toomey}, {Onaka}, {Denault},
  {Stahlberger}, {Vacca}, {Cushing}, \& {Wang}}]{rayner03}
{Rayner}, J.~T., {Toomey}, D.~W., {Onaka}, P.~M., {Denault}, A.~J.,
  {Stahlberger}, W.~E., {Vacca}, W.~D., {Cushing}, M.~C., \& {Wang}, S. 2003,
  \pasp, 115, 362

\bibitem[{{Rest} {et~al.}(2005){Rest}, {Stubbs}, {Becker}, {Miknaitis},
  {Miceli}, {Covarrubias}, {Hawley}, {Smith}, {Suntzeff}, {Olsen}, {Prieto},
  {Hiriart}, {Welch}, {Cook}, {Nikolaev}, {Huber}, {Prochtor}, {Clocchiatti},
  {Minniti}, {Garg}, {Challis}, {Keller}, \& {Schmidt}}]{rest05}
{Rest}, A., et~al.\ 2005,  \apj, 634, 1103

\bibitem[{{Richardson}(2009)}]{richardson09}
{Richardson}, D. 2009, \aj, 137, 347

\bibitem[{{Richardson} {et~al.}(2006){Richardson}, {Branch}, \&
  {Baron}}]{richardson06}
{Richardson}, D., {Branch}, D., \& {Baron}, E. 2006, \aj, 131, 2233

\bibitem[{{Richmond} {et~al.}(1994){Richmond}, {Treffers}, {Filippenko},
  {Paik}, {Leibundgut}, {Schulman}, \& {Cox}}]{richmond94}
{Richmond}, M.~W., {Treffers}, R.~R., {Filippenko}, A.~V., {Paik}, Y.,
  {Leibundgut}, B., {Schulman}, E., \& {Cox}, C.~V. 1994, \aj, 107, 1022

\bibitem[{{Riess} {et~al.}(1998){Riess}, {Filippenko}, {Challis},
  {Clocchiatti}, {Diercks}, {Garnavich}, {Gilliland}, {Hogan}, {Jha},
  {Kirshner}, {Leibundgut}, {Phillips}, {Reiss}, {Schmidt}, {Schommer},
  {Smith}, {Spyromilio}, {Stubbs}, {Suntzeff}, \& {Tonry}}]{riess98}
{Riess}, A.~G., et~al.\ 1998, \aj, 116, 1009

\bibitem[{{Roming} {et~al.}(2005){Roming}, {Kennedy}, {Mason}, {Nousek}, {Ahr},
  {Bingham}, {Broos}, {Carter}, {Hancock}, {Huckle}, {Hunsberger}, {Kawakami},
  {Killough}, {Koch}, {McLelland}, {Smith}, {Smith}, {Soto}, {Boyd},
  {Breeveld}, {Holland}, {Ivanushkina}, {Pryzby}, {Still}, \&
  {Stock}}]{roming05}
{Roming}, P.~W.~A., et~al.\ 2005, Space Science Reviews,  120, 95

\bibitem[{{Sakamoto} {et~al.}(2008){Sakamoto}, {Hullinger}, {Sato}, {Yamazaki},
  {Barbier}, {Barthelmy}, {Cummings}, {Fenimore}, {Gehrels}, {Krimm}, {Lamb},
  {Markwardt}, {Osborne}, {Palmer}, {Parsons}, {Stamatikos}, \&
  {Tueller}}]{sakamoto08}
{Sakamoto}, T., et~al.\ 2008,  \apj, 679, 570

\bibitem[{{Sakamoto} {et~al.}(2005){Sakamoto}, {Lamb}, {Kawai}, {Yoshida},
  {Graziani}, {Fenimore}, {Donaghy}, {Matsuoka}, {Suzuki}, {Ricker}, {Atteia},
  {Shirasaki}, {Tamagawa}, {Torii}, {Galassi}, {Doty}, {Vanderspek}, {Crew},
  {Villasenor}, {Butler}, {Prigozhin}, {Jernigan}, {Barraud}, {Boer},
  {Dezalay}, {Olive}, {Hurley}, {Levine}, {Monnelly}, {Martel}, {Morgan},
  {Woosley}, {Cline}, {Braga}, {Manchanda}, {Pizzichini}, {Takagishi}, \&
  {Yamauchi}}]{sakamoto05}
{Sakamoto}, T., et~al.\ 2005, \apj, 629, 311

\bibitem[{{Sauer} {et~al.}(2006){Sauer}, {Mazzali}, {Deng}, {Valenti},
  {Nomoto}, \& {Filippenko}}]{sauer06}
{Sauer}, D.~N., {Mazzali}, P.~A., {Deng}, J., {Valenti}, S., {Nomoto}, K., \&
  {Filippenko}, A.~V. 2006, \mnras, 369, 1939

\bibitem[{{Schawinski} {et~al.}(2008){Schawinski}, {Justham}, {Wolf},
  {Podsiadlowski}, {Sullivan}, {Steenbrugge}, {Bell}, {R{\"o}ser}, {Walker},
  {Astier}, {Balam}, {Balland}, {Carlberg}, {Conley}, {Fouchez}, {Guy},
  {Hardin}, {Hook}, {Howell}, {Pain}, {Perrett}, {Pritchet}, {Regnault}, \&
  {Yi}}]{schawinski08}
{Schawinski}, K., et~al.\ 2008,  Science, 321, 223

\bibitem[{{Schechter} {et~al.}(1993){Schechter}, {Mateo}, \&
  {Saha}}]{schechter93}
{Schechter}, P.~L., {Mateo}, M., \& {Saha}, A. 1993, \pasp, 105, 1342

\bibitem[{{Scheck} {et~al.}(2006){Scheck}, {Kifonidis}, {Janka}, \&
  {M{\"u}ller}}]{scheck06}
{Scheck}, L., {Kifonidis}, K., {Janka}, H.-T., \& {M{\"u}ller}, E. 2006, \aap,
  457, 963

\bibitem[{{Schmidt} {et~al.}(1993){Schmidt}, {Kirshner}, {Eastman}, {Grashuis},
  {dell'Antonio}, {Caldwell}, {Foltz}, {Huchra}, \& {Milone}}]{schmidt93}
{Schmidt}, B.~P., et~al.\ 1993, \nat, 364, 600

\bibitem[{{Schmidt} {et~al.}(1989){Schmidt}, {Weymann}, \& {Foltz}}]{schmidt89}
{Schmidt}, G.~D., {Weymann}, R.~J., \& {Foltz}, C.~B. 1989, \pasp, 101, 713

\bibitem[{{Skrutskie} {et~al.}(2006){Skrutskie}, {Cutri}, {Stiening},
  {Weinberg}, {Schneider}, {Carpenter}, {Beichman}, {Capps}, {Chester},
  {Elias}, {Huchra}, {Liebert}, {Lonsdale}, {Monet}, {Price}, {Seitzer},
  {Jarrett}, {Kirkpatrick}, {Gizis}, {Howard}, {Evans}, {Fowler}, {Fullmer},
  {Hurt}, {Light}, {Kopan}, {Marsh}, {McCallon}, {Tam}, {Van Dyk}, \&
  {Wheelock}}]{skrutskie06}
{Skrutskie}, M.~F., et~al.\ 2006, \aj, 131, 1163

\bibitem[{{Smith} \& {Conti}(2008)}]{smith08_wnh}
{Smith}, N. \& {Conti}, P.~S. 2008, \apj, 679, 1467

\bibitem[{{Smith} {et~al.}(2008){Smith}, {Foley}, \&
  {Filippenko}}]{smith08_dust}
{Smith}, N., {Foley}, R.~J., \& {Filippenko}, A.~V. 2008, \apj, 680, 568

\bibitem[{{Smith} \& {Owocki}(2006)}]{smith06}
{Smith}, N. \& {Owocki}, S.~P. 2006, \apjl, 645, L45

\bibitem[{{Soderberg} {et~al.}(2008{\natexlab{a}}){Soderberg}, {Berger}, {Fox},
  {Cucchiara}, {Rau}, {Ofek}, {Kasliwal}, \& {Cenko}}]{soderberg08_gcn}
{Soderberg}, A., et~al.\ 2008{\natexlab{a}}, GRB Coordinates  Network, 7165, 1

\bibitem[{{Soderberg} {et~al.}(2008{\natexlab{b}}){Soderberg}, {Berger},
  {Page}, {Schady}, {Parrent}, {Pooley}, {Wang}, {Ofek}, {Cucchiara}, {Rau},
  {Waxman}, {Simon}, {Bock}, {Milne}, {Page}, {Barentine}, {Barthelmy},
  {Beardmore}, {Bietenholz}, {Brown}, {Burrows}, {Burrows}, {Byrngelson},
  {Cenko}, {Chandra}, {Cummings}, {Fox}, {Gal-Yam}, {Gehrels}, {Immler},
  {Kasliwal}, {Kong}, {Krimm}, {Kulkarni}, {Maccarone}, {M{\'e}sz{\'a}ros},
  {Nakar}, {O'Brien}, {Overzier}, {de Pasquale}, {Racusin}, {Rea}, \&
  {York}}]{soderberg08}
{Soderberg}, A.~M., et~al.\ 2008{\natexlab{b}},  \nat, 453, 469

\bibitem[{{Soderberg} {et~al.}(2006){Soderberg}, {Kulkarni}, {Nakar}, {Berger},
  {Cameron}, {Fox}, {Frail}, {Gal-Yam}, {Sari}, {Cenko}, {Kasliwal},
  {Chevalier}, {Piran}, {Price}, {Schmidt}, {Pooley}, {Moon}, {Penprase},
  {Ofek}, {Rau}, {Gehrels}, {Nousek}, {Burrows}, {Persson}, \&
  {McCarthy}}]{soderberg06_06aj}
{Soderberg}, A.~M., et~al.\ 2006, \nat, 442, 1014

\bibitem[{{Sollerman} {et~al.}(1998){Sollerman}, {Leibundgut}, \&
  {Spyromilio}}]{sollerman98}
{Sollerman}, J., {Leibundgut}, B., \& {Spyromilio}, J. 1998, \aap, 337, 207

\bibitem[{{Sollerman} {et~al.}(2006)}]{sollerman06}
{Sollerman}, J. {et~al.} 2006, \aap, 454, 503

\bibitem[{{Spyromilio}(1991)}]{spyromilio91}
{Spyromilio}, J. 1991, \mnras, 253, 25P

\bibitem[{{Spyromilio}(1994)}]{spyromilio94_93j}
---. 1994, \mnras, 266, L61+

\bibitem[{{Stanek} {et~al.}(2003)}]{stanek03}
{Stanek}, K.~Z. {et~al.} 2003, \apjl, 591, L17

\bibitem[{{Stritzinger} {et~al.}(2002)}]{stritzinger02}
{Stritzinger}, M. {et~al.} 2002, \aj, 124, 2100

\bibitem[{{Tanaka} {et~al.}(2009){Tanaka}, {Tominaga}, {Nomoto}, {Valenti},
  {Sahu}, {Minezaki}, {Yoshii}, {Yoshida}, {Anupama}, {Benetti}, {Chincarini},
  {Valle}, {Mazzali}, \& {Pian}}]{tanaka08}
{Tanaka}, M., et~al.\ 2009, \apj,  692, 1131

\bibitem[{{Taubenberger} {et~al.}(2006){Taubenberger}, {Pastorello}, {Mazzali},
  {Valenti}, {Pignata}, {Sauer}, {Arbey}, {B{\"a}rnbantner}, {Benetti}, {Della
  Valle}, {Deng}, {Elias-Rosa}, {Filippenko}, {Foley}, {Goobar}, {Kotak}, {Li},
  {Meikle}, {Mendez}, {Patat}, {Pian}, {Ries}, {Ruiz-Lapuente}, {Salvo},
  {Stanishev}, {Turatto}, \& {Hillebrandt}}]{taubenberger06}
{Taubenberger}, S., et~al.\ 2006, \mnras, 371, 1459

\bibitem[{{Taubenberger} {et~al.}(2009){Taubenberger}, {Valenti}, {Benetti},
  {Cappellaro}, {Della Valle}, {Elias-Rosa}, {Hachinger}, {Hillebrandt},
  {Maeda}, {Mazzali}, {Pastorello}, {Patat}, {Sim}, \&
  {Turatto}}]{taubenberger09}
{Taubenberger}, S., et~al.\ 2009, MNRS, in press (arXiv:0904.4632)

\bibitem[{{Th{\"o}ne} {et~al.}(2009){Th{\"o}ne}, {Micha{\l}owski}, {Leloudas},
  {Cox}, {Fynbo}, {Sollerman}, {Hjorth}, \& {Vreeswijk}}]{thoene08}
{Th{\"o}ne}, C.~C., {Micha{\l}owski}, M.~J., {Leloudas}, G., {Cox}, N.~L.~J.,
  {Fynbo}, J.~P.~U., {Sollerman}, J., {Hjorth}, J., \& {Vreeswijk}, P.~M. 2009,
  \apj, 698, 1307

\bibitem[{{Tominaga} {et~al.}(2005)}]{tominaga05}
{Tominaga}, N. {et~al.} 2005, \apjl, 633, L97

\bibitem[{{Tomita} {et~al.}(2006){Tomita}, {Deng}, {Maeda}, {Yoshii}, {Nomoto},
  {Mazzali}, {Suzuki}, {Kobayashi}, {Minezaki}, {Aoki}, {Enya}, \&
  {Suganuma}}]{tomita06}
{Tomita}, H., et~al.\ 2006, \apj, 644, 400

\bibitem[{{Uomoto} \& {Kirshner}(1986)}]{uomoto86}
{Uomoto}, A. \& {Kirshner}, R.~P. 1986, \apj, 308, 685

\bibitem[{{Vacca} {et~al.}(2003){Vacca}, {Cushing}, \& {Rayner}}]{Vacca03}
{Vacca}, W.~D., {Cushing}, M.~C., \& {Rayner}, J.~T. 2003, \pasp, 115, 389

\bibitem[{{Valenti} {et~al.}(2008{\natexlab{a}}){Valenti}, {Fugazza},
  {Maiorano}, {D'Elia}, {Antonelli}, {Covino}, {Magazzu'}, {Pinilla-Alonso},
  {Della Valle}, {Chincarini}, {Pian}, {Mazzali}, {Harutyunyan}, \&
  {Benetti}}]{valenti08_08dhe}
{Valenti}, S., et~al.\ 2008{\natexlab{a}}, GRB Coordinates Network, 7171, 1

\bibitem[{{Valenti} {et~al.}(2008{\natexlab{b}})}]{valenti08_idiauc}
{Valenti}, S. {et~al.} 2008{\natexlab{b}}, Central Bureau Electronic Telegrams,
  1205, 1

\bibitem[{{Vogt} {et~al.}(1994){Vogt}, {Allen}, {Bigelow}, {Bresee}, {Brown},
  {Cantrall}, {Conrad}, {Couture}, {Delaney}, {Epps}, {Hilyard}, {Hilyard},
  {Horn}, {Jern}, {Kanto}, {Keane}, {Kibrick}, {Lewis}, {Osborne},
  {Pardeilhan}, {Pfister}, {Ricketts}, {Robinson}, {Stover}, {Tucker}, {Ward},
  \& {Wei}}]{vogt94}
{Vogt}, S.~S., et~al.\ 1994, in Presented  at the Society of Photo-Optical Instrumentation Engineers (SPIE) Conference,  Vol. 2198, Proc. SPIE Instrumentation in Astronomy VIII, David L. Crawford;  Eric R. Craine; Eds., Volume 2198, p. 362, ed. D.~L. {Crawford} \& E.~R.  {Craine}, 362--+

\bibitem[{{Wade} \& {Horne}(1988)}]{wade88}
{Wade}, R.~A. \& {Horne}, K. 1988, \apj, 324, 411

\bibitem[{{Wang} {et~al.}(2006){Wang}, {Lai}, \& {Han}}]{wang06_nskicks}
{Wang}, C., {Lai}, D., \& {Han}, J.~L. 2006, \apj, 639, 1007

\bibitem[{{Wang} {et~al.}(2007){Wang}, {Li}, {Waxman}, \&
  {M{\'e}sz{\'a}ros}}]{wang07_06aj}
{Wang}, X.-Y., {Li}, Z., {Waxman}, E., \& {M{\'e}sz{\'a}ros}, P. 2007, \apj,
  664, 1026

\bibitem[{{Waxman} {et~al.}(2007){Waxman}, {M{\'e}sz{\'a}ros}, \&
  {Campana}}]{waxman07}
{Waxman}, E., {M{\'e}sz{\'a}ros}, P., \& {Campana}, S. 2007, \apj, 667, 351

\bibitem[{{Wheeler} {et~al.}(1993){Wheeler}, {Barker}, {Benjamin}, {Boisseau},
  {Clocchiatti}, {de Vaucouleurs}, {Gaffney}, {Harkness}, {Khokhlov}, {Lester},
  {Smith}, {Smith}, \& {Tomkin}}]{wheeler93}
{Wheeler}, J.~C., et~al.\ 1993,  \apjl, 417, L71+

\bibitem[{{Wood-Vasey} {et~al.}(2008){Wood-Vasey}, {Friedman}, {Bloom},
  {Hicken}, {Modjaz}, {Kirshner}, {Starr}, {Blake}, {Falco}, {Szentgyorgyi},
  {Challis}, {Blondin}, {Mandel}, \& {Rest}}]{wood-vasey08}
{Wood-Vasey}, W.~M., et~al.\ 2008, \apj, 689, 377

\bibitem[{{Woosley} \& {Bloom}(2006)}]{woosley06_rev}
{Woosley}, S.~E. \& {Bloom}, J.~S. 2006, \araa, 44, 507

\bibitem[{{Woosley} {et~al.}(1993){Woosley}, {Langer}, \& {Weaver}}]{woosley93}
{Woosley}, S.~E., {Langer}, N., \& {Weaver}, T.~A. 1993, \apj, 411, 823

\bibitem[{{Woosley} {et~al.}(1987){Woosley}, {Pinto}, {Martin}, \&
  {Weaver}}]{woosley87}
{Woosley}, S.~E., {Pinto}, P.~A., {Martin}, P.~G., \& {Weaver}, T.~A. 1987,
  \apj, 318, 664

\bibitem[{{Xu} {et~al.}(2008){Xu}, {Watson}, {Fynbo}, {Fan}, {Zou}, \&
  {Hjorth}}]{xu08}
{Xu}, D., {Watson}, D., {Fynbo}, J., {Fan}, Y., {Zou}, Y.-C., \& {Hjorth}, J.
  2008, in COSPAR, Plenary Meeting, Vol.~37, 37th COSPAR Scientific Assembly,
  3512


\end{thebibliography}

\begin{deluxetable}{cccc}
\tablecolumns{4} 
\tablewidth{0pc}
\tablecaption{ X-ray Light Curve of \xtransient }
\tabletypesize{\scriptsize}
\tablehead{
\colhead{$t$ [s]} & 
\colhead{$dt$ [s]} & 
\colhead{Luminosity $L_{\rm{x}}$ [$10^{42}$ \ergs]} &
\colhead{$\sigma_{(L_{\rm{X}})}$ [$10^{42}$ \ergs]}
} 
\startdata
       11.29   &      11.28  &    3.60   &     1.06 \\
       27.59   &      5.02   &    15.38  &     4.53 \\
       37.62   &      5.02   &    13.65  &     4.34 \\
       47.65   &      5.02   &    12.43  &     4.14 \\
       55.17   &      2.51   &    28.02  &     8.67 \\
       60.19   &      2.51   &    28.02  &     8.67 \\
       65.20   &      2.51   &    38.47  &     10.11 \\
       71.47   &      3.76   &    18.49  &     5.78 \\ 
       78.99   &      3.76   &    20.32  &     6.04 \\
       86.51   &      3.76   &    21.68  &     6.29 \\ 
       92.78   &      2.51   &    25.26  &     8.28 \\
       99.05   &      3.76   &    16.84  &     5.52 \\
       109.08  &      6.27   &    13.88  &     3.92 \\
       119.11  &      3.76   &    20.62  &     6.04 \\
       125.38  &      2.51   &    30.36  &     9.06 \\
       131.65  &      3.76   &    17.03  &     5.50 \\
       137.91  &      2.51   &    25.70  &     8.23 \\
       149.20  &      8.78   &    8.04   &     2.48 \\
       164.24  &      6.27   &    14.51  &     3.91 \\
       175.52  &      5.01   &    14.37  &     4.34 \\
       186.81  &      6.27   &    10.17  &     3.30 \\
       200.60  &      7.52   &    11.18  &     3.14 \\
       215.64  &      7.52   &    8.57   &     2.75 \\
       228.18  &      5.01   &    12.71  &     4.13 \\
       244.46  &      11.28   &    5.68  &     1.83 \\
       263.28  &      7.52   &    9.53  &     2.89 \\
       278.32  &      7.52   &    9.34  &     2.89 \\
       294.62  &      8.78   &    7.34 &      2.36 \\
       317.19  &      13.79   &    4.64 &     1.57 \\
       346.02  &      15.04   &    2.69  &    0.86 \\
       369.84  &      8.78   &    3.54  &     1.18 \\
       383.63  &      5.01   &    6.19 &     2.06 \\
       398.68  &      10.03   &    3.42  &     1.09 \\
       427.51  &      18.81   &    1.79  &    0.58 \\
       472.64  &      26.33   &    1.29  &   0.41  \\
       514.01  &      15.04   &    2.06  &    0.69 \\ 
       778.53  &      249.47   &   0.26  &   0.060 \\
       14944.11  &      9184.50   & 0.0040  &  0.0055 \\
       677205.64  &     376223.75 &  $<$ 0.0016 & \ldots \\
       898641.0\tablenotemark{a}  &     17900. &  0.00032   & 0.00017
\enddata
\tablecomments{Using a count-to-luminosity conversion of $4.3 \times
  10^{42}$ \ergs\ per count~s$^{-1}$ based on the pure PL spectral
  fit, for a distance of \distd . {\it Swift} data unless noted otherwise.}
\tablenotetext{a}{{\it Chanda} data.}
\label{xraylc_fulltable}
\end{deluxetable}


\begin{deluxetable}{cccc}
\tablecolumns{4} 
\tablewidth{0pc}
\tablecaption{Model Parameters for X-ray Spectral Fits of \xtransient }
\tabletypesize{\scriptsize}
\tablehead{
\colhead{Parameter} & 
\colhead{Power-Law (PL) Fit} & 
\colhead{Blackbody (BB) Fit} &
\colhead{Combined PL+BB Fit\tablenotemark{a}}
} 
\startdata
Quality of Fit ($\chi^2/dof$) & 12.73/24 &28.56/25 & 12.57/23 \\
Implied $N_H$ [$10^{21}$ \cm ] & $5.2^{+2.1}_{-1.8}  $ & $<1.5 $ & $5.2^{+2.1}_{-1.8}$  \\
Time-integrated $\Gamma$, $kT$, and [$\Gamma$;$kT$] respectively &   $2.1^{+0.3}_{-0.4}$  &$ 0.75 \pm 0.07$ & $2.1^{+0.3}_{-0.4}$; $0.10 \pm 0.01$ \\
$L_{\rm X,{\rm iso}}$\tablenotemark{b} [ $10^{42}$ erg s$^{-1}$] & $12^{+5}_{-2}$  & $4.6 \pm 0.6 $ &  $13 \pm 2$\tablenotemark{c} \\
$E_{\rm X,{\rm iso}}$\tablenotemark{b} [$10^{45}$ erg] &  $5.8^{+2}_{-1}$  & 2.4$\pm$ 3 &  $6.7 \pm 4$\tablenotemark{c} 
\enddata
\tablenotetext{a}{Assuming the same (factor unity) ratio of PL to BB flux as in XRF~060218 \citep{butler06}.}
\tablenotetext{b}{Unabsorbed and time-averaged value over the range 0.3$-$10.0 keV (for PL fit) and bolometric (for BB fit), respectively, assuming \distd .}
\tablenotetext{c}{Listed only for the BB component. $L_{\rm X,{\rm iso}}$ and $E_{\rm X,{\rm iso}}$ are computed assuming $R_{\rm{BB}}^{\rm X}=10^{11}$ cm as required from the combined PL and BB fit. }
\label{xrayvals_table}
\end{deluxetable}


\begin{deluxetable}{ccccccccccc}
\tablewidth{0pc}
\tablecaption{Photometry of comparison stars in the field of \snd\ by KAIT and FLWO 1.2 m}
\tabletypesize{\scriptsize}
\tablehead{
\colhead{ID}&\colhead{$U$}& \colhead{$N_U$} 
&\colhead{$B$}& \colhead{$N_B$} &
\colhead{$V$}&\colhead{$N_V$} & \colhead{$R/r'$}& \colhead{$N_{R/r'}$} &
\colhead{$I/i'$} &\colhead{$N_{I/i'}$}\\
\colhead{ } &
\colhead{[mag]} &
\colhead{ } &
\colhead{[mag]}  &
\colhead{ } &
\colhead{[mag]}  &
\colhead{ } &
\colhead{[mag]}  &
\colhead{ } &
\colhead{[mag]}  &
\colhead{ } 
}
\startdata
1&16.771(029)& 2& 16.438(002)& 3& 15.563(006)& 2& 15.117(013)& 3& 14.640(016)& 3 \\
2&17.703(037)& 2& 17.732(003)& 2& 16.997(002)& 4& 16.590(005)& 2& 16.274(007)& 6 \\
3&13.611(005)& 3& 13.745(008)& 3& 13.234(010)& 4& 12.933(010)& 4& 12.639(015)& 3 \\
4&17.822(069)& 2& 17.865(009)& 3& 17.409(012)& 4& 17.061(010)& 5& 16.781(002)& 6 \\
5&$-$        &$-$&17.994(011)& 2& 17.316(010)& 4& 16.841(013)& 4& 16.498(009)& 5 \\
6&$-$        &$-$&18.132(014)& 4& 17.550(016)& 3& 17.115(005)& 4& 16.856(011)& 6 \\
7&12.466(005)& 3& 12.530(002)& 3& $-$        &$-$& $-$       & $-$& $-$      & $-$ \\
8&$-$        &$-$&18.977(012)& 4& 17.924(011)& 4& 17.227(006)& 5& 16.789(003)& 3 \\
9&15.003(007)& 3& 14.969(003)& 3& 14.394(016)& 3& 14.041(010)& 3& 13.734(013)& 3 \\
\\
\hline
\\
1&$-$        &$-$&16.415(049)& 1& 15.586(033)& 1& 15.326(045)& 1& 15.088(046)& 1 \\
3&$-$        &$-$&13.657(031)& 1& 13.191(022)& 1& 13.089(031)& 1& 12.999(031)& 1 \\
10&$-$       &$-$&16.498(049)& 1& 15.594(033)& 1& 15.298(045)& 1& 15.047(045)& 1 \\
11&$-$       &$-$&16.908(051)& 1& 15.764(035)& 1& 15.344(047)& 1& 15.004(048)& 1 \\
12&$-$       &$-$&18.068(089)& 1& 17.237(060)& 1& 17.038(079)& 1& 16.805(081)& 1 \\
13&$-$       &$-$&15.659(040)& 1& 14.827(027)& 1& 14.585(037)& 1& 14.391(037)& 1 \\
14&$-$       &$-$&14.364(033)& 1& 13.631(023)& 1& 13.446(031)& 1& 13.311(031)& 1 \\
15&$-$       &$-$&17.584(086)& 1& 17.164(058)& 1& 17.087(076)& 1& 17.040(078)& 1 \\
16&$-$       &$-$&16.054(045)& 1& 15.320(031)& 1& 15.105(041)& 1& 14.946(042)& 1 \\
17&$-$       &$-$&17.309(055)& 1& 15.966(037)& 1& 15.486(049)& 1& 15.059(051)& 1 
\enddata
\tablecomments{ Uncertainties (standard deviation of the mean) are
  indicated in parentheses. Stars (see Fig.~\ref{fchart_fig}) in the
  top half of the table were observed with KAIT in $UBVRI$ and the
  stars in the bottom half were observed with the FLWO 1.2-m telescope
  in $BVr'i'$.}
\label{standard_table}
\end{deluxetable}

\begin{deluxetable}{lcccccccccl}
\tablewidth{0pc}
\tabletypesize{\scriptsize}
\tablecaption{Optical Photometry of SN 2008D}
\tablehead{
\colhead{MJD} &
\colhead{$\Delta t$\tablenotemark{a}}
&\colhead{$B$}& 
\colhead{$V$}& \colhead{$R$}& \colhead{$R(c)$\tablenotemark{b}} &
 \colhead{$r'$}& \colhead{$i'$}& 
\colhead{$I$} &\colhead{Tel.\tablenotemark{c}} \\
\colhead{[day]} &
\colhead{[day]} &
\colhead{[mag]}  &
\colhead{[mag]}  &
\colhead{[mag]}  &
\colhead{[mag]}  &
\colhead{[mag]}  &
\colhead{[mag]}  &
\colhead{[mag]}
}
\startdata
54474.45 & $-$0.12 & $>$   20.10 & $>$ 19.80 & $-$ &  $-$ & $>$ 20.20 & $>$   20.20& $-$ & FLWO   \\
54475.41 & 0.84 & 19.08(03) & 18.52(04) & $-$  &  $-$  & 18.31(04) & 18.17(03) &  $-$ & FLWO \\   
54476.41 & 1.85 & 19.05(07)&18.33(04)&17.80(03)&17.78(05)& $-$ &  $-$& 17.34(05)& KAIT \\
54476.43 & 1.87 & 18.98(03) & 18.38(03) & $-$  &  $-$  & 18.11(03) & 17.86(03) &  $-$ & FLWO \\   
54477.35 & 2.78 &19.01(03)  & 18.39(05)&17.84(03)&$-$      & $-$ &  $-$ & 17.36(02)&Nickel\\
54477.36 & 2.79 & 19.11(02) & 18.44(03) & $-$  &  $-$ & 18.15(02) & 17.86(02) &  $-$ & FLWO \\   
54477.46 & 2.89 &19.15(08)  & 18.43(05) & 17.85(04) & 17.82(05) & $-$ &  $-$&17.38(04)&KAIT\\
54478.34 & 3.78 & 19.09(03) & 18.37(02) & 17.89(02) & $-$      & $-$ &  $-$ &17.37(02) & Nickel\\
54478.39 & 3.83 & 19.24(02) & 18.47(02) & $-$  &  $-$ & 18.17(02) & 17.89(02) &  $-$ & FLWO \\   
54478.44 & 3.87 & 19.22(09) & 18.43(05) & 17.83(04) & 17.81(05) & $-$ & $-$ & 17.40(04) & KAIT\\ 
54479.43 & 4.86 & 19.14(06) & 18.30(04) & 17.78(04) & 17.76(05) & $-$ & $-$ & 17.19(04) & KAIT\\ 
54480.39 & 5.83 & 19.34(06) & 18.22(04) & $-$  &  $-$ & 18.02(03) & 17.72(04) &  $-$ & FLWO \\   
54480.40 & 5.84 & 19.23(07) & 18.22(04) & 17.68(03) & 17.64(05) & $-$ & $-$ & 17.17(03) & KAIT\\ 
54481.43 & 6.86 & 18.95(11) & 17.97(05) & 17.47(06) & 17.48(05) & $-$ & $-$ & 16.97(05) & KAIT\\ 
54481.51 & 6.94 & 19.10(04) & 18.05(04) & $-$  &  $-$ & 17.75(05) & 17.58(09) &  $-$ & FLWO \\   
54482.39 & 7.82 & 18.88(11) & 17.93(05) & 17.34(03) & 17.31(05) & $-$ & $-$ & 16.81(04) & KAIT\\ 
54482.43 & 7.86 & 18.98(02) & 17.94(03) & $-$  &  $-$ & 17.65(03) & 17.34(03) &  $-$ & FLWO \\   
54483.27 & 8.71 & 18.89(05) & 17.70(03) & $-$  &  $-$ & 17.51(02) & 17.25(03) &  $-$ & FLWO \\   
54483.36 & 8.79 & 18.90(10) & 17.73(04) & 17.27(02) & 17.21(06) & $-$ & $-$ & 16.72(03) & KAIT\\ 
54484.36 & 9.79 & 18.69(07) & 17.69(04) & 17.13(02) & 17.07(06) & $-$ & $-$ & 16.58(02) & KAIT\\ 
54485.34 & 10.77 & 18.54(08) & 17.61(05) & 17.07(02) & 16.97(08)  & $-$ & $-$ & 16.41(02) & KAIT\\ 
54489.41 & 19.90 & 18.76(05) & 17.63(03) & $-$  &  $-$ & 17.00(02) & 16.87(02) &  $-$ & FLWO \\   
54495.36 & 20.80 & 18.36(04) & 17.38(03) & $-$  &  $-$ & 17.00(02) & 16.65(02) &  $-$ & FLWO \\   
54496.34 & 21.77 & 18.52(03) & 17.41(02) & $-$  &  $-$ & 16.95(02) & 16.59(02) &  $-$ & FLWO \\   
54497.39 & 22.83 & 18.63(03) & 17.40(02) & $-$  &  $-$ & 17.06(03) & 16.68(03) &  $-$ & FLWO \\   
54498.30 & 23.74 & 18.63(05) & 17.50(04) & $-$  &  $-$ & 17.05(02) & 16.65(02) &  $-$ & FLWO \\   
54499.35 & 24.79 & 18.72(02) & 17.49(02) & $-$  &  $-$ & 17.07(02) & 16.70(02) &  $-$ & FLWO \\   
54501.42 & 26.85 & 18.92(07) & 17.54(03) & 16.82(02) & $-$ & $-$ & $-$ & 16.20(02) & KAIT\\ 
54502.36 & 27.79 & 19.01(05) & 17.58(03) & 16.86(02) & $-$ & $-$ & $-$ & 16.22(02) & KAIT\\ 
54502.44 & 27.87 & 18.93(04) & 17.51(02) & $-$  &  $-$  &17.18(02) & 16.77(02) &  $-$ & FLWO \\   
54503.34 & 28.78 & 19.15(03) & 17.67(02) & $-$  &  $-$  &17.22(02) & 16.78(02) &  $-$ & FLWO \\   
54503.38 & 28.81 & 19.13(06) & 17.64(03) & 16.86(02) & $-$ & $-$ & $-$ & 16.24(02) & KAIT\\ 
54504.35 & 29.78 & 19.19(05) & 17.74(03) & 16.95(02) & $-$ & $-$ & $-$ & 16.27(03) & KAIT\\ 
54504.37 & 29.81 & 19.25(04) & 17.71(02) & $-$  &  $-$  & 17.29(02) & 16.83(02) &  $-$ & FLWO \\   
54505.33 & 30.76 & 19.28(06) & 17.82(03) & 17.01(02) & $-$ & $-$ & $-$ & 16.28(02) & KAIT\\ 
54507.32 & 32.75 & 19.52(08) & 17.92(03) & 17.11(02) & $-$ & $-$ & $-$ & 16.38(02) & KAIT\\ 
54509.29 & 34.72 & 19.70(06) & 18.06(03) & 17.21(02) & $-$ & $-$ & $-$ & 16.42(02) & KAIT
\enddata
\tablecomments{Uncertainties of the measurements are
indicated in parentheses.}
\tablenotetext{a}{Time in days since $t_0$.}
\tablenotetext{b}{$R (c)$-band magnitudes derived from unfiltered observations.}
\tablenotetext{c}{Telescope used: KAIT = 0.76-m Katzman Automatic Imaging
Telescope; Nickel = Lick Observatory 1-m Nickel telescope; FLWO = FLWO 1.2-m telescope.}
\label{opt_table}
\end{deluxetable}


\begin{deluxetable}{cccccc}
\tablewidth{0pc}
\tablecaption{ {\it Swift}/UVOT photometry of SN~2008D}
\tablehead{
\colhead{$\Delta t$\tablenotemark{a}}&
\colhead{$\delta t$\tablenotemark{b}}& 
\colhead{Exp \tablenotemark{c}}&
\colhead{Mag} & 
\colhead{$\sigma$} &
\colhead{Filter} \\
\colhead{[day]} &
\colhead{[day]} &
\colhead{[s]} &
\colhead{} &
\colhead{[mag]} &
\colhead{} 
}
\startdata
0.1359&0.2688&412.27&20.24&0.21& $U$ \\
2.0047&1.0058&2677.07&18.62&0.03& $U$ \\
2.9769&0.8591&1001.81&19.07&0.06& $U$ \\
3.9479&0.9349&1314.54&19.36&0.07& $U$ \\
4.9872&0.9346&1347.46&19.37&0.07& $U$ \\
5.9550&0.9326&921.90&19.45&0.08& $U$ 
\enddata
\tablecomments{Table \ref{uvot_table} is published in its entirety in
  the electronic edition of the {\it Astrophysical Journal}. A portion
  is shown here for guidance regarding its form and content. Listed
  uncertainties do no include systematic uncertainties.}  
\tablenotetext{a}{Time in days (middle point of the  combined image) since $t_0$= 2008-01-09 13:32:49.}
\tablenotetext{b}{Time bin in days (from the start of the first
  exposure to the end of the last exposure in the combined image).}
\tablenotetext{c}{Total exposure time (in seconds).  }
\label{uvot_table}
\end{deluxetable}


\begin{deluxetable*}{lcccc}
\singlespace
\tablewidth{0pt}
\tablecaption{ Near-Infrared Photometry of SN~2008D }
\tablehead{\colhead{MJD} & 
\colhead{$\Delta t$} & 
\colhead{$J $} &
\colhead{$H $}  &
\colhead{$K_s $} \\
\colhead{[day]} &
\colhead{[day]} &
\colhead{[mag]}  &
\colhead{[mag]}  &
\colhead{[mag]}
 }
\startdata
54474.35 &    -0.21 & $>$   18.15 & $>$   17.12 & $>$   16.18\\
54475.27 &     0.71 &  17.00(03) &  16.70(05) &  17.09(12) \\
54476.33 &     1.77 &  16.54(07) &  16.22(09) &  16.22(10) \\
54477.34 &     2.78 &  16.45(07) &  16.15(09) &  16.13(15) \\
54478.34 &     3.78 &  16.44(04) &  16.05(09) &  15.99(12) \\
54479.38 &     4.82 &  16.38(10) &  15.96(10) &  15.81(10) \\
54480.34 &     5.78 &  16.23(10) &  15.86(10) &  15.63(07) \\
54481.32 &     6.76 &  15.99(10) &  15.72(07) &  15.55(02) \\
54482.26 &     7.70 &  15.88(04) &  15.55(00) &  15.46(05) \\
54483.32 &     8.76 &  15.74(08) &  15.41(13) &  15.21(06) \\
54484.35 &     9.79 &  15.62(05) &  15.31(05) &  15.11(07) \\
54485.31 &    10.75 &  15.51(02) &  15.17(16) &  15.05(13) \\
54486.35 &    11.79 &  15.43(01) &  15.12(05) &  14.92(05) \\
54497.19 &    22.63 &  15.00(04) &  14.76(08) &  14.53(03) \\
54499.26 &    24.70 &  15.07(21) &  14.73(03) &  14.44(04) \\
54505.20 &    30.64 &  15.11(05) &  14.76(02) &  14.54(11) \\
54508.24 &    33.68 &  15.25(02) &  14.86(01) &  14.68(13) \\
54509.22 &    34.66 &  15.24(05) &  14.88(03) &  14.68(05) \\
54510.25 &    35.69 &  15.32(04) &  14.89(06) &  14.68(03)
\enddata
\label{nir_table}
\end{deluxetable*}


\begin{deluxetable}{lccccccl}[!ht]
\tabletypesize{\scriptsize}
\tablewidth{0pt}
\tablecaption{Journal of Spectroscopic Observations of \snd }
\tablehead{ 
\colhead{UT Date} &
\colhead{$\Delta t$\tablenotemark{a}} &
\colhead{$t_{\rm{Vmax}}$\tablenotemark{b} } &
\colhead{Telescope\tablenotemark{c}} &
\colhead{Range\tablenotemark{d}}  &
\colhead{Airmass\tablenotemark{e}} & 
\colhead{Slit} &
\colhead{Exp.} \\
\colhead{} &
\colhead{[days]} &
\colhead{[days]} &
\colhead{} &
\colhead{[\AA]} &
\colhead{} &
\colhead{[$^{\prime\prime}$]} &
\colhead{[sec]} }
\startdata
   2008 Jan. 11.27 & 1.70 & $-$16.7   & GMOS-S &   4000$-$8500 & 2.25 & 0.75 & 3 $\times$600 \\
   2008 Jan. 11.41 & 1.84 & $-$16.6   & MMT    &   3200$-$8400 & 1.12 & 1.00 & 1200 \\
   2008 Jan. 12.36 & 2.80 & $-$15.6   & Lick   &   3300$-$8100 & 1.04 & 2.00 & 2$\times$ 1200 \\
   2008 Jan. 12.55 & 2.99 & $-$15.4   & MMT    &   3300$-$8100 & 1.62 & 1.00 & 2$\times$ 300 \\
   2008 Jan. 12.67 & 3.11 & $-$15.3   & Keck II  &   5000$-$10000 & 1.64 & 1.00 & 300 \\
   2008 Jan. 13.30 & 3.74 & $-$14.7   & MMT    &   3300$-$8100 & 1.08 & 1.00 & 1200 \\
   2008 Jan. 13.38 & 3.82 & $-$14.6   & Lick   &   3300$-$8100 & 1.01 & 2.00 & 2$\times$ 1200 \\
   2008 Jan. 14.36 & 4.79 & $-$13.6   & MMT    &   3300$-$8400 & 1.00 & 2.00 & 1200 \\
   2008 Jan. 14.40 & 4.84 & $-$13.6   & Lick   &   3300$-$8100 & 1.01 & 2.00 & 4$\times$ 1200 \\
   2008 Jan. 14.45 & 4.89 & $-$13.6   & IRTF   &   8000$-$25,000 & 1.07 & 0.5 & 10 $\times$ 150\\
   2008 Jan. 15.31 & 5.75 & $-$12.7   & MMT    &   3300$-$8400 & 1.05 & 3.50 & 1800 \\
   2008 Jan. 15.37 & 5.80 & $-$12.6   & Lick   &   3300$-$10700 & 1.02 & 2.00 & 2$\times$ 1800 \\
   2008 Jan. 16.34 & 6.77 & $-$11.7   & MMT    &   3300$-$6500 & 1.05 & 1.00 & 3$\times$ 900 \\
   2008 Jan. 18.57 & 9.00  &$-$9.4    & Lick   &   3500$-$10700 & 1.80 & 2.00 & 2$\times$ 1800 \\
   2008 Jan. 20.27 & 10.70 & $-$7.7   & APO    &   3500$-$5600 & 1.13 & 1.50 & 2$\times$ 1200 \\
   2008 Jan. 21.22 & 11.66 & $-$6.8   & APO    &   3300$-$9800 & 1.30 & 1.50 & 3$\times$ 1200 \\
   2008 Jan. 21.51 & 11.95 & $-$6.5   & IRTF   &   8000$-$25,000 & 1.04 & 0.5 &  11 $\times$ 150\\
   2008 Jan. 28.51 & 18.95 & 0.5      & IRTF   &   8000$-$25,000 & 1.07 & 0.5 & 11 $\times$ 150 \\
   2008 Feb. 01.35 & 22.78 & 4.3   & FLWO   &   3500$-$7400 & 1.00 & 3.00 &  2$\times$ 1800 \\
   2008 Feb. 02.44 & 23.88 & 5.4   & APO     &   3300$-$9800 & 1.25 & 1.50 & 2$\times$ 1200 \\
   2008 Feb. 09.12 & 30.56 & 12.1  & MMT     &   3300$-$8400 & 1.05 & 1.00 & 3$\times$ 900 \\
   2008 Feb. 11.18 & 32.61 & 14.2  & MMT     &   3300$-$8400 & 1.05 & 1.00 & 3$\times$ 900 \\
   2008 Feb. 12.43 & 33.89 & 15.4  & Keck    &   3100$-$9300 & 1.03 & 1.00 & 2$\times$ 900 \\
   2008 Feb. 16.36 & 37.79 & 19.4  & Lick    &  3300$-$10700 & 1.02 & 2.00 &  1800 \\
   2008 Feb. 27.08 & 48.52 & 30.1  & LCO     &   3800$-$10800 & 2.57 & 1.0 & 3$\times$300 \\
   2008 Feb. 28.37 & 49.81 & 31.4  & APO     &   3300$-$9800 & 1.24 & 1.50 & 2$\times$ 1200 \\
   2008 Mar. 10.23 & 60.67  & 42.2  & Keck    &   3100$-$9100 & 1.30 & 1.00 &  600 \\ 
   2008 Mar. 14.48 & 63.92 &  45.5 & IRTF    &   8000$-$25,000 & 1.54 & 0.5 & 5 $\times$ 150 \\
   2008 Apr. 28.33 & 109.77 & 91.4  & Keck    &   3100$-$9300 & 1.30 & 1.00 & 2$\times$ 400 \\
   2008 Jun. 07.26 & 149.69 & 132.4  & Keck    &   3100$-$9300 & 1.50 & 1.00 & 2$\times$ 900
\enddata
\tablecomments{This Table is published in its entirety in the electronic edition of the {\it Astrophysical Journal}. A portion is shown here for guidance regarding its form and content.}
\tablenotetext{a}{Days after X-ray outburst, $t_0 $= 2008-01-09 13:32:49 UT.}
\tablenotetext{b}{Days with respect to date of $V$-band maximum.}
\tablenotetext{c}{APO = 3.5m Telescope at Apache Point Observatory/DIS; FLWO = Tillinghast 1.5m/FAST; GMOS-S = Gemini-South 8.4m/GMOS; IRTF= IRTF 3m/SpeX; Keck I = Keck I 10m/LRIS; Keck II = Keck II 10m/DEIMOS; LCO = 6.5m Clay Magellan Telescope at Las Campanas Observatory/LDSS-3; Lick = Lick 3m/Tillinghast; MMT = 6.5m MMT/Bluechannel.}
\tablenotetext{d}{Airmass at the middle of the set of observations.}
\tablenotetext{e}{Observed wavelength range of spectrum. In some cases the blue and red ends of the spectrum are extremely noisy and are not shown in the figures.}
\label{spec_table}
\end{deluxetable}

\begin{deluxetable}{lcccc}
\tablewidth{0pt}
\tablecaption{Blackbody Fits and Bolometric Light Curves of SN~2008D from $UVW1 \rightarrow K_s$ Data}
\tablehead{\colhead{$\Delta t$} & 
\colhead{$T_{\rm{BB}}$} & 
\colhead{$R_{\rm{BB}}$} & 
\colhead{log($L_{\rm{BB}}$)} &
\colhead{log($L_{UVW1-K_s}$)} \\
\colhead{[days]} &
\colhead{[K]} &
\colhead{[$10^{14}$ cm]} &
\colhead{[\ergs]} &
\colhead{[\ergs]} 
}
\startdata
  0.80 &  12000$^{+  3000}_{-  2000}$ &   2.75$^{+  0.36}_{-  0.37}$ &  42.04$^{+  0.27}_{-  0.21}$ &  41.92$^{+  0.18}_{ -0.17}$ \\
  1.80 &  10000$^{+  1600}_{-  1200}$ &   3.78$^{+  0.30}_{-  0.30}$ &  42.05$^{+  0.19}_{-  0.17}$ &  42.00$^{+  0.17}_{ -0.16}$ \\
  2.80 &   8700$^{+  1300}_{-  1000}$ &   4.64$^{+  0.47}_{-  0.46}$ &  41.93$^{+  0.17}_{-  0.15}$ &  41.89$^{+  0.15}_{ -0.14}$ \\
  3.80 &   7900$^{+  1000}_{-   830}$ &   5.19$^{+  0.47}_{-  0.47}$ &  41.86$^{+  0.15}_{-  0.13}$ &  41.83$^{+  0.14}_{ -0.13}$ \\
  4.80 &   7600$^{+   970}_{-   780}$ &   5.62$^{+  0.52}_{-  0.50}$ &  41.87$^{+  0.14}_{-  0.13}$ &  41.85$^{+  0.14}_{ -0.12}$ \\
  5.80 &   7400$^{+   910}_{-   740}$ &   6.12$^{+  0.56}_{-  0.54}$ &  41.90$^{+  0.14}_{-  0.12}$ &  41.89$^{+  0.13}_{ -0.12}$ \\
  6.80 &   7200$^{+   850}_{-   690}$ &   6.92$^{+  0.61}_{-  0.60}$ &  41.95$^{+  0.13}_{-  0.12}$ &  41.93$^{+  0.13}_{ -0.12}$ \\
  7.80 &   7000$^{+   800}_{-   660}$ &   7.64$^{+  0.65}_{-  0.64}$ &  42.00$^{+  0.13}_{-  0.11}$ &  41.98$^{+  0.13}_{ -0.12}$ \\
  8.80 &   6900$^{+   770}_{-   630}$ &   8.30$^{+  0.69}_{-  0.69}$ &  42.04$^{+  0.13}_{-  0.11}$ &  42.02$^{+  0.12}_{ -0.11}$ \\
  9.80 &   6800$^{+   740}_{-   610}$ &   8.89$^{+  0.74}_{-  0.73}$ &  42.08$^{+  0.12}_{-  0.11}$ &  42.05$^{+  0.12}_{ -0.11}$ \\
 10.80 &   6700$^{+   720}_{-   590}$ &   9.45$^{+  0.78}_{-  0.76}$ &  42.10$^{+  0.12}_{-  0.11}$ &  42.08$^{+  0.12}_{ -0.11}$ \\
 11.80 &   6600$^{+   700}_{-   580}$ &   9.95$^{+  0.79}_{-  0.81}$ &  42.13$^{+  0.12}_{-  0.11}$ &  42.11$^{+  0.12}_{ -0.11}$ \\
 12.80 &   6600$^{+   680}_{-   570}$ &  10.40$^{+  0.84}_{-  0.83}$ &  42.15$^{+  0.12}_{-  0.10}$ &  42.13$^{+  0.12}_{ -0.11}$ \\
 13.80 &   6500$^{+   660}_{-   550}$ &  10.80$^{+  0.92}_{-  0.84}$ &  42.17$^{+  0.11}_{-  0.10}$ &  42.15$^{+  0.12}_{ -0.11}$ \\
 14.80 &   6400$^{+   650}_{-   540}$ &  11.20$^{+  0.94}_{-  0.86}$ &  42.18$^{+  0.11}_{-  0.10}$ &  42.17$^{+  0.12}_{ -0.11}$ \\
 15.80 &   6400$^{+   630}_{-   530}$ &  11.60$^{+  0.90}_{-  0.93}$ &  42.19$^{+  0.11}_{-  0.10}$ &  42.18$^{+  0.12}_{ -0.11}$ \\
 16.80 &   6300$^{+   620}_{-   520}$ &  11.90$^{+  0.97}_{-  0.89}$ &  42.20$^{+  0.11}_{-  0.10}$ &  42.19$^{+  0.11}_{ -0.11}$ \\
 17.80 &   6200$^{+   600}_{-   510}$ &  12.30$^{+  0.98}_{-  0.96}$ &  42.21$^{+  0.11}_{-  0.10}$ &  42.20$^{+  0.11}_{ -0.10}$ \\
 18.80 &   6200$^{+   590}_{-   500}$ &  12.60$^{+  1.00}_{-  0.97}$ &  42.21$^{+  0.11}_{-  0.10}$ &  42.20$^{+  0.11}_{ -0.10}$ \\
 19.80 &   6100$^{+   570}_{-   480}$ &  12.90$^{+  1.00}_{-  0.98}$ &  42.21$^{+  0.11}_{-  0.10}$ &  42.20$^{+  0.11}_{ -0.10}$ \\
 20.80 &   6000$^{+   560}_{-   470}$ &  13.20$^{+  1.00}_{-  0.99}$ &  42.21$^{+  0.10}_{-  0.10}$ &  42.20$^{+  0.11}_{ -0.10}$ \\
 21.80 &   6000$^{+   540}_{-   460}$ &  13.40$^{+  1.10}_{-  1.00}$ &  42.20$^{+  0.10}_{-  0.09}$ &  42.20$^{+  0.11}_{ -0.10}$ \\
 22.80 &   5900$^{+   530}_{-   450}$ &  13.70$^{+  1.00}_{-  1.00}$ &  42.19$^{+  0.10}_{-  0.09}$ &  42.19$^{+  0.11}_{ -0.10}$ \\
 23.80 &   5800$^{+   520}_{-   440}$ &  13.90$^{+  1.10}_{-  1.00}$ &  42.18$^{+  0.10}_{-  0.09}$ &  42.18$^{+  0.10}_{ -0.10}$ \\
 24.80 &   5700$^{+   500}_{-   430}$ &  14.20$^{+  1.10}_{-  1.10}$ &  42.17$^{+  0.10}_{-  0.09}$ &  42.16$^{+  0.10}_{ -0.10}$ \\
 25.80 &   5600$^{+   480}_{-   410}$ &  14.50$^{+  1.10}_{-  1.10}$ &  42.15$^{+  0.09}_{-  0.09}$ &  42.15$^{+  0.10}_{ -0.09}$ \\
 26.80 &   5400$^{+   460}_{-   400}$ &  14.80$^{+  1.20}_{-  1.10}$ &  42.13$^{+  0.09}_{-  0.08}$ &  42.13$^{+  0.10}_{ -0.09}$ \\
 27.80 &   5300$^{+   440}_{-   380}$ &  15.20$^{+  1.20}_{-  1.20}$ &  42.11$^{+  0.09}_{-  0.08}$ &  42.11$^{+  0.10}_{ -0.09}$ \\
 28.80 &   5200$^{+   420}_{-   360}$ &  15.50$^{+  1.20}_{-  1.20}$ &  42.09$^{+  0.09}_{-  0.08}$ &  42.09$^{+  0.10}_{ -0.09}$ \\
 29.80 &   5100$^{+   400}_{-   350}$ &  15.70$^{+  1.20}_{-  1.20}$ &  42.08$^{+  0.09}_{-  0.08}$ &  42.08$^{+  0.10}_{ -0.09}$ \\
 30.80 &   5100$^{+   400}_{-   350}$ &  15.80$^{+  1.20}_{-  1.20}$ &  42.06$^{+  0.08}_{-  0.07}$ &  42.07$^{+  0.09}_{ -0.09}$ \\

\enddata
\tablecomments{Adopted values: \ebvhost , \distd , and $t_0 $= 2008-01-09 13:32:49 UT. The uncertainties include the uncertainty in reddening, but not the systematic uncertainty in distance.}
\label{BBfit_table}
\end{deluxetable}



\begin{deluxetable}{lccccccc}
\tablewidth{0pt}
\tabletypesize{\scriptsize}
\singlespace
\tablecaption{Optical and Near-Infrared \He\ Absorption-Line Velocities of SN~2008D }
\tablehead{
\colhead{JD$-$2,400,000} & 
\colhead{$\Delta t$} & 
\colhead{$ t_{V\rm{max}}$} & 
\colhead{\HeFive} &
\colhead{\HeSix}  &
\colhead{\HeSeven} &
\colhead{\ion{He}{1} 1.0830 $\mu$m} &
\colhead{\ion{He}{1} 2.0581 $\mu$m} \\
\colhead{[day]} &
\colhead{[day]} &
\colhead{[day]} &
\colhead{[\kms]}  &
\colhead{[\kms]}  &
\colhead{[\kms]}  &
\colhead{[\kms]}  &
\colhead{[\kms]}
 }
\startdata

54479.96 &     4.89  &   $-$13.6 & $-$     & $-$     & $-$      &  $-$21,700\tablenotemark{a}  &    $-$14,200  \\
54480.82 &     5.75 &    $-$12.7 & $-$14200 & $-$11800 & $-$10500  & $-$ & $-$ \\
54481.84 &     6.77 &    $-$11.7 & $-$13300 & $-$12700 & $-$13000 & $-$ & $-$  \\
54484.07 &     9.00 &     $-$9.4 & $-$12300 & $-$12400 & $-$10900 & $-$ & $-$  \\
54486.72 &    11.66 &     $-$6.8 & $-$11500 & $-$11100 & $-$10500  & $-$ & $-$ \\
54487.01 &    11.95 &     $-$6.5 & $-$     & $-$     & $-$  &    $-$14,900  &    $-$12,200   \\
54494.01 &    18.95 &        0.5 & $-$     & $-$     & $-$  &    $-$13,900  &    $-$11,700   \\
54497.85 &    22.78 &      4.3 & $-$10600 & $-$10400 & $-$10100  & $-$ & $-$ \\
54498.94 &    23.88 &      5.4 & $-$10600 & $-$10200 & $-$10200  & $-$ & $-$ \\
54505.62 &    30.56 &     12.1 & $-$10500 & $-$10200 & $-$9900   & $-$ & $-$\\
54507.68 &    32.61 &     14.2 & $-$10400 & $-$10000 & $-$9900   & $-$ & $-$\\
54508.93 &    33.87 &     15.4 & $-$10300 & $-$9900 & $-$9700   & $-$ & $-$ \\
54512.86 &    37.79 &     19.4 & $-$10000 & $-$9600 & $-$9500   & $-$ & $-$ \\
54524.87 &    49.81 &     31.4 & $-$9300 & $-$8600 & $-$8700  & $-$ & $-$ \\
54535.73 &    60.67 &     42.2 & $-$8600 & $-$7900 & $-$8100  & $-$ & $-$ \\
54539.0  &    63.92  &    45.5 & $-$     & $-$     & $-$      & $-$13900  &   $-$10200           
\enddata
\tablenotetext{a}{Likely due to a blend and not traced by He I.}

\label{hevels_table}
\end{deluxetable}

\end{document}